\documentclass[A4paper,useAMS,usenatbib,fleqn]{mn2e}
\usepackage{graphicx}
\usepackage{amssymb}
\usepackage{times}
\usepackage{enumerate}
\usepackage{longtable}
\usepackage{color}

\title[Ultra-stripped SNe: progenitors and fate]{Ultra-stripped supernovae: progenitors and fate}

\author[Tauris, Langer \& Podsiadlowski]
{Thomas M. Tauris$^{1,2}$\thanks{E-mail: tauris@astro.uni-bonn.de},
Norbert Langer$^{1}$,
Philipp Podsiadlowski$^{3}$
\\
$^{1}$ Argelander-Institut f\"ur Astronomie, Universit\"at Bonn, Auf dem H\"ugel 71, D-53121 Bonn, Germany\\
$^{2}$ Max-Planck-Institut f\"ur Radioastronomie, Auf dem H\"ugel 69, D-53121 Bonn, Germany\\
$^{3}$ Department of Astronomy, Oxford University, Oxford OX1~3RH, UK
}
\let\leqslant=\leq 
\date{Accepted 2015 May 1}
\begin{document}

\maketitle

\begin{abstract}
The explosion of ultra-stripped stars in close binaries can lead to ejecta masses $< 0.1\;M_{\odot}$ and may explain 
some of the recent discoveries of weak and fast optical transients. 
In Tauris~et~al.~(2013), it was demonstrated that helium star companions to neutron stars (NSs) may experience mass transfer and evolve into
naked $\sim\!1.5\;M_{\odot}$ metal cores, barely above the Chandrasekhar mass limit. 
Here we elaborate on this work and present a systematic investigation of the progenitor evolution leading to ultra-stripped supernovae (SNe).
In particular, we examine the binary parameter space leading to electron-capture (EC~SNe) and iron core-collapse SNe (Fe~CCSNe), respectively,
and determine the amount of helium ejected with applications to their observational classification as Type~Ib or Type~Ic.
We mainly evolve systems where the SN progenitors are helium star donors of initial mass $M_{\rm He}=2.5-3.5\;M_{\odot}$ in tight binaries 
with orbital periods of $P_{\rm orb}=0.06-2.0\;{\rm days}$, and hosting an accreting NS,
but we also discuss the evolution of wider systems and of both more massive and lighter -- as well as single -- helium stars. 
In some cases we are able to follow the evolution until the onset of silicon burning, just a few days prior to the SN explosion.
We find that ultra-stripped SNe are possible for both EC~SNe and Fe~CCSNe.  
EC~SNe only occur for $M_{\rm He}=2.60-2.95\;M_{\odot}$ depending on $P_{\rm orb}$.
The general outcome, however, is an Fe~CCSN above this mass interval and an ONeMg or CO white dwarf for smaller masses. 
For the exploding stars, the amount of helium ejected is correlated with $P_{\rm orb}$ 
-- the tightest systems even having donors being stripped down to envelopes of less than $0.01\;M_{\odot}$. 
We estimate the rise time of ultra-stripped SNe to be in the range $12\;{\rm hr}-8\;{\rm days}$, and
light curve decay times between 1 and 50~days. 
A number of fitting formulae for our models are provided with applications to population synthesis.
Ultra-stripped SNe may produce NSs in the mass range $1.10-1.80\;M_{\odot}$ and are highly relevant for LIGO/VIRGO 
since most (possibly all) merging double NS systems have evolved through this phase. Finally, we discuss the low-velocity kicks
which might be imparted on these resulting NSs at birth. 
\end{abstract}

\begin{keywords}
supernovae: general ---  binaries: close --- X-rays: binaries --- stars: mass-loss --- stars: neutron --- white dwarfs 
\end{keywords}

\section{Introduction}\label{sec:intro}
The nature of a supernova (SN) in a close binary determines if the system will remain bound or dissolve into two runaway components after the explosion. 
The two main decisive factors are: 
  i) the kick velocity added to the newborn neutron star (NS) relative to the pre-SN orbital velocity of its progenitor star, and 
 ii) the amount of mass ejected during the SN.
For bound systems the post-SN orbital characteristics and systemic velocities depend on these quantities which are therefore
related to whether the explosion is an iron core-collapse SN (Fe~CCSN)
or an electron-capture SN (EC~SN). 
It has been demonstrated that Fe~CCSNe may produce kicks up to, and even above, $1000\;{\rm km\,s}^{-1}$
\citep{skjm06,jan12,wjm13} whereas EC~SNe are expected to be almost symmetric (i.e. fast explosions) and produce very small kicks of less 
than, perhaps, a few $10\;{\rm km\,s}^{-1}$ \citep[e.g.][]{plp+04,kjh06,dbo+06}. 

It is well known that the outcome of stellar evolution in close binaries differs significantly from that of single stars.
The main effects, besides from mass loss/gain, are changes in the stellar rotation rate, the nuclear burning scheme and
the wind mass-loss rate \citep{deM10,lan12}.
As a result, binary interactions affect the final core mass prior to the collapse \citep{bhl+01,plp+04},
and therefore possibly the nature of the SN event, the type of compact remnant left behind, the amount of envelope mass ejected, 
and thus the kinematics of the post-SN binary.
Given that almost all massive stars are members of close binaries \citep{chn+12} and 70~per~cent of them interact
with their companion star \citep{sdd+12}, it is important to probe the evolution leading to SNe
in close binaries.

In recent years, high-cadence surveys and dedicated SN searches have increased the 
discovery rate of unusual optical transients with low peak 
luminosities ($\la 10^{42}\;{\rm erg}\,{\rm s}^{-1}$) and/or rapidly decaying light curves.  
Examples comprise SN~2005ek \citep{dsm+13}, SN~2010X \citep{kkg+10} and SN~2005E \citep{pgm+10}.
These SNe show diverse spectroscopic signatures \citep[cf.][]{kk14} and thus 
their explanation may require a diversity of models. \citet{dsm+13} concluded that
SN~2005ek represents the smallest ratio of ejecta to remnant mass observed for a
core-collapse SN to date. \citet{tlm+13} recently suggested that ultra-stripped SNe may be
responsible for such events and demonstrated that 
synthetic light curves of these SNe, in which only $\sim\!0.1\;M_{\odot}$ of material
is ejected from an exploding star barely above the Chandrasekhar mass limit, are consistent with observations of SN~2005ek-like events.  

Furthermore, merging NS/black-hole binaries are the prime candidate sources of high frequency gravitational waves 
to be detected by the LIGO/VIRGO network within the next 2--3~yr \citep{aaa+13}.
In this connection, a search for their electromagnetic counterparts \citep[e.g.][]{kn14}
is of great interest. A good understanding of the formation of such systems, 
the nature of the SN explosion which created the last formed NS and its expected light curve, as well as the distribution of the resulting NS masses,
is therefore of uttermost importance. 

Most Type~Ib/Ic SNe are expected to originate in binary systems from the initially more massive star  
which has lost its hydrogen envelope by mass-transfer to its companion \citep{pjh92,eit08};  
these pre-SN stars typically have an envelope mass of $1\;M_{\odot}$ or more \citep{ywl10}. 
In close X-ray binaries, however, a second mass-transfer stage from a helium star to a NS 
can strip the helium star further prior to the SN \citep[][and references therein]{hab86a,dpsv02,dp03,ibk+03,tlm+13}. 
The fast decay of the Type~Ic SN~1994I with an absolute magnitude of $M_{\rm V}\approx -18$ was explained by \citet{nyp+94,smd+06} 
from such a model, for which they suggested a carbon-oxygen star progenitor with a mass of  
$\sim\!2\;M_{\odot}$ and a corresponding ejecta mass of $\sim\!0.9\;M_{\odot}$.
As mentioned above, an even more extreme case was demonstrated by \citet{tlm+13}, hereafter Paper~I, 
in which only $\sim 0.1\;M_{\odot}$ of material was ejected.
A similarly small ejecta mass was previously suggested by \citet{ps04,ps05} and \citet{pdl+05}
for the second formed pulsar in PSR~J0737$-$3039. 

A common envelope (CE) phase \citep[e.g.][]{ijc+13} resulting from high-mass X-ray binaries (HMXBs) is thought to have formed
the tight helium star--NS binaries, which are the starting point for the above-mentioned calculations, leading to a second SN explosion in these systems. 
For a general review on the formation and evolution of HMXBs, see e.g. \citet{tv06}.
Of particular interest for the scenario discussed in this paper is mass transfer by Roche-lobe overflow (RLO) initiated by the 
helium-star expansion after core helium depletion, i.e. so-called Case~BB RLO, first discussed by \citet{st76,dd77,dt81}.

Here we investigate the progenitor evolution in such close binaries leading to a SN event. The aim is to
determine which stellar systems lead to EC~SNe, which systems produce Fe~CCSN, and which systems
avoid a SN altogether and leave behind a massive white dwarf (WD) remnant.
We accomplish this by a systematic and detailed modelling of about 70 helium star--NS binaries in close orbits. 
In Section~\ref{sec:binary_evol}, we briefly refer to the stellar evolution code behind our
numerical computations, and in Section~\ref{sec:results} we present our models with detailed examples of four different outcomes.
We highlight a peculiar model in which extensive mass loss results in the complete quenching of the central nuclear burning. 
In Section~\ref{sec:obs}, we discuss the expected observational properties of our exploding stars
and in Section~\ref{sec:HeMdot} we discuss the mass-transfer rates from the helium star donors. 
In Section~\ref{sec:DNS}, we discuss the importance of ultra-stripped SNe for the formation of double NS systems 
and comment on their masses and imparted kicks. In addition, we estimate their merger times which are relevant for LIGO/VIRGO. 
Further discussions of our findings are given in Section~\ref{sec:discussions}, including a comparison with previous work and a mapping of
initial parameter space to final outcome. We summarize our conclusions in Section~\ref{sec:summary}, and in the Appendix
we analyse the timescale for NS in-spiral in the dilute envelope of a pre-SN star.

\section{Binary stellar evolution code and initial parameter space}\label{sec:binary_evol}

For our detailed calculations of the evolution of helium star--NS binaries, we applied the binary stellar evolution code BEC \citep[e.g.][and references therein]{ywl10}.
This code was originally developed by \citet{wlb01} on the basis of a single-star code \citep[][and references therein]{lan98}.
It is a one-dimensional implicit Lagrangian code which solves the hydrodynamic form of the stellar structure and evolution equations \citep{kw90,kyl14}.
The evolution of the donor star, the mass-transfer rate, and the orbital separation are computed simultaneously through an implicit coupling scheme 
\citep[see also][]{wl99} using the Roche approximation in the formulation of \citet{egg83}. 
To compute the mass-transfer rate, we use the prescription of \citet{rit88}. 
We employ the radiative opacities of \cite{ir96}, which we interpolated in tables as function of density, temperature, and chemical element mass fractions, 
including carbon and oxygen. For the electron conduction opacity, we follow \cite{hl69} in the non-relativistic case, and \cite{can70} in the relativistic case.
The stellar models are computed using extended nuclear networks \citep{hlw00}.

Our helium stars are calculated for a metallicity of $Z=0.02$ (mainly isotopes of C, N, O and Ne).
In our default models, we assumed a mixing-length parameter of $\alpha=l/H_{\rm p}=2.0$ and
treated semiconvection as in \citet{lfs83} adopting an efficiency parameter $\alpha _{\rm SEM}=1$ \citep{lan91}.
Convective core-overshooting was not applied to our models, because the convective core mass in
core helium burning stars increases as function of time, which results in a chemical discontinuity
at the upper edge of the convective core, thereby hindering mixing (cf. Section~\ref{sec:results}). 
We tested a number of models using $\alpha=l/H_{\rm p}=1.5$ \citep{lan91} which resulted
in only slightly smaller final metal-core masses (by $< 2\;{\rm per~cent}$). However, a smaller value of $\alpha$ 
did result in more stripping from donor stars in very wide orbits (cf. Section~\ref{subsec:mixing_length}).
In our computations, we did not include effects of X-ray irradiation of the donor star. Although X-ray irradiation may lead to a cyclic mass-transfer behavior 
(at least for low-mass, hydrogen-rich donor stars), previous studies show that these effects are relatively unimportant 
for the final donor mass and the orbital evolution of X-ray binaries with RLO \citep[e.g.][]{bdh15}.

For evolved helium stars, the typical mass-transfer rates are 3--4 orders of magnitude larger than the Eddington accretion 
limit of a NS ($\sim\!4\times 10^{-8}\;M_{\odot}\,{\rm yr}^{-1}$, see Section~\ref{sec:HeMdot}). This implies that 
$>99.9$~per~cent of the transferred material is lost from the system, presumably in the form of
a jet, or a wind from the disk, and we assume that it carries the specific orbital angular momentum of the NS, i.e. following
the so-called isotropic re-emission model \citep{bv91,tau96}. 
For recent applications of BEC to X-ray binaries and further description of Case~BB RLO, see e.g. \citet{tlk11,tlk12,tlm+13,ltk+14,itl14}.  

To probe the systems leading to Fe~CCSNe, EC~SNe and massive WDs, and to study the effect of mass transfer 
initiated at different evolutionary stages, we
investigated systems with helium star donors of initial masses of $M_{\rm He,i}=2.5-10.0\;M_{\odot}$ 
and initial orbital periods, $P_{\rm orb,i}=0.06-120\;{\rm days}$. For simplicity, we assumed in all cases an
initial mass of $M_{\rm NS}=1.35\;M_{\odot}$ for the accreting NS. 
We also evolved isolated helium stars for comparison with our binary stars. All configurations are summarized
in Table~\ref{table:models} (see Section~\ref{sec:results} for a description of the parameters).

The effect of stellar wind mass loss is less important for our results. The reason is that orbital dynamics, and more importantly,
the stripping of the donor star, is completely dominated by the Roche-lobe overflow. 
For a compilation of different mass-loss rates of Wolf-Rayet (naked helium) stars, see \citet{ywl10}.
Here we followed \citet{wl99} and applied the stellar wind mass-loss rates of \citet{hsh82} or \citet{hkw95},
depending on the luminosities of our relatively low-mass helium stars. 
For the lightest helium stars with an initial mass of $M_{\rm He,i}\le 3.0\;M_{\odot}$ we neglected wind mass loss in
most of our models, except for the widest systems. 
Applying different prescriptions for the wind mass-loss rate would in effect partly be mimicked by choosing a different
initial value of $P_{\rm orb}$.

\section{Results}\label{sec:results}
\begin{figure*}
\centering
\includegraphics[width=1.15\columnwidth,angle=-90]{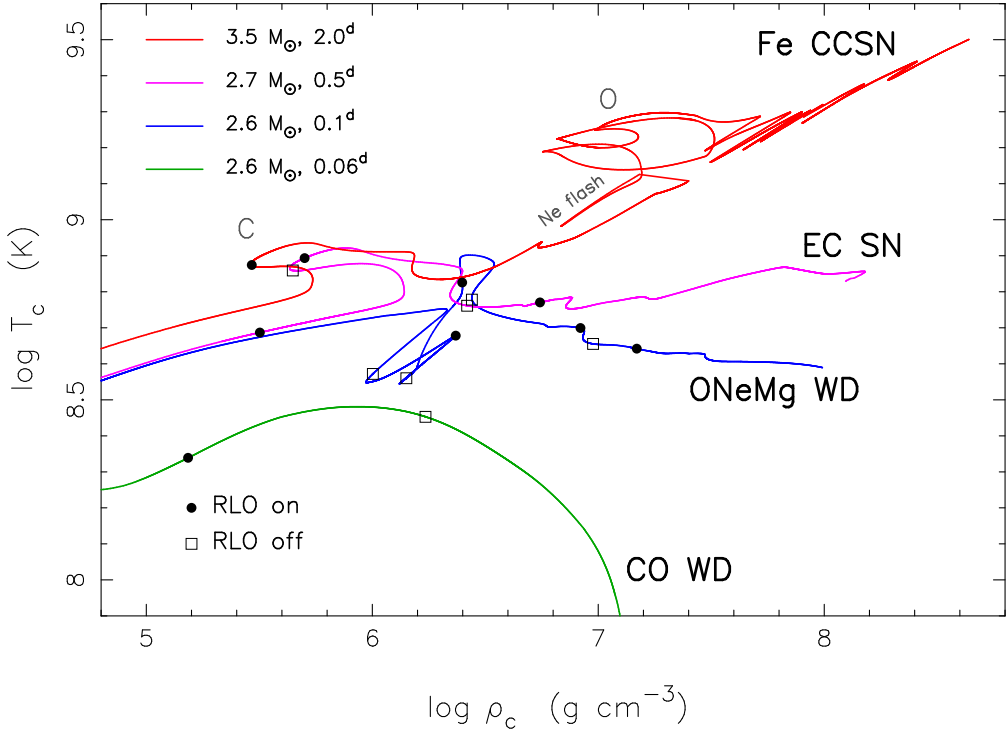}
\caption{
  Central temperature versus central density for four selected helium star--NS models which 
  terminate their evolution as an Fe~CCSN, EC~SN, ONeMg~WD and CO~WD, respectively, cf. Sections~\ref{subsec:FeCCSN}--\ref{subsec:COWD}.
  Carbon burning (marked by C) is seen in the first three cases, and oxygen burning
  (marked by O) occurs in the most massive star. 
  Epochs of RLO onset (solid circles) and RLO termination (open squares) are shown
  along each track. (Note, in some cases the first episode of mass transfer
  occurs for $\log \rho _c < 4.8$.)
  }
\label{fig:rho_c-T_c}
\end{figure*}
\begin{figure*}
\centering
\includegraphics[width=1.15\columnwidth,angle=-90]{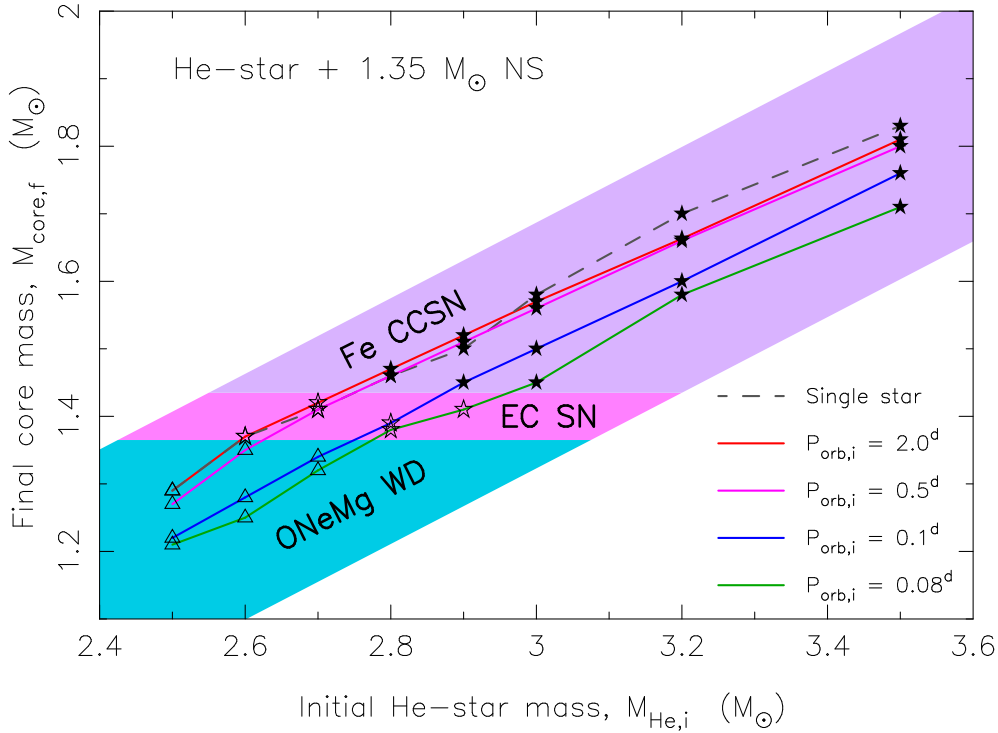}
\caption{
  Sequences of final core mass, $M_{\rm core,f}$ as a function of initial helium star mass, $M_{\rm He,i}$ for
  four different values of the initial orbital period. Also shown (dashed line) are the estimated final core masses
  for single helium stars.
  $M_{\rm core,f}$ is defined as the metal core (i.e. where the mass fraction of helium is less than 10~per~cent). 
  The coloured regions indicate the final destiny of the helium stars:
  light blue (with open triangle symbols) $\rightarrow$ ONeMg~WDs; pink (open stars) $\rightarrow$ EC~SNe; 
  and purple (solid stars) $\rightarrow$ Fe~CCSNe. In all cases the accreting companion star is (initially) a $1.35\;M_{\odot}$ NS. 
  }
\label{fig:Mcore_final}
\end{figure*}

In Fig.~\ref{fig:rho_c-T_c}, we show four examples of our modelling with very different outcomes.
By gradually changing the initial helium star mass, $M_{\rm He,i}$ from 2.6 to $3.5\;M_{\odot}$,
and the initial orbital period, $P_{\rm orb,i}$ from 0.06 to $2.0\;{\rm days}$, the outcome of the
resulting remnant varies from a CO~WD, or an ONeMg~WD, to a NS produced via an EC~SN or an Fe~CCSN.
In Sections~\ref{subsec:FeCCSN}--\ref{subsec:COWD} we discuss in detail the results of each of these four different models.
The final outcome of most of our calculations with $M_{\rm He,i}=2.5-3.5\;M_{\odot}$ is plotted in Fig.~\ref{fig:Mcore_final}.

\begin{table*}
\scriptsize
  \caption[]{Stellar and binary parameters of 68 helium star--NS systems, and 8 single helium stars, evolved to the pre-SN (or WD) stage -- see Section~\ref{sec:results}.}
  \begin{center}
    \begin{tabular}{rrrclllrccccrcl}
      \hline
      \noalign{\smallskip}
Wind & $M_{\rm He,i}$ & $P_{\rm orb,i}$ & RLO  & $M_{\rm core,f}$ & $M_{\rm He,f}^{\rm env}$ & $M_{\star\rm f}$ & $P_{\rm orb,f}$ & $\Delta M_{\rm NS}$ & $\Delta t_x$ & $|\dot{M}_{\rm He}^{\rm max}|$        & $|\dot{M}_{\rm f}|$           & final fate & $\tau _{\rm grw}$ & comment \\
     & $(M_{\odot})$  & (days)          & Case & $(M_{\odot})$    & $(M_{\odot})$            & $(M_{\odot})$    & (days)          & $(M_{\odot})$  & (yr)         & $(M_{\odot}\;{\rm yr}^{-1})$          & $(M_{\odot}\;{\rm yr}^{-1})$  &            & (Myr)             & \\
      \noalign{\smallskip}
      \hline
      \noalign{\smallskip}
yes  & 2.5 & --  & -- & 1.29  & 1.099 & 2.41  & --    & --     &  --    & --     & --        & ONeMg  & --   & single star\\
yes  & 2.6 & --  & -- & 1.37  & 0.961 & 2.37  & --    & --     &  --    & --     & --        & EC-SN  & --   & single star\\
yes  & 2.7 & --  & -- & 1.41  & 1.010 & 2.46  & --    & --     &  --    & --     & --        & EC-SN  & --   & single star\\
yes  & 2.8 & --  & -- & 1.46  & 0.984 & 2.49  & --    & --     &  --    & --     & --        & FeCCSN & --   & single star\\
yes  & 2.9 & --  & -- & 1.50  & 0.982 & 2.54  & --    & --     &  --    & --     & --        & FeCCSN & --   & single star\\
yes  & 3.0 & --  & -- & 1.58  & 0.970 & 2.60  & --    & --     &  --    & --     & --        & FeCCSN & --   & single star\\
yes  & 3.2 & --  & -- & 1.70  & 0.973 & 2.72  & --    & --     &  --    & --     & --        & FeCCSN & --   & single star\\
yes  & 3.5 & --  & -- & 1.83  & 1.003 & 2.91  & --    & --     &  --    & --     & --        & FeCCSN & --   & single star\\
yes  & 3.0 &120  & -- & 1.58  & 0.968 & 2.59  & 145.5 & --     &  --    & --     & --        & FeCCSN & 1.e11& no RLO at all\\
yes  & 3.0 &100  & BC & 1.57  & 0.750 & 2.38  & 109.7 & 4.8e-5 &  1.2e3 & 2.7e-4 &  2.7e-4   & FeCCSN & 5.e10 & \\
yes  & 3.0 & 80  & BC & 1.57  & 0.736 & 2.37  & 87.1  & 5.2e-5 &  1.3e3 & 2.6e-4 &  1.4e-4   & FeCCSN & 3.e10& \\
yes  & 3.0 & 50  & BC & 1.58  & 0.689 & 2.32  & 53.2  & 6.3e-5 &  1.5e3 & 2.7e-4 &  1.1e-3   & FeCCSN & 6.6e9& \\
yes  & 2.5 & 20  & BC & 1.29  & 0.137 & 1.45  & 20.3  & 6.1e-4 &  1.4e4 & 2.2e-4 &  6.7e-5   & ONeMg  & 1.1e8& \\
yes  & 2.6 & 20  & BC & 1.37  & 0.189 & 1.56  & 19.3  & 6.3e-4 &  1.5e4 & 8.8e-5 &  8.1e-5   & EC-SN  & 1.9e8& \\
yes  & 2.7 & 20  & BC & 1.42  & 0.150 & 1.59  & 19.7  & 4.5e-4 &  1.1e4 & 2.2e-4 &  7.7e-5   & EC-SN  & 1.9e8& \\
yes  & 2.8 & 20  & BC & 1.47  & 0.217 & 1.72  & 19.4  & 3.2e-4 &  7.8e3 & 3.0e-4 &  detached & FeCCSN & 2.1e8& \\
yes  & 2.9 & 20  & BC & 1.52  & 0.342 & 1.91  & 19.3  & 2.2e-4 &  5.4e3 & 3.2e-4 &  3.9e-8   & FeCCSN & 2.6e8& \\
yes  & 3.0 & 20  & BC & 1.58  & 0.548 & 2.17  & 20.0  & 1.4e-4 &  3.5e3 & 3.0e-4 &  8.1e-5   & FeCCSN & 3.8e8& \\
yes  & 3.2 & 20  & BC & 1.70  & 0.919 & 2.67  & 23.8  & 2.8e-5 &  7.0e2 & 1.6e-4 &  1.6e-4   & FeCCSN & 1.0e9& \\
yes  & 3.5 & 20  & -- & 1.83  & 1.007 & 2.91  & 25.7  & --     &  --    & --     &  --       & FeCCSN & 1.4e9& no RLO at all\\ 
yes  & 3.0 & 10  & BC & 1.57  & 0.375 & 2.00  & 9.51  & 2.3e-4 &  5.6e3 & 2.9e-4 &  4.e-10   & FeCCSN & 4.1e7& \\
yes  & 3.0 & 5.0 & BC & 1.57  & 0.246 & 1.86  & 4.61  & 4.1e-4 &  1.0e4 & 2.0e-4 &  detached & FeCCSN & 4.7e6& \\
no   & 2.5 & 2.0 & BC & 1.29  & 0.093 & 1.40  & 1.94  & 6.8e-4 &  2.2e4 & 2.5e-4 &  1.5e-5   & ONeMg  & 2.2e5& \\
no   & 2.6 & 2.0 & BC & 1.37  & 0.079 & 1.46  & 1.78  & 8.3e-4 &  2.2e4 & 2.9e-4 &  3.8e-5   & EC-SN  & 2.8e5& \\
no   & 2.7 & 2.0 & BC & 1.42  & 0.092 & 1.53  & 1.62  & 6.9e-4 &  1.8e4 & 2.5e-4 &  3.7e-6   & EC-SN  & 2.2e5& \\
no   & 2.8 & 2.0 & BC & 1.47  & 0.121 & 1.62  & 1.47  & 6.2e-4 &  1.6e4 & 2.0e-4 &  4.2e-5   & FeCCSN & 1.8e5& \\
no   & 2.9 & 2.0 & BC & 1.52  & 0.152 & 1.72  & 1.34  & 5.5e-4 &  1.4e4 & 1.7e-4 &  3.9e-5   & FeCCSN & 1.5e5& \\
no   & 3.0 & 2.0 & BC & 1.57  & 0.176 & 1.80  & 1.23  & 4.8e-4 &  1.2e4 & 1.8e-4 &  2.8e-6   & FeCCSN & 1.2e5& \\
yes  & 3.2 & 2.0 & BC & 1.66  & 0.244 & 1.98  & 1.71  & 4.0e-4 &  9.7e3 & 1.5e-4 &  7.7e-8   & FeCCSN & 3.4e5& \\
yes  & 3.5 & 2.0 & BC & 1.81  & 0.488 & 2.39  & 1.78  & 2.6e-4 &  6.5e3 & 1.9e-4 &  1.1e-7   & FeCCSN & 5.4e5& Section~\ref{subsec:FeCCSN}\\
yes  & 3.7 & 2.0 & BC & 1.93  & 0.754 & 2.75  & 2.10  & 1.8e-4 &  4.5e3 & 1.6e-4 &  1.5e-4   & FeCCSN & 1.1e6& \\
yes  & 4.0 & 2.0 & BC & 2.02  & 0.850 & 2.95  & 2.45  & 1.5e-4 &  3.6e3 & 1.0e-4 &  8.9e-5   & FeCCSN & 1.9e6& \\
yes  & 5.0 & 2.0 & -- & 2.19  & 0.861 & 3.15  & 3.91  & --     &  --    & --     &  --       & FeCCSN & 6.4e6& no RLO at all\\
no   & 3.2 & 2.0 & BC & 1.70  & 0.232 & 1.98  & 1.05  & 3.5e-4 &  8.7e3 & 2.2e-4 &  4.7e-5   & FeCCSN & 8.5e4& \\
no   & 3.5 & 2.0 & BC & 1.91  & 1.00  & 2.99  & 1.30  & 1.8e-4 &  4.4e3 & 2.8e-4 &  1.6e-3   & FeCCSN & 4.5e5& \\
no   & 2.5 & 0.5 & BB & 1.27  & 0.053 & 1.41  & 0.500 & 1.0e-3 &  4.2e4 & 1.0e-4 &  2.3e-5   & ONeMg  & 5960 & \\
no   & 2.6 & 0.5 & BB & 1.35  & 0.003 & 1.35  & 0.475 & 9.0e-4 &  3.9e4 & 1.2e-4 &  2.8e-4   & ONeMg  & 4940 & \\
no   & 2.7 & 0.5 & BB & 1.41  & 0.048 & 1.48  & 0.415 & 8.4e-4 &  3.2e4 & 1.4e-4 &  7.7e-5   & EC-SN  & 5430 & Section~\ref{subsec:ECSN}\\
no   & 2.8 & 0.5 & BB & 1.46  & 0.072 & 1.56  & 0.376 & 8.2e-4 &  2.9e4 & 1.6e-4 &  8.1e-4   & FeCCSN & 4350 & \\
no   & 2.9 & 0.5 & BB & 1.51  & 0.090 & 1.64  & 0.341 & 7.8e-4 &  2.6e4 & 1.8e-4 &  1.4e-7   & FeCCSN & 3480 & \\
no   & 3.0 & 0.5 & BB & 1.56  & 0.112 & 1.73  & 0.311 & 7.2e-4 &  2.2e4 & 1.8e-4 &  6.6e-5   & FeCCSN & 2880 & \\
yes  & 3.2 & 0.5 & BB & 1.66  & 0.142 & 1.86  & 0.387 & 6.1e-4 &  1.7e4 & 1.7e-4 &  detached & FeCCSN & 5270 & \\
yes  & 3.5 & 0.5 & BB & 1.80  & 0.183 & 2.07  & 0.365 & 5.0e-4 &  1.2e4 & 2.0e-4 &  2.4e-5   & FeCCSN & 4860 & \\
yes  &10.0 & 0.5 & BC & 2.17  & 0.446 & 3.02  & 2.94  & 2.7e-5 &  9.3e2 & 5.7e-5 &  2.4e-8   & FeCCSN & 2.6e6& \\
no   & 3.2 & 0.5 & BB & 1.68  & 0.134 & 1.86  & 0.261 & 6.3e-4 &  1.8e4 & 2.7e-4 &  9.3e-6   & FeCCSN & 1770 & \\
no   & 3.5 & 0.5 & BB & 1.88  & 0.203 & 2.15  & 0.207 & 4.9e-4 &  1.2e4 & 3.9e-4 &  1.0e-4   & FeCCSN & 1050 & \\
no   & 2.5 & 0.1 & BB & 1.22  & 0.015 & 1.24  & 0.108 & 4.2e-3 &  1.2e5 & 4.1e-5 &  2.1e-5   & ONeMg  &  104 & \\
no   & 2.6 & 0.1 & BB & 1.28  & 0.004 & 1.29  & 0.098 & 3.6e-3 &  1.1e5 & 5.3e-5 &  1.1e-4   & ONeMg  & 76.8 & Section~\ref{subsec:ONeMgWD}\\
no   & 2.7 & 0.1 & BB & 1.34  & 0.004 & 1.35  & 0.088 & 3.1e-3 &  9.2e4 & 6.6e-5 &  7.4e-4   & ONeMg  & 55.5 & \\
no   & 2.8 & 0.1 & BB & 1.39  & 0.018 & 1.43  & 0.079 & 2.7e-3 &  8.2e4 & 8.4e-5 &  8.4e-3   & EC-SN  & 61.9 & \\
no   & 2.9 & 0.1 & BB & 1.45  & 0.033 & 1.50  & 0.070 & 2.1e-3 &  7.3e4 & 1.1e-4 &  1.1e-2   & FeCCSN & 44.8 & Paper~I\\
no   & 3.0 & 0.1 & BB & 1.50  & 0.040 & 1.58  & 0.063 & 2.0e-3 &  6.5e4 & 1.3e-4 &  1.9e-2   & FeCCSN & 35.3 & \\
yes  & 3.2 & 0.1 & BB & 1.60  & 0.054 & 1.71  & 0.056 & 1.7e-3 &  5.3e4 & 1.7e-4 &  4.6e-4   & FeCCSN & 26.5 & \\
yes  & 3.5 & 0.1 & BB & 1.76  & 0.077 & 1.88  & 0.048 & 1.2e-3 &  4.1e4 & 2.2e-4 &  detached & FeCCSN & 17.2 & \\
yes  & 3.7 & 0.1 & BB & 1.86  & 0.079 & 2.00  & 0.047 & 1.1e-3 &  3.4e4 & 2.2e-4 &  detached & FeCCSN & 16.4 & \\
yes  & 4.0 & 0.1 & BB & 1.96  & 0.094 & 2.12  & 0.048 & 9.4e-4 &  2.9e4 & 2.1e-4 &  detached & FeCCSN & 17.6 & \\
yes  & 5.0 & 0.1 & BB & 2.10  & 0.112 & 2.36  & 0.062 & 6.6e-4 &  1.9e4 & 1.1e-4 &  6.0e-3   & FeCCSN & 39.5 & \\
yes  & 6.0 & 0.1 & BB & 2.15  & 0.109 & 2.37  & 0.064 & 6.7e-4 &  1.9e4 & 1.0e-4 &  3.0e-5   & FeCCSN & 40.2 & \\
no   & 3.2 & 0.1 & BB & 1.61  & 0.057 & 1.71  & 0.052 & 1.7e-3 &  5.4e4 & 2.0e-4 &  4.6e-4   & FeCCSN & 21.3 & \\
no   & 3.5 & 0.1 & BB & 1.81  & 0.074 & 1.93  & 0.039 & 1.0e-3 &  3.9e4 & 3.9e-4 &  detached & FeCCSN & 9.76 & \\
no   & 2.5 & 0.08& BB & 1.21  & 0.017 & 1.24  & 0.085 & 4.6e-3 &  1.3e5 & 4.7e-5 &  detached & ONeMg  & 55.1 & \\
no   & 2.6 & 0.08& BB & 1.25  & 0.097 & 1.27  & 0.078 & 4.2e-3 &  1.2e5 & 5.9e-5 &  3.4e-5   & ONeMg  & 42.6 & \\
no   & 2.7 & 0.08& BB & 1.32  & 0.005 & 1.32  & 0.071 & 3.6e-3 &  1.1e5 & 7.5e-5 &  1.9e-4   & ONeMg  & 31.7 & \\
no   & 2.8 & 0.08& BB & 1.38  & 0.005 & 1.38  & 0.064 & 2.9e-3 &  9.1e4 & 9.6e-5 &  2.8e-3   & EC-SN  & 32.7 & \\
no   & 2.9 & 0.08& BB & 1.41  & 0.016 & 1.45  & 0.057 & 2.6e-3 &  8.5e4 & 1.2e-4 &  3.2e-4   & EC-SN  & 25.7 & \\
no   & 3.0 & 0.08& BB & 1.45  & 0.021 & 1.49  & 0.052 & 2.3e-3 &  8.0e4 & 1.6e-4 &  6.0e-2   & FeCCSN & 19.9 & \\
yes  & 3.2 & 0.08& BB & 1.58  & 0.046 & 1.68  & 0.043 & 1.6e-3 &  5.9e4 & 2.3e-4 &  2.3e-3   & FeCCSN & 12.9 & \\
yes  & 3.5 & 0.08& BB & 1.71  & 0.055 & 1.81  & 0.035 & 8.5e-4 &  5.1e4 & 4.3e-4 &  detached & FeCCSN & 7.23 & \\
no   & 2.5 & 0.06& BA & 0.80  & 0.036 & 0.87  & 0.102 & 3.3e-2 &  5.1e7 & 5.9e-5 &  detached & CO     &  127 & \\
no   & 2.6 & 0.06& BA & 0.76  & 0.045 & 0.84  & 0.104 & 2.5e-2 &  4.9e7 & 7.9e-5 &  detached & CO     &  140 & Section~\ref{subsec:COWD}\\
no   & 2.7 & 0.06& BA & 0.67  & 0.064 & 0.78  & 0.118 & 1.3e-3 &  6.0e4 & 1.1e-4 &  detached & CO     &  220 & \\
no   & 2.8 & 0.06& BA & 0.0   & 1.79  & 2.22  & ?     & 1.1e-3 &  4.7e4 & 1.6e-4 &  2.3e-4   & CE(?)  & --   & fate?\\
no   & 2.9 & 0.06& BA & 0.0   & 1.88  & 2.30  & ?     & 9.3e-4 &  4.3e4 & 2.3e-4 &  9.5e-4   & CE(?)  & --   & fate?\\ 
no   & 3.0 & 0.06& BA & 0.0   & 2.13  & 2.30  & ?     & 8.0e-4 &  3.9e4 & 2.6e-4 &  4.4e-4   & CE(?)  & --   & fate?\\
yes  & 3.2 & 0.06& BA & 0.0   & --    & 2.67  & --    & 6.7e-4 &  3.4e4 & 3.6e-4 &  988.     & CE     & --   & merger\\
yes  & 3.5 & 0.06& BA & 0.0   & --    & 3.06  & --    & 5.2e-4 &  2.8e4 & 1.1e-3 &  1004.    & CE     & --   & merger\\
      \noalign{\smallskip}
      \hline
    \end{tabular}
    \end{center}
  \label{table:models}
\end{table*}

In Table~\ref{table:models}, we list all relevant parameters for the outcome of each calculated system.
The parameters listed are the following: 
$M_{\rm He,i}$ is the initial mass of the naked helium star; 
$P_{\rm orb,i}$ is the initial orbital period (always assuming a $1.35\;M_{\odot}$ NS companion);
The RLO Case BA, BB or BC indicates if mass transfer is initiated during core helium burning, helium shell burning, or beyond;
$M_{\rm core,f}$ is the final mass of the metal core of the helium star prior to the SN or formation of a WD;
$M_{\rm He,f}^{\rm env}$ is the final mass of helium in the envelope;
$M_{\star \rm f}$ is the final total mass of the helium star;
$P_{\rm orb,f}$ is the final orbital period prior to the SN or formation of a WD;
$\Delta M_{\rm NS}$ is the amount of material accreted by the $1.35\;M_{\odot}$ NS (assuming Eddington-limited accretion, cf. Section~\ref{sec:HeMdot}); 
$\Delta t_x$ is the total duration of the X-ray phase, including detached epochs in-between;
$|\dot{M}_{\rm He}^{\rm max}|$ is the maximum mass-loss rate of the helium star donor during stable Case~BA, Case~BB or Case~BC RLO;
$|\dot{M}_{\rm f}|$ is the mass-transfer rate in our last calculated model.
The last three columns yield the final fate of the helium star, the merger time of the resulting double degenerate system (Section~\ref{subsec:DNSmerger}) 
and additional comments, respectively. 
The very first column to the left indicates whether or not wind mass loss is included in the evolution of the helium star.

\subsection{Evolution leading to an iron core-collapse SN}\label{subsec:FeCCSN}
\begin{figure}
\centering
\includegraphics[width=1.08\columnwidth]{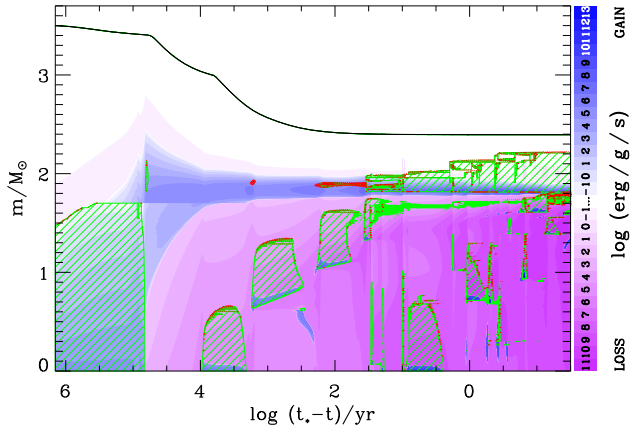}
\caption{
  Kippenhahn diagram of the $3.5\,M_{\odot}$ helium star plotted in Fig.~\ref{fig:rho_c-T_c} and 
  leading to the final structure shown in the upper panel of Fig.~\ref{fig:abundances}.
  The plot shows cross-sections of the helium star in mass coordinates
  from the center to the surface of the star, along the $y$-axis, as a function of stellar age on the $x$-axis.
  The value $(t_{*}-t)$ is the remaining time of our calculations, spanning a total time of $t_*= 1.433~{\rm Myr}$.  
  The green hatched areas denote zones with convection; red colour indicates semiconvection.
  The intensity of the blue/purple colour indicates the net energy-production rate.
  Carbon burning was ignited at $\log (t_{*}-t)\approx 4.0$ and oxygen burning was ignited at $\log (t_{*}-t)\approx 1.0$.
  Our calculations ended just before silicon ignition, a few days prior to the gravitational collapse.
 }
\label{fig:FeCCSN-kippenhahn}
\end{figure}

Using our stellar evolution code, we were able to evolve the helium stars until the onset of silicon burning,
a few days prior to the gravitational core collapse. In Fig.~\ref{fig:FeCCSN-kippenhahn}, we have plotted as an example
the Kippenhahn diagram for the evolution of an initial $3.5\;M_{\odot}$ helium star with $P_{\rm orb,i}=2.0\;{\rm days}$.
The mass loss of the helium star is initially caused only by its stellar wind (cf. Section~\ref{sec:binary_evol}).
At an age of $t=1.426\;{\rm Myr}$ (at the second hump in the black curve, shortly after the onset of central carbon burning), the expanding helium star fills its Roche lobe
with a radius of $R=5.68\;R_{\odot}$ and initiates mass transfer (Case~BC RLO) to the NS at a rate of $\sim\!10^{-4}\;M_{\odot}\;{\rm yr}^{-1}$ (see Section~\ref{sec:HeMdot}). 
At this stage the remaining stellar lifetime of our calculation is $\log(t_\star-t) \simeq 3.75$, i.e. some $5600\;{\rm yr}$
prior to the core collapse. The preceding hump seen in the total mass profile of the donor star at $\log(t_\star-t) \simeq 4.70$
is caused by a sudden change in the calculated wind mass-loss rate when the luminosity of the star increases
significantly following the depletion of core helium burning.
After a sequence of carbon shell burning stages, central oxygen burning is ignited about 10~yr prior
to core collapse (when $\log(t_\star-t) \simeq 1.0$).
During the subsequent oxygen shell burning stages, silicon is efficiently mixed via convection out to a mass coordinate of $\sim\!1.55\;M_{\odot}$.
This is clearly seen in Fig.~\ref{fig:abundances} (upper panel) where we have plotted the chemical abundance profile of our final stellar model. 
At this stage (a few days prior to core collapse during which silicon burning would take place, producing an iron core)
the exploding star has a total mass of $M_{\rm \star,f}=2.39\;M_{\odot}$ with a metal core of $M_{\rm core,f}=1.81\;M_{\odot}$.

In the example illustrated here, there is no doubt that the final fate of the donor star is an Fe~CCSN. 
However, in other cases it is less trivial to predict the nature of the SN explosion. This is the case for 
stars with a final metal core mass close to the Chandrasekhar mass.
We could not evolve our stars until the very onset of the collapse.
However, as argued in Paper~I, we can compare our evolutionary tracks in the ($\rho_c,\,T_c$)--plane  
with those available in the literature for isolated massive stars, e.g. \citet{uyt12} and \citet{jhn+13}.
As a simple rule of thumb, we find that evolutionary tracks where the post-carbon burning central temperature, $T_c$ rises
above the value of $T_c$ during carbon burning, will eventually ignite oxygen (typically at $T_c \ga 10^9\;{\rm K}$, depending on $\rho_c$), 
and later burn silicon to produce an iron core and finally undergo an Fe~CCSN. This is the case for our models
which have a metal core, $M_{\rm core,f}>1.43\;M_{\odot}$. We therefore adopt this mass as an approximate threshold limit
separating Fe~CCSNe from EC~SNe. 

For low-mass metal cores the exact boundary between Fe~CCSNe and EC~SNe depends on the location of ignition of the off-centre  
neon and oxygen burning \citep{twt94}, and also on the propagation of the neon--oxygen flame which is sensitive to mixing 
in the convective layers across the flame front \citep{jhn14}.
The latter process may affect the electron fraction and the density in the central region, and thereby the electron-capture efficiency, 
and thus the nature of the SN.

\subsection{Evolution leading to an electron-capture SN}\label{subsec:ECSN}
For stars which do not ignite oxygen, but for which their ONeMg cores are more massive than $1.37\;M_{\odot}$,  
the final fate is an EC~SN \citep{nom87,plp+04,tyu13}.
We find that ten of our models (including six with $P_{\rm orb}\le 2.0\;{\rm days}$, cf. Fig.~\ref{fig:Mcore_final}) 
produce EC~SNe. All these stars have metal cores with a mass, $M_{\rm core,f}\simeq 1.37-1.43\;M_{\odot}$.

In the central panel of Fig.~\ref{fig:abundances}, we have plotted the chemical abundance profile of our final stellar model resulting from a helium star donor 
with $M_{\rm He,i}=2.7\;M_{\odot}$ and $P_{\rm orb,i}=0.5\;{\rm days}$.
The plot shows the chemical composition shortly before the loss of pressure support, due to inverse $\beta$-decay, leading to the EC~SN event.
In this particular example, the exploding star has an envelope containing less than $5\times 10^{-2}\;M_{\odot}$ of helium.
Therefore, this EC~SN would most likely be classified as a Type~Ic SN, see discussion in Section~\ref{subsec:spectra}.

\subsection{Evolution leading to an ONeMg~WD}\label{subsec:ONeMgWD}
\begin{figure}
\centering
\includegraphics[width=1.00\columnwidth,angle=0]{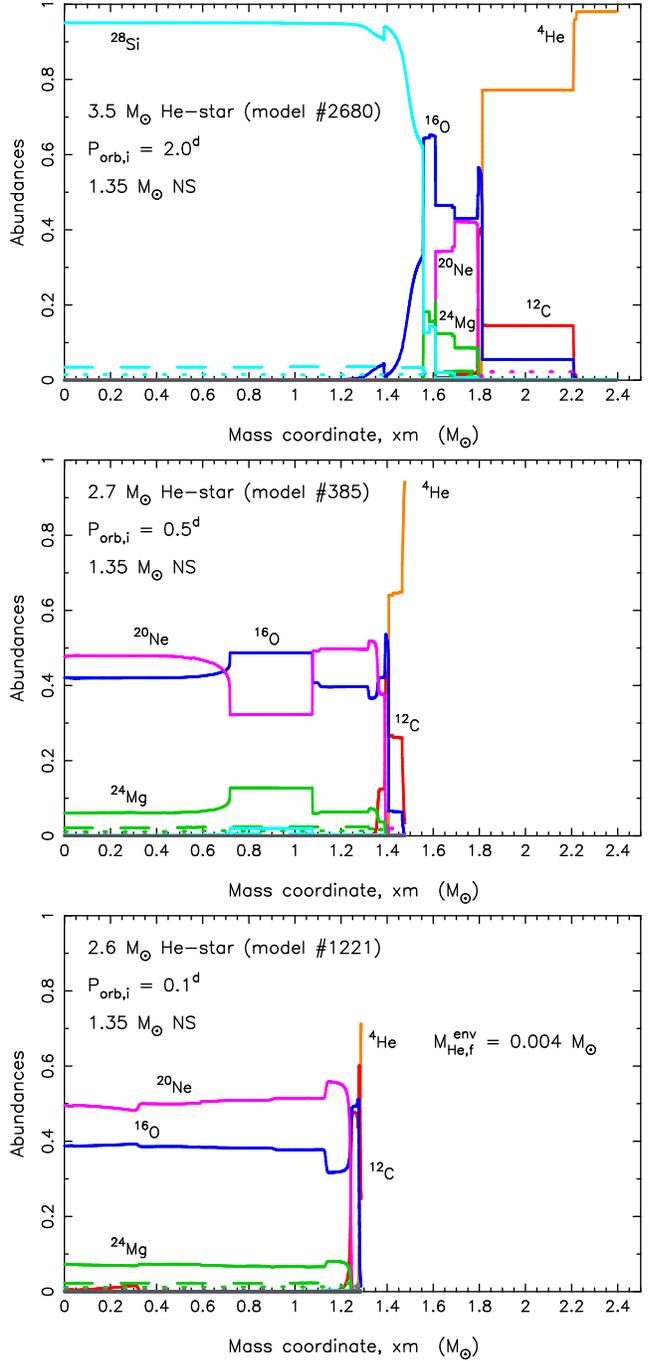}
\caption{
  Final chemical abundance structure of three evolved helium stars. 
  {\it Upper panel:} a helium star with an initial mass of $3.5\;M_{\odot}$ placed in a binary 
  with a $1.35\;M_{\odot}$ NS and an orbital period of 2.0~days. The final mass of the exploding star is
  $2.39\;M_{\odot}$, with a metal core of $1.81\;M_{\odot}$, leading to an Fe~CCSN of Type~Ib. 
  Our last calculated model \#2680 is just before the onset of silicon burning, a few days prior to the explosion, at  
  a helium star age of $t=1.433\;{\rm Myr}$.
  {\it Central panel:} a helium star with an initial mass of $2.7\;M_{\odot}$ placed in a binary 
  with a $1.35\;M_{\odot}$ NS and an orbital period of 0.5~days. After Case~BB RLO the final mass of the exploding star is
  $1.48\;M_{\odot}$, with a metal core of $1.41\;M_{\odot}$, leading to an EC~SN, probably of Type~Ic (see text). 
  Three magnesium isotopes ($^{24}$Mg,~$^{25}$Mg,~$^{26}$Mg) are plotted as a solid, dashed and a dotted green line, respectively, 
  yielding a total mass fraction of 10.9~per~cent.
  {\it Lower panel:} a helium star with an initial mass of $2.6\;M_{\odot}$ placed in a binary 
  with a $1.35\;M_{\odot}$ NS and an orbital period of 0.1~days. The final remnant is an almost pure
  ONeMg WD with a mass of $1.29\;M_{\odot}$.
  The total mass fraction of Mg is 10.2~per~cent. 
  }
\label{fig:abundances}
\end{figure}

In the lower panel of Fig.~\ref{fig:abundances}, we have plotted the chemical abundance profile of a stripped donor star 
whose progenitor was a $2.6\;M_{\odot}$ helium star with $P_{\rm orb,i}=0.1\;{\rm days}$.
Following Case~BB RLO, the final mass is $1.29\;M_{\odot}$ and this star will leave behind an ONeMg~WD. 
It is interesting to note the pureness of this ONeMg WD. Almost the entire WD consists of only three elements: 
oxygen (38.0~per~cent), neon (48.9~per~cent) and magnesium (10.2~per~cent). It is surrounded with a tiny envelope of 
less than $4\times 10^{-3}\;M_{\odot}$ helium and $<3\times 10^{-2}\;M_{\odot}$ carbon.
If such a WD were to undergo pulsations \citep{wk08}, one might be able to verify its structure observationally.

\subsection{Evolution leading to a CO~WD}\label{subsec:COWD}
\begin{figure}
\centering
\includegraphics[width=1.08\columnwidth]{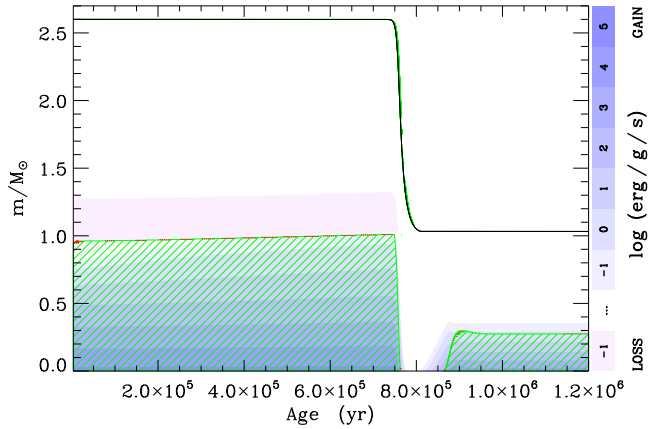}
\caption{
Kippenhahn diagram of models~1 to 554 of the $2.6\,M_{\odot}$ helium star plotted in Figs.~\ref{fig:rho_c-T_c} and \ref{fig:HR}. 
The energy production is indicated in blue shades. The hatched green colour indicates convection.
The central helium burning is completely shut-off as a result of the intense mass loss from the
outer layers of the donor star during the first mass-transfer phase (Case~BA RLO). 
Central helium burning is only reignited after the RLO has terminated and the central temperature has risen again.
See also Fig.~\ref{fig:rho_c-T_c_COWD}.
}
\label{fig:kippenhahn_COWD}
\end{figure}

\begin{figure*}
\centering
\includegraphics[width=1.15\columnwidth,angle=-90]{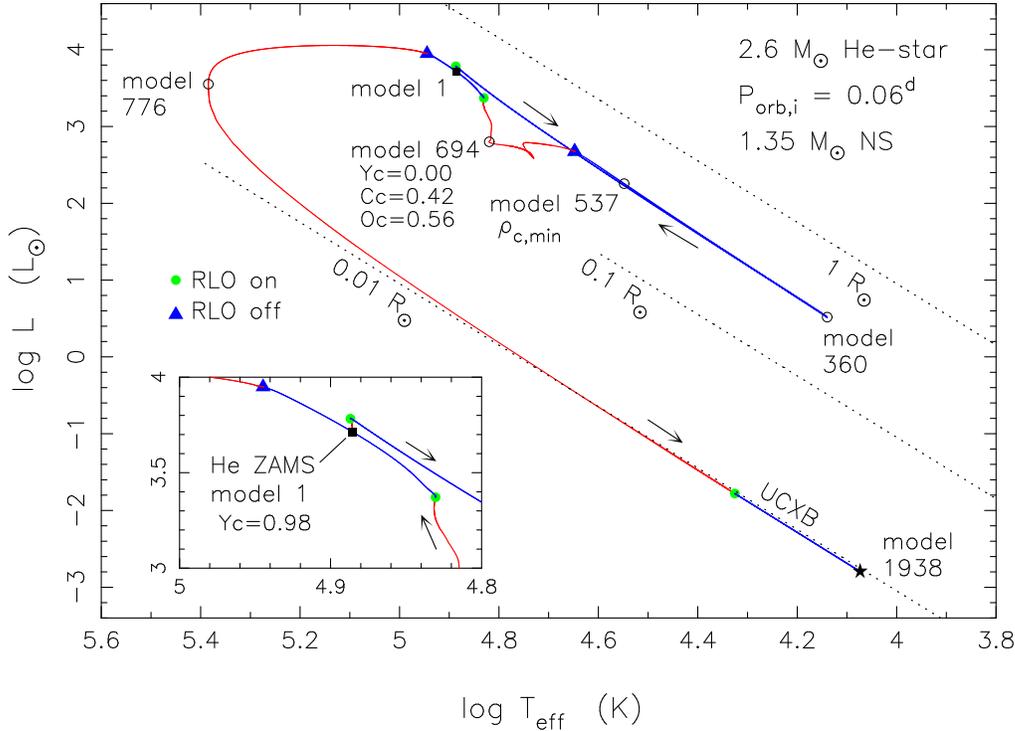}
\caption{
HR-diagram of the $2.6\;M_{\odot}$ helium star donor discussed in Section~\ref{subsec:COWD}.
The evolutionary calculations start on the helium~zero-age main sequence (ZAMS, model~1 marked with a black square). 
Three phases of RLO mass transfer are indicated with blue colour along the evolutionary track.
Central helium burning is temporarily shut off during the first RLO. It is reignited shortly after model~537, and at
model~694 ($t=6.8\;{\rm Myr}$) central helium is depleted, yielding a CO~core made of
42~per~cent carbon and 56~per~cent oxygen. 
The subsequent expansion of the star during helium shell burning causes a second phase of mass transfer.
Thereafter, the $0.84\;M_{\odot}$ CO~WD settles on the cooling track before gravitational wave radiation
causes it to fill its Roche~lobe again ($t\simeq 130\;{\rm Myr}$) at $P_{\rm orb}=32\;{\rm sec}$,
leading to a short UCXB-phase before the system merges (model~1938, see text).
The inset is a zoom-in of the area where the track crosses itself near the location of model~1 ($t=0$).
}
\label{fig:HR}
\end{figure*}

\begin{figure}
\centering
\includegraphics[width=0.68\columnwidth,angle=-90]{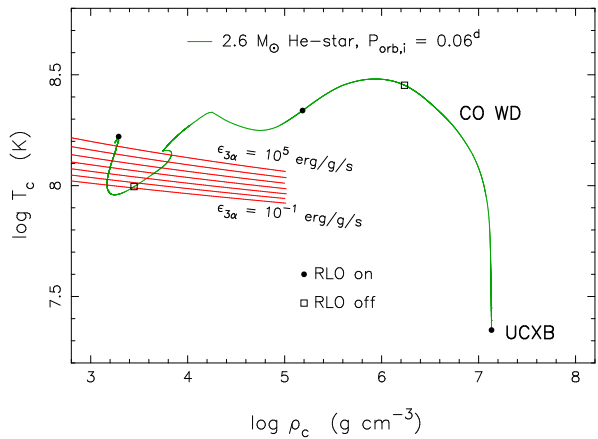}
\caption{
Central temperature versus central mass density for the donor star plotted in Figs.~\ref{fig:rho_c-T_c}, \ref{fig:kippenhahn_COWD} and \ref{fig:HR}, leading to a $0.84\;M_{\odot}$ CO~WD.
The red lines indicate rates of energy production from central helium burning (the difference
between adjacent lines is one order of magnitude).
Shortly into the first mass-transfer episode (Case~BA), central helium burning is temporarily shut off 
as a result of the decreasing core temperature. The reason for this is work required
to expand the inner region while replacing the outer layers of the star which are lost via RLO.
}
\label{fig:rho_c-T_c_COWD}
\end{figure}

Last but not least, we present the evolution of a similar $2.6\;M_{\odot}$ helium star, but now placed in
a tighter orbit with $P_{\rm orb,i}=0.06~{\rm days}$. This star fills its Roche lobe already during early core
helium burning. The resulting mass transfer causes the star to lose its outer $\sim\!1.6\;M_{\odot}$
of material in less than $50\,000~\rm{yr}$, cf. Fig.~\ref{fig:kippenhahn_COWD}.
As a direct consequence of this mass loss, the star is driven out of thermal equilibrium. In order for
the star to replace the lost envelope with material from further below and to remain in hydrostatic equilibrium, 
an endothermic expansion of its inner region (requiring work against gravity) causes the surface luminosity to decrease by more than 3~orders of magnitude,
cf. Fig.~\ref{fig:HR}.
This effect has an important consequence for the nuclear burning in the star. The core expansion causes the
central temperature to decrease (Fig.~\ref{fig:rho_c-T_c_COWD}), and given the extreme dependence of the
triple-$\alpha$ process on temperature, the result is that all fusion processes are completely quenched
for about $80\,000~{\rm yr}$. Thereafter, when the central temperature has risen again,
core helium burning is reignited. 

\begin{figure}
\centering
\includegraphics[width=0.68\columnwidth,angle=-90]{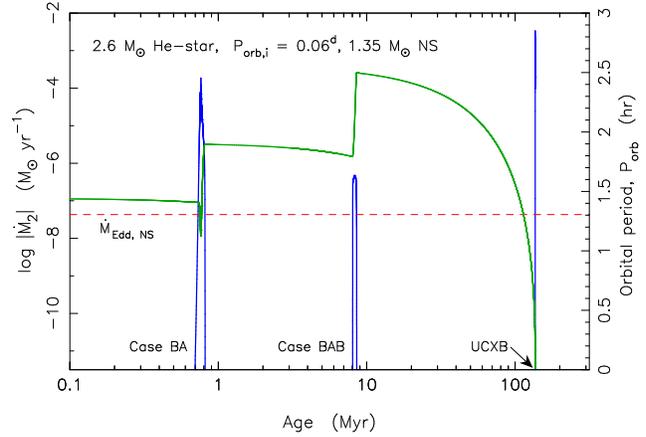}
\caption{
Mass-transfer rate, $|\dot{M}_2|$ (blue peaks) and orbital period, $P_{\rm orb}$ (green curve) as a function of stellar age for the full evolution of
the $2.60\;M_{\odot}$ helium star orbiting a NS with an initial orbital period of 0.06~days 
(also shown in Figs.~\ref{fig:kippenhahn_COWD}--\ref{fig:rho_c-T_c_COWD}).
There are three phases of mass transfer: Case~BA (onset during core helium burning);
Case~BAB (a subsequent episode of mass transfer when the remaining part of the donor star expands during helium shell burning);
and finally mass transfer as an UCXB when the detached CO~WD fills its Roche~lobe as a result of
strong loss of orbital angular momentum via gravitational wave radiation.
The dashed red line indicates the Eddington accretion rate of the NS ($\sim\!4\times10^{-8}\;M_{\odot}\,{\rm yr}^{-1}$
for accretion of helium).
The evolution of the orbital period is dominated by either mass transfer (RLO) or gravitational wave radiation. 
}
\label{fig:Mdot_COWD}
\end{figure}

In Fig.~\ref{fig:Mdot_COWD}, we have plotted the full evolution of the $2.6\;M_{\odot}$ helium star
(with $P_{\rm orb,i}=0.06~{\rm days}$) as a function of age. There are three phases of mass transfer:
Case~BA was described in the previous section; Case~BAB is a result of stellar expansion during helium shell burning;
and finally the star fills its Roche lobe again, $130\;{\rm Myr}$ after it has become a $0.84\;M_{\odot}$ CO~WD, 
as a result of orbital shrinkage due to gravitational wave radiation. At this point $P_{\rm orb}=32\;{\rm sec}$ and the system becomes
an ultra-compact X-ray binary (UCXB). 
However, as a result of the relatively large mass ratio between the WD and the NS,  
this UCXB is short lived and most likely leads to a merger event \citep{vnv+12}.
Depending on the NS equation-of-state, the outcome of this merger is either a massive NS
or a black~hole \citep[possibly accompanied with a $\gamma$-ray burst--like event, see e.g.][]{da06}.
It has been shown that a significant fraction (50 to 80~per~cent) of the WD mass could be ejected during the merger process \citep{met12}; however
the details of the modelling of such events still remain uncertain \citep{ples11}.
Even if assuming fully conservative mass transfer, the gravitational mass of the resulting compact object formed by the merger considered here is at most $\sim\!2.1\;M_{\odot}$,
i.e. close to the value of the recently measured NS mass of $\sim\!2.01\pm0.04\;M_{\odot}$ for PSR~J0348+0432 \citep{afw+13}.
Therefore, we conclude that in this case the final product is most likely still a NS. 

\subsection{Evolution of massive helium star--NS binaries}\label{subsec:massive}
We explored the evolution of a $10\;M_{\odot}$ helium star orbiting a NS with an
initial orbital period of either $P_{\rm orb,i}=0.1$, 0.5 or $2.0\;{\rm days}$. 
Cygnus~X-3 is a potential observational example of such a system.
It contains a compact object with a Wolf-Rayet star companion \citep{vcg+92} and has $P_{\rm orb}=4.8\;{\rm hours}$.
The exact nature of the compact object (NS or BH), however, is still unknown \citep[][and references therein]{zmb13}.

The orbital evolution, and thus the fate, of such a system is determined by the competition between gravitational wave radiation
and stellar wind mass loss from the massive helium (Wolf-Rayet) star, which causes the orbit to shrink and widen, respectively.
Also spin-orbit coupling may become important in such systems \citep{dlpi08}, which is, however, not considered here. 
For $P_{\rm orb,i}=0.1\;{\rm days}$, the orbital damping by gravitational wave radiation is efficient and
causes the helium star to initiate RLO at $P_{\rm orb}=0.073\;{\rm days}$ while it is still rather massive (about $8.90\;M_{\odot}$).
As a result of the high mass ratio between the helium star donor and the accreting NS at this point, the system becomes dynamically unstable,
enters a CE phase, and merges.
For $P_{\rm orb,i}=2.0\;{\rm days}$, and with our applied wind mass-loss prescription, the system widens significantly and reaches 
a large dimension, preventing the radius of the helium star from filling its Roche lobe at any time. 
For orbital periods in-between, there are solutions which result in stable RLO. We find, for example, a solution for
$P_{\rm orb,i}=0.5\;{\rm days}$ (Table~\ref{table:models}). 

For all such massive helium star systems which undergo stable RLO, the mass-transfer phase is very short lasting ($<10^3\;{\rm yr}$)
since these stars are evolved to an advanced stage, close to carbon shell burning, before they expand sufficiently to 
compensate for the orbital widening by wind mass loss and initiate RLO. Hence, their remaining nuclear lifetime is short
before they collapse and explode (a situation which is similar for low-mass helium stars initiating RLO in very wide-orbit systems, cf. Section~\ref{subsec:wideDNS}).
For the particular case with $P_{\rm orb,i}=0.5\;{\rm days}$, our adopted stellar wind mass loss \citep{hkw95} reduces the mass of the helium star
down to $3.04\;M_{\odot}$ prior to RLO. During the subsequent mass transfer, the evolved helium star only loses $0.02\;M_{\odot}$ 
before it explodes. 

We caution that different applied wind mass-loss rates of massive Wolf-Rayet stars can result in significantly different pre-SN stellar masses.
For example, reducing the wind mass-loss rate of \citet{hkw95} by a factor of 10, or applying the prescription of \citet{nl00},
yields a pre-SN mass of about $7.0\;M_{\odot}$ in the above-mentioned case.
Therefore, we notice that our helium stars models with $M_{\rm He,i}=5.0$, 6.0 and $10.0\;M_{\odot}$, which were found to have very similar
metal core masses between 2.10 and $2.20\;M_{\odot}$ by the time of their explosion, could be more massive if the wind mass-loss rate
is reduced. 
For a discussion and a comparison between various prescriptions of wind mass loss-rates, see \citet{ywl10}.

\section{Observational properties of ultra-stripped supernovae}\label{sec:obs}

\subsection{Remaining helium envelope mass prior to the SN}\label{subsec:spectra} 
The modelling of synthetic SN spectra by \citet{hmt+12} suggests that more than $0.06\;M_{\odot}$ of helium is needed
for helium lines to become visible in optical/IR spectra, and thus for classification as a Type~Ib rather than a Type~Ic SN event. 
However, this question also depends on the amount of nickel synthesized during the explosion and how well it is mixed 
into the helium-rich material of the ejected envelope \citep{dhl+11}, given that the detectable helium is excited by the radioactive decay of nickel. 
Furthermore, explosive helium burning during the SN explosion may also play a role in reducing the amount of helium \citep{ww95}. 

\begin{figure}
\centering
\includegraphics[width=0.72\columnwidth,angle=-90]{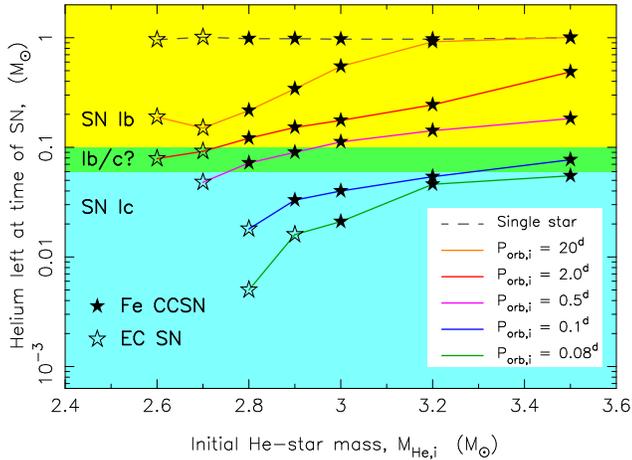}
\caption{
  Total amount of helium prior to explosion for the models from Fig.~\ref{fig:Mcore_final} leading to
  EC~SNe (open stars) and Fe~CCSNe (solid stars). 
  The solid lines connect systems with equal values of $P_{\rm orb,i}$. 
  According to \citet{hmt+12},  
  ejection of at least $0.06\;M_{\odot}$ of helium is needed to detect a SN as Type~Ib (yellow and green region). 
  Below this limit the SN is likely to be classified as a Type~Ic (light blue region).   
  However, this also depends on the amount of nickel synthesized -- see text. 
  }
\label{fig:He-env}
\end{figure}

In Fig.~\ref{fig:He-env}, we show the amount of helium, $M_{\rm He,f}^{\rm env}$ in the envelopes of most of our pre-SN models with $M_{\rm He,i}\le 3.5\;M_{\odot}$.
We find that, in general, EC~SNe in binaries (particularly close binaries) are most likely to be observed as ultra-stripped Type~Ic SNe,
whereas our resulting Fe~CCSNe can be observed as both ultra-stripped (faint) Type~Ic and Type~Ib SNe, as well as more regular Type~Ib SNe 
(the latter still with relatively small helium ejecta masses, typically less than $0.5\;M_{\odot}$). 

In comparison to the remaining amount of helium, $M_{\rm He,f}^{\rm env}$, we find from our binary models that the total envelope mass
$M_{\rm env, f}=M_{\star \rm f}-M_{\rm core,f}$ (i.e. all material outside the CO core) is larger by $0.03\;M_{\odot}$  on average.
This additional material is mainly composed of carbon which was mixed into the envelope. 

\subsubsection{Definition of ultra-stripped SNe}\label{subsubsec:definition}    
\begin{figure}
\centering
\includegraphics[width=0.72\columnwidth,angle=-90]{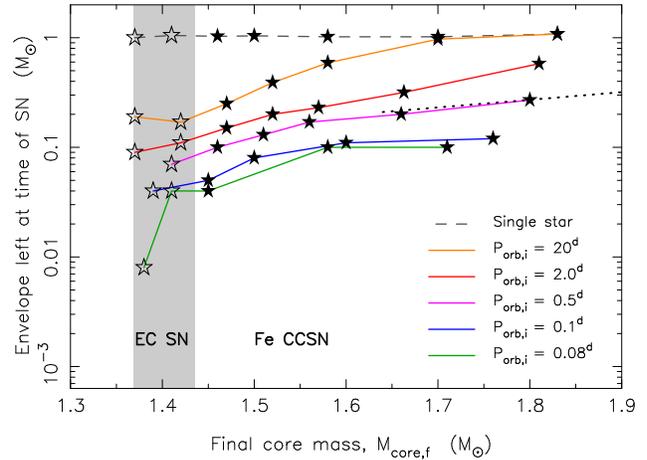}
\caption{
  Total envelope mass versus final core mass prior to the explosion for the models shown in Figs.~\ref{fig:Mcore_final} and \ref{fig:He-env}. 
  EC~SNe (limited to the grey-shaded area) and Fe~CCSNe are plotted with open and solid stars, respectively. The dotted line shows
  the remaining envelope mass in pre-SN stars which explode first in massive binaries (i.e. after RLO to a main-sequence star companion)
  according to the calculations by \citet{ywl10}. Hence, we use this division to define ultra-stripped SNe as exploding stars
  in compact binaries which have $M_{\rm env,f}\la 0.2\;M_{\odot}$ -- see text. 
  }
\label{fig:He-Mcore}
\label{fig:Mcore-Menv}
\end{figure}

In Fig.~\ref{fig:Mcore-Menv}, we show the total amount of envelope mass, $M_{\rm env, f}$ for most of our pre-SN models 
as a function of the final core mass of the exploding star, $M_{\rm core,f}$. The dotted line above $M_{\rm env,f}\approx 0.2\;M_{\odot}$ shows the minimum envelope mass expected
for the stars which explode first in close massive binaries \citep{ywl10}. These stars lose mass to a main-sequence (or slightly evolved) companion star 
via stable RLO prior to the SN. 

In the following we assume that the envelope mass surrounding the pre-SN metal core can be taken as a rough estimate of the amount of material ejected during the SN explosion.
(The determination of the exact location of the mass cut requires detailed computations of the explosion event.) 
{\it We therefore suggest to define ultra-stripped SNe as exploding stars whose progenitors are stripped more than what is possible
with a non-degenerate companion. In other words, ultra-stripped SNe are exploding stars which contain envelope masses $\la 0.2\;M_{\odot}$ 
and having a compact star companion.} The compact nature of their companion allows for extreme stripping in a tight
binary, in some cases yielding an envelope mass $\le 0.008\;M_{\odot}$. 

In addition to forming low-mass ($1.1-1.4\;M_{\odot}$) NSs from (almost) naked metal cores slightly above the 
Chandrasekhar mass\footnote{\citet{tww96,whw02}.}, we also expect ultra-stripped SNe
to produce more massive NSs. From Table~\ref{table:models}, it is seen that ultra-stripped SNe occur in metal cores as massive as $2.15\;M_{\odot}$
(cf. the model with $M_{\rm He,i}=6.0\;M_{\odot}$ and $P_{\rm orb,i}=0.1$~days, and which leads to a post-Case~BB RLO stellar mass of $M_{\star,\rm f}=2.37\;M_{\odot}$
and a helium and carbon-rich envelope of only $M_{\rm env,f}=0.22\;M_{\odot}$). 
Therefore, it seems plausible that ultra-stripped SNe may produce NSs with a wide range of masses between $1.10-1.80\;M_{\odot}$,
see Section~\ref{subsec:NSmass} for further discussion. 

We emphasize that ultra-stripped SNe only constitute a small fraction of all SNe~Ic ($\la 1$~per~cent, Paper~I), 
and also note that the ejected envelopes of exploding massive Wolf-Rayet stars may contain considerable amounts of carbon and oxygen. 
Whereas the pre-SN orbital periods of ultra-stripped SNe are typically between 1~hr and 2~days,
the models of \citet{ywl10} resulted in pre-SN orbital periods, at the moment of the first explosion in close massive binaries experiencing stable RLO, 
of at least $\sim\!10\;{\rm days}$. 

As also seen from Figs.~\ref{fig:He-env} and \ref{fig:Mcore-Menv}, besides core mass, the final amount of helium and the total envelope mass is also correlated 
with the initial orbital period of the binary prior to Case~BB RLO. Therefore, we also find that $M_{\rm He,f}^{\rm env}$ and $M_{\rm env,f}$ are correlated
with the final orbital period, $P_{\rm orb,f}$ just prior to the SN, cf. Table~\ref{table:models}. 

\subsection{Expected light curve properties}\label{subsec:spectra}
The relatively recent advent of all-sky, high-cadence transient surveys has led to the discovery of many rapidly evolving, luminous
transients \citep[see][for details]{dcs+14}. \citet{dcs+14} have estimated the rate of rapid transients (whose luminosity
decreases by at least 50~per~cent in 12~days) to be 4--8~per~cent of the core-collapse rate (at a redshift $z=0.2$). Observationally, these
newly discovered luminous transients form a heterogeneous group of objects, possibly suggesting a variety of explosion mechanisms. 

For comparison, the rate of less luminous SN~2005ek-like events is at least 1--3~per~cent of the SN~Ia rate \citep{dsm+13}. 
As we have argued in Paper~I, the oxygen-rich, Type~Ic supernova SN 2005ek is probably consistent with an ultra-stripped SN, i.e. a helium star stripped of most of 
its helium by binary interactions, as predicted by our models \citep[see also][]{dsm+13}. In Paper~I, we estimated
that the fraction of ultra-stripped SNe to all SNe could be as high as 0.1--1~per~cent.

It is beyond the scope of this paper to compute the SN explosions from our progenitor models.
However, the corresponding events can be expected to be fast and, in most cases, probably quite faint. Based on the photon diffusion time 
through a homologously expanding SN envelope \citep{arn79,arn82,kk14}, we estimate the rise time $\tau_{\rm r}$ from the
light diffusion time through the SN as:  
\begin{equation}\label{eq:SNrise}
  \tau_{\rm r} = 5.0\;{\rm days} \;\;M_{0.1}^{3/4} \,\kappa_{0.1}^{1/2} \,E_{50}^{-1/4}, 
\end{equation}
which for the parameter range of our ultra-stripped SN progenitor models
results in rise times between 12~hr and 8~days (cf. Table~\ref{table:SN}). Here, $M_{0.1}$ is the ejecta mass in units of
$0.1\;M_{\odot}$, $E_{50}$ the SN kinetic energy in units of $10^{50}\;{\rm erg}$, and $\kappa_{0.1}$ the opacity
in units of $0.1\;{\rm cm}^2\,{\rm g}^{-1}$. 
Here we assumed $\kappa_{0.1}=1$, which is suitable for electron scattering in singly ionized helium  
\citep[although the opacity of the SN ejecta does depend on the temperature and the composition of the expelled material,][]{kk14}.

The peak brightness can be estimated according to Arnett's
rule \citep{arn82}, which results in absolute magnitudes of $M=-11.8$, $-13.4$, $-15.0$, and $-16.6$ for nickel masses of
0.001, 0.005, 0.01, and $0.05\;M_{\odot}$, respectively, where we gauged the relation with the models 
of Paper~I. As the diffusion time is shorter than the radioactive decay timescale, the light curves should
follow the radioactive decay law immediately after the light curve peak, until the SN becomes optically thin
to gamma-rays at time $\tau_{\rm d}$, when the decay of the optical light curve should accelerate. 
By assuming a constant density in the SN ejecta, and a constant velocity of the ejecta of
$v_{\rm exp} = \left( 2 E_{\rm SN} / M_{\rm ej} \right)^{0.5}$, we can estimate the time when the 
electron scattering optical depth of the SN light curve becomes equal to one as:
\begin{equation}\label{eq:SNdecay}
  \tau_{\rm d} = 25\;{\rm days} \;\; M_{0.1} \,\kappa_{0.1}^{1/2} \,E_{50}^{-1/2},
\end{equation}
which results in timescales of the order of 1 to 50~days for our models (cf. Table~\ref{table:SN}).

Fast and faint SNe have been suggested to occur also based on other models: thermonuclear explosions \citep{bswn07,dh15}, 
accretion-induced collapse of massive WDs \citep{dbo+06} as well as core collapse \citep{kk14}, 
where for some of them it remains to be seen whether they are
produced by Nature. As the observed short orbital period NS systems give evidence to the scenario elaborated here (cf. Section~\ref{sec:DNS}),
we hope that the spectroscopic features of these ultra-stripped SNe --- which
remain to be worked out --- will allow an unambiguous identification of these events which could then
serve to measure the NS merger rate directly (Section~\ref{subsec:DNSmerger}). 

\begin{table}
\caption[]{Expected properties of the light curves resulting from ultra-stripped SNe. The assumed explosion energy ($E_{\rm SN}$)
           is given in the left column. For each of the different SN ejecta masses ($M_{\rm ej}$) in the following four columns, estimated values 
           of the rise time ($\tau_{\rm r}$) and the decay time ($\tau_{\rm d}$) of the SN light curve are stated.}
\label{tau}
\begin{tabular}{ccccc}
\hline
  $E_{\rm SN}\;$(erg)        & $M_{\rm ej}= 0.2\;M_{\odot}$  & $0.1\;M_{\odot}$   & $0.03\;M_{\odot}$ & $0.01\;M_{\odot}$\\
\hline  
\noalign{\smallskip}
                   $10^{50}$ & $\tau_{\rm r}= 8.4$~days      &         5.0~days   &         2.0~days  &      0.9~days    \\
 ~                           & $\tau_{\rm d}= 50$~days       &         25~days    &         7.5~days  &      2.5~days    \\   
\noalign{\smallskip}
            $3\times10^{50}$ & $\tau_{\rm r}= 6.4$~days      &         3.8~days   &         1.5~days  &      0.7~days    \\
 ~                           & $\tau_{\rm d}= 29$~days       &         14~days    &         4.3~days  &      1.4~days    \\   
\noalign{\smallskip}
                   $10^{51}$ & $\tau_{\rm r}= 4.7$~days      &         2.8~days   &         1.1~days  &      0.5~days    \\
 ~                           & $\tau_{\rm d}= 16$~days       &         7.9~days   &         2.4~days  &      0.8~days    \\   
\hline
\end{tabular}
\label{table:SN}
\end{table}

\subsection{Location of ultra-stripped SNe in host galaxies}\label{subsec:location}
To further help identifying optical transients as ultra-stripped SNe it is important to know their expected location
with respect to star-forming regions in host galaxies. An ultra-stripped SN is the second SN in a binary and thus
the travel distance of the system, $d$, with respect to its original birth location as a ZAMS binary, 
is the product of the systemic velocity resulting from the first SN, 
$v_{\rm sys}$, and the lifetime of the
secondary star following mass accretion from the primary star, $t_2$. Given that the progenitor mass of the exploding ultra-stripped star
is expected within the interval $8-25\;M_{\odot}$, we have that $t_2\simeq 7-40\;{\rm Myr}$. (This is the original mass before this secondary star 
loses its hydrogen envelope in a CE, which creates the
naked helium star orbiting the NS, and which subsequently undergoes Case~BB RLO.)

For the production of double neutron star (DNS) systems, it was argued by \citet{vt03} that the progenitor binaries must have been
fairly wide following the first SN explosion to avoid a merger during the subsequent CE and spiral-in phase when 
the secondary star evolved. The reason is that only massive stars which are evolved (i.e. requiring a {\it wide} binary) have small
enough envelope binding energies to allow for a successful ejection of their envelope leading to survival of the CE-phase \citep{dt00,dt01,pod01}.
As a consequence of this criterion, also the pre-SN systems (prior to the first explosion) must, in general, have been wide. Therefore, 
to ensure survival of these systems, the first SN must have imparted a small kick on the newborn NS, less than $200\;{\rm km}\,{\rm s}^{-1}$ and 
with an average of only $\sim 50\;{\rm km}\,{\rm s}^{-1}$ \citep{vt03}. The resulting systemic velocities are therefore also small
with an average of $v_{\rm sys}\simeq 10\;{\rm km}\,{\rm s}^{-1}$.
This picture seems to have observational support when considering the typical small migration distances ($\sim\!0.1\;{\rm kpc}$) of HMXBs from the spiral arms in the Milky~Way \citep{cc13}.

To conclude, we find that the typical expected value of the travel distance is $d=v_{\rm sys}\cdot t_2 \simeq 200-300\;{\rm pc}$ (the entire
interval spanning between $100-800\;{\rm pc}$). Therefore, we predict ultra-stripped SNe to occur relatively close to the
star-forming regions in host galaxies.

\subsubsection{Comparison to SN~2005ek}\label{subsec:location}
In Paper~I, we argued that SN~2005ek is a good candidate for an ultra-stripped SN. 
This SN was discovered by \citet{dsm+13}. It exploded in the star-forming galaxy UGC~2526, but with a large projected offset
to areas with strong H$\alpha$~emission. \citet{dsm+13} argued that, although the upper limit of the H$\alpha$~line~flux ($3.3\times 10^{37}\;{\rm erg}\,{\rm s}^{-1}$)
is an order of magnitude below the mean value for H\textsc{ii} regions associated with core-collapse SNe \citep{cro13},
such a low value at the explosion site of core-collapse SNe has been measured previously.
Furthermore, they argue that, if the progenitor star was older than $\sim\!20\;{\rm Myr}$, the lack of observed H$\alpha$~emission
could be caused by a shorter lifetime for giant H\textsc{ii} regions. Given that our DNS modelling predicts progenitor
lifetimes up to 40~Myr for ultra-stripped exploding stars (especially for EC~SNe), we find that SN~2005ek remains a good candidate for
an observed ultra-stripped SN.

\section{Helium star mass-transfer and NS accretion rates}\label{sec:HeMdot}
\begin{figure}
\centering
\includegraphics[width=0.72\columnwidth,angle=-90]{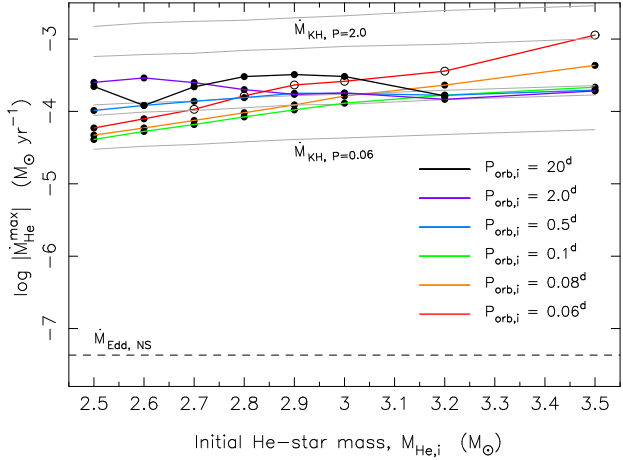}
\caption{
The maximum mass-transfer rate, $|\dot{M}_{\rm He}^{\rm max}|$, from the helium star donor during Case~BB (or Case~BA/BC) RLO as a function
of its initial mass, $M_{\rm He,i}$, for binaries with different initial orbital periods (coloured lines).
The light grey-coloured lines represent Kelvin-Helmholz mass-transfer rates, $\dot{M}_{\rm KH}\equiv M_{\rm He}/\tau _{\rm KH}$,
where $\tau _{\rm KH}$ is the Kelvin-Helmholz (thermal) timescale of the donor star at the onset of the RLO. 
It is notable that $|\dot{M}_{\rm He}^{\rm max}|\approx 10^{-4}\;M_{\odot}\,{\rm yr}^{-1}$ for almost all 
values of $M_{\rm He,i}$ and different initial orbital periods, and also that these rates are fairly close 
to the values of $\dot{M}_{\rm KH}$. The Eddington accretion rate of the NS is shown as a dashed line, some 3--4 orders of magnitude
smaller than $|\dot{M}_{\rm He}^{\rm max}|$. See text.
}
\label{fig:Mdot_max}
\end{figure}

The calculated mass-transfer rates from our models are plotted in Fig.~\ref{fig:Mdot_max}.
This plot shows how the peak value of the mass-transfer rate, $|\dot{M}_{\rm He}^{\rm max}|$ during stable RLO (Case~BA, BB or BC)
depends on $P_{\rm orb,i}$ and $M_{\rm He,i}$ of the helium star. For the smallest value of $P_{\rm orb,i}$ (Case~BA RLO), $|\dot{M}_{\rm He}^{\rm max}|$
increases by more than an order of magnitude when $M_{\rm He,i}$ is increased from $2.5$ to $3.5\;M_{\odot}$. 
However, for the wider systems with initial $P_{\rm orb,i}\ge2.0\;{\rm days}$ (Case~BC RLO), $|\dot{M}_{\rm He}^{\rm max}|$ 
is roughly constant or slightly decreasing with $M_{\rm He,i}$.  
In general, $|\dot{M}_{\rm He}^{\rm max}|$ is always close to $10^{-4}\;M_{\odot}\,{\rm yr}^{-1}$ within a factor of a few.
We have also plotted the Kelvin-Helmholtz mass-transfer rate, 
$\dot{M}_{\rm KH}$, defined as the ratio between the stellar mass
and the Kelvin-Helmholtz (thermal) timescale of the star at the onset of the RLO, i.e.
$\dot{M}_{\rm KH}\equiv M_{\rm He}/\tau _{\rm KH} = RL/(GM_{\rm He})$, where $R$ and $L$ are the radius and the luminosity of
the star at the onset of the RLO, and $G$ is the gravitational constant. The range of values of $|\dot{M}_{\rm He}^{\rm max}|$
is fairly consistent with that of $\dot{M}_{\rm KH}$ although the latter range has a larger spread. 
Only for very evolved helium stars in wide binaries with $P_{\rm orb,i}=20\;{\rm days}$ is the value of
$\dot{M}_{\rm KH}\simeq 10^{-2}\;M_{\odot}\,{\rm yr}^{-1}$ (not shown on plot) significantly larger than $|\dot{M}_{\rm He}^{\rm max}|$. 
However, whereas $\dot{M}_{\rm KH}$  
increases with $P_{\rm orb,i}$, the calculated values of $|\dot{M}_{\rm He}^{\rm max}|$ show the opposite trend for larger
values of $M_{\rm He,i}$. One reason for this is that the value of $|\dot{M}_{\rm He}^{\rm max}|$ is influenced by the shrinking of
the orbit when mass is transferred from the (at least initially) more massive donor star to the lighter accreting NS,
thereby leading to sub-thermal timescale mass transfer.

\begin{figure}
\centering
\includegraphics[width=0.72\columnwidth,angle=-90]{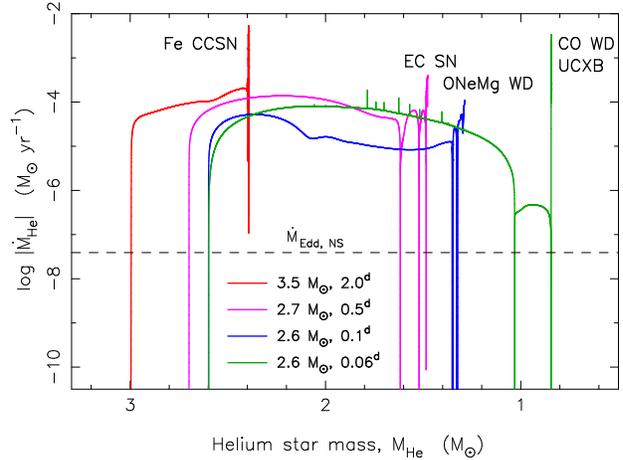}
\caption{
The mass-transfer rate, $|\dot{M}_{\rm He}|$ from the four helium star binaries discussed in 
Sections~\ref{subsec:FeCCSN}, \ref{subsec:ECSN}, \ref{subsec:ONeMgWD} and \ref{subsec:COWD},
leading to an Fe~CCSN, EC~SN, ONeMg~WD and a CO~WD, respectively.
The mass-transfer rates all peak close to $10^{-4}\;M_{\odot}\,{\rm yr}^{-1}$.
The narrow peaks in $|\dot{M}_{\rm He}|$ at the end of the computations are caused by numerical
instabilities and/or shell flashes. For the helium star with an initial mass of $M_{\rm He,i}=2.6\;M_{\odot}$
and $P_{\rm orb,i}=0.06\;{\rm days}$, the system leads to a merger in the UCXB-phase. 
The helium star with an initial mass of $M_{\rm He,i}=3.5\;M_{\odot}$ has reduced its mass to $3.0\;M_{\odot}$ 
at the onset of RLO as a result of stellar wind mass loss (Fig.~\ref{fig:FeCCSN-kippenhahn}). 
For the other (low-mass) helium stars, stellar wind mass loss was neglected.
}
\label{fig:Mdot_four}
\end{figure}

\begin{figure}
\centering
\includegraphics[width=0.72\columnwidth,angle=-90]{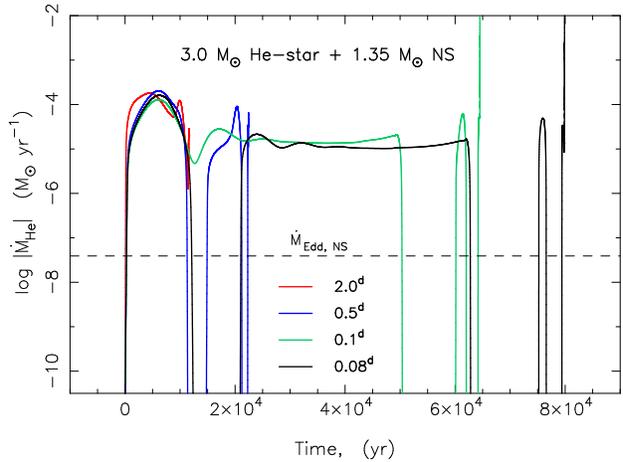}
\caption{
The mass-transfer rate, $|\dot{M}_{\rm He}|$ as a function of time for $3.0\;M_{\odot}$ helium star donors 
with different values of $P_{\rm orb,i}$. The time $t=0$ is defined at RLO onset for all systems.
For the two narrow systems the calculations are terminated by a vigorous helium shell flash following
off-centre oxygen ignition, similar to the case discussed in Paper~I.
It is clearly seen how the duration of the mass-transfer phase, $\Delta t_x$, increases with decreasing value of $P_{\rm orb,i}$,
resulting in more rapidly spinning recycled pulsars in DNS systems in close orbits. In all cases shown the outcome is an 
ultra-stripped Fe~CCSN with $M_{\rm env,f}=0.04-0.23\;M_{\odot}$. For wider systems with $P_{\rm orb,i}=5-120\;{\rm days}$,
see Fig.~\ref{fig:wideDNS}. 
}
\label{fig:Mdot_3.0}
\end{figure}

Fig.~\ref{fig:Mdot_four} shows examples of more detailed mass-transfer rates for the four systems discussed in Section~\ref{sec:results}.
In general, the computed mass-transfer rates
are close to $10^{-4}\;M_{\odot}\,{\rm yr}^{-1}$. This value is 3--4 orders of magnitude larger than the Eddington accretion 
rate for NSs and therefore implies that $>99.9$~per~cent of the transferred material is lost from the system (presumably in the form of
a jet, or a wind from the disk, with the specific orbital angular momentum of the NS, as assumed in the isotropic re-emission model applied to
our computations).
Depending on the exact assumed accretion efficiency for such Case~BB RLO calculations \citep[see][for discussions]{ltk+14}, 
the NS thus typically accretes $\sim\! 10^{-3}\;M_{\odot}$ of material (see also Fig.~\ref{fig:Mdot_3.0}). For the values quoted in
Table~\ref{table:models} we assumed an accretion efficiency of 100~per~cent of the Eddington accretion rate. However, it is possible
that the amount of material accreted by the NS, $\Delta M_{\rm NS}$, is larger by at least 
a factor of three, based on evidence for a high accretion efficiency from PSR~J1952+2630 \citep{ltk+14}. 

Super-Eddington accretion by a factor of a few is easily obtained by a thick accretion disk with a central funnel \citep{acn80}, or a
reduction in the electron scattering cross-section caused by a strong B-field \citep{bs76,pac92}.

\section{Double neutron star (DNS) systems}\label{sec:DNS}
\subsection{Correlation between $P_{\rm orb}$ and $P_{\rm spin}$ for mildly recycled pulsars in DNS systems?}\label{subsec:PorbPspin}
In Fig.~\ref{fig:Mdot_3.0}, one can see from our computations how the duration of the mass-transfer phase, $\Delta t_x$, increases as a function of 
decreasing $P_{\rm orb,i}$ for a given binary (see also Fig.~\ref{fig:wideDNS} for wider systems).
In other words, the larger $P_{\rm orb}$ prior to Case~BB RLO, the shorter is the 
duration of the mass-transfer phase (due to the more evolved stage of the donor star at the onset of the RLO) and thus
the less efficient is the recycling process and thus the longer is the expected spin period, $P_{\rm spin}$, of the mildly recycled pulsar. 
Assuming this relation to survive the SN explosion,
one might expect a correlation between $P_{\rm orb}$ and $P_{\rm spin}$ of the recycled pulsars in DNS systems  
(see further discussion in Tauris~et~al., in preparation).
A correlation between $P_{\rm orb}$ and $P_{\rm spin}$ has previously been hinted
indirectly on similar qualitative arguments \citep[e.g.][]{cb05,dpp05}.
However, we caution that the value of $P_{\rm spin}$ depends on the location of the magnetospheric boundary of the accreting NS 
and therefore it depends on its B-field \citep{pr72,lpp73}.
This dependence is difficult to correct for given the current poor understanding of B-field decay via accretion and --- probably worse ---
convolved with the scatter in NS B-fields from their birth and the unknown magnetic inclination angle which affects the braking torque and thus the observed 
time derivative of $P_{\rm spin}$. 

\subsubsection{Spin periods of mildly recycled pulsars in DNS systems}\label{subsubsec:DNSPspin}
As demonstrated in \citet{tlk12}, the calculated amount of material accreted on to NSs during Case~BB RLO in post-CE binaries 
matches well with the observed spin period values for mildly recycled pulsars.
Using their equation~(14), we find that the values of $\Delta M_{\rm NS}$ calculated here
(between $10^{-4}\;M_{\odot}$ and $3\times 10^{-3}\;M_{\odot}$, if assuming an accretion efficiency of 100~per~cent up to the Eddington limit,
cf. discussion in Section~\ref{sec:HeMdot}) are adequate to produce a mildly recycled 
pulsar with a spin period roughly between 25--350~ms. If the accretion efficiency can be super-Eddington by a factor of 1.5, 2.0
or 3.0 \citep{ltk+14}, the fastest spin period of a DNS pulsar becomes 19, 15 and 11~ms, respectively.
For comparison we note that the DNS system with the fastest spinning recycled NS known to date is the double pulsar PSR~J0737$-$3039 \citep{bdp+03} which 
has a spin period of $P_{\rm spin}=23\;{\rm ms}$.

The formation and properties of very wide-orbit DNS systems (with post-SN $P_{\rm orb}>20\;{\rm days}$) are discussed in Section~\ref{subsec:wideDNS}.
 
\vspace{1.5cm}

\subsection{Neutron star masses and kicks}\label{subsec:NSmass}
\begin{figure}
\centering
\includegraphics[width=0.72\columnwidth,angle=-90]{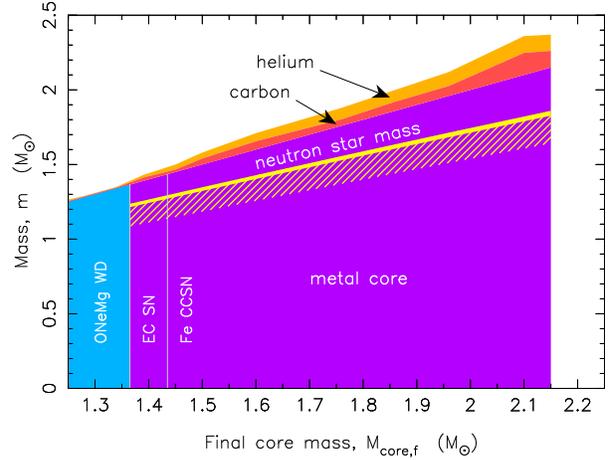}
\caption{
   Cross sections of pre-SN stars versus their final core mass. The plot is based on evolved helium stars with initial masses
   of $2.8-6.0\;M_{\odot}$ and an initial orbital period of 0.1 days (cf. Table~\ref{table:models}). The total mass is divided into three regions representing
   the amount of helium (orange), the remaining carbon-rich envelope (red), and the metal core (purple). The blue shaded region represents stars leaving
   ONeMg~WD remnants. 
   The yellow line indicates the gravitational masses of the NS remnants ($\sim\!1.20-1.80\;M_{\odot}$) calculated for a SN mass cut 
   at the metal core boundary, applying the gravitational binding energy following \citet{ly89}. 
   The hollow yellow hatching indicates that lower NS masses are possible if the mass cut occurs deeper in the exploding stars. 
   Helium stars evolved in X-ray binaries with shorter initial orbital periods will be further stripped, resulting in even smaller envelopes 
   (Figs.~\ref{fig:He-env} and \ref{fig:Mcore-Menv}).  
  }
\label{fig:cross-section-SN}
\end{figure}
In Fig.~\ref{fig:cross-section-SN}, we present the interior composition of our pre-SN models as cross sections by mass, which were calculated for helium stars
with initial masses between 2.5 and $6.0\;M_{\odot}$, in binaries with an initial orbital period of 0.1~days and an accreting $1.35\;M_{\odot}$ NS.
(Note, the helium stars with the smallest initial masses do not produce a SN, cf. Table~\ref{table:models}). 
This figure illustrates the extreme degree of stripping for, in particular, the lowest mass helium stars.
We estimate that the NSs produced from our ultra-stripped SNe could potentially have masses in the range $1.10-1.80\;M_{\odot}$.
The exact mass of the NS depends on the mass cut during the explosion and the NS equation of state, both of which are still uncertain. 
We notice that the second formed NS in a DNS system with the lowest known mass is PSR~J0453+1559 \citep{mar15} which 
has a precisely measured mass of $1.17\;M_{\odot}$ (less than $\sim\!1.18\;M_{\odot}$ at the 99.7~per~cent confidence level). 
For recent discussions on the upper limit of NS birth masses in binaries, see e.g. \citet{tlk11,opns12,fbw+12}. 

\begin{figure}
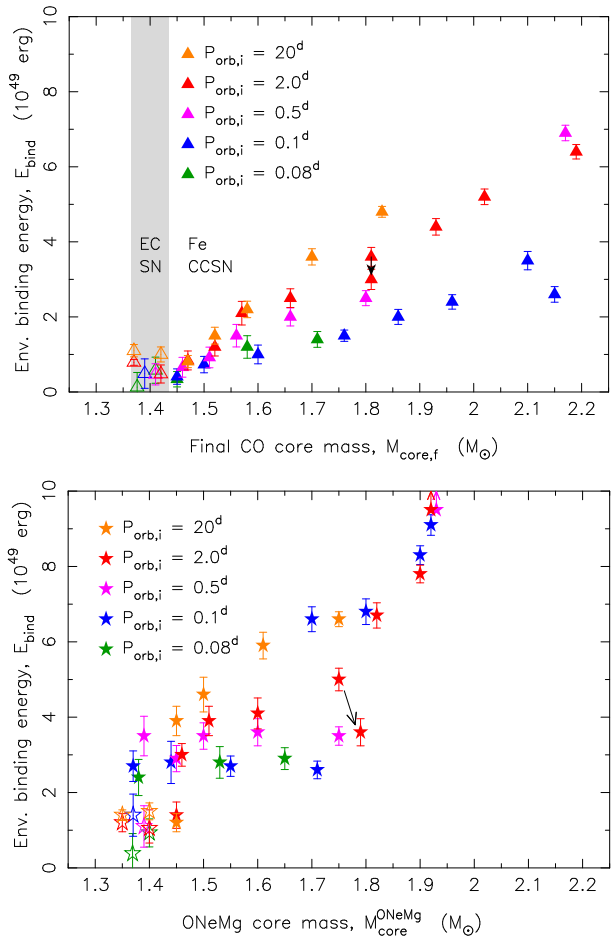

\centering
\includegraphics[width=0.72\columnwidth,angle=-90]{Ebind_CO.ps}

\vspace{0.2cm}
\includegraphics[width=0.72\columnwidth,angle=-90]{Ebind_ONeMg.ps}
\caption{
   Gravitational binding energy of the envelope as a function of CO core mass (upper panel) and ONeMg core mass (lower panel) of our pre-SN models. 
   Open and filled symbols correspond to models leading to an EC~SN and an Fe~CCSN, respectively (cf. Section~\ref{sec:results}). 
   The different colours represent $P_{\rm orb,i}$ of the original helium star.
   The error bars were calculated by varying the core/envelope boundary by $0.01\;M_{\odot}$. 
   The black arrow indicates the change in location in the diagram for one particular model, from the stage of early core oxygen burning
   to a more evolved stage near the end of core oxygen burning. Two data points (shown with arrows at the top of the lower panel) are located at $E_{\rm bind}\simeq1.4\times10^{50}\;{\rm erg}$.  
  }
\label{fig:Ebind}
\end{figure}

As a result of the SN explosion, a significant momentum kick is often imparted on a newborn NS.
Such kicks are associated with explosion asymmetries and may arise from non-radial hydrodynamic instabilities in the collapsing stellar core \citep[i.e. neutrino-driven
convection and the standing accretion-shock instability,][]{bm06,skjm06,fgsj07,mj09,jan12,jan13,blh+15,fkg+15}. These instabilities lead to large-scale anisotropies
of the innermost SN ejecta, which interact gravitationally with the proto-NS and accelerate the nascent NS on a timescale of several seconds \citep[e.g.][]{jan12,wjm13}.

The momentum kick imparted to a newborn NS via an EC~SN is expected to be small.
This follows from detailed simulations which imply: i) explosion energies significantly smaller than those inferred for 
standard Fe~CCSNe \citep{kjh06,dbo+06}; and ii) short timescales to revive the stalled supernova shock,
compared to the timescales of the non-radial hydrodynamic instabilities which are required to produce strong 
anisotropies, e.g. \citet{plp+04,jan12}.
Simulations of EC~SNe by \citet{kjh06,dbo+06} predict explosion energies of about (or even less than) $10^{50}\;{\rm erg}$, and 
thus possibly kick velocities less than $50\;{\rm km\,s}^{-1}$, i.e. significantly smaller
than the average kick velocities of the order $400-500\;{\rm km\,s}^{-1}$ which are evidently imparted on young pulsars in general \citep{ll94,hllk05}.

One may ask what would be the magnitude of a kick imparted to a NS born in an ultra-stripped SN, which can be either an EC~SN or an Fe~CCSN.
We see two factors which imply that in these objects, the NS kicks may by small.
First, from our modelling of the progenitor stars of ultra-stripped SNe, we have demonstrated (Section~\ref{sec:obs}) that
the amount of ejecta is extremely small compared to standard SN explosions. 
This may likely lead to a weaker gravitational tug on the proto-NS \citep[caused by asphericity of the 
ejecta, e.g.][]{jan12} and thus a small kick.  Second, the binding energies of the envelopes of our models,
even including the CO-layer, are often only a few $10^{49}\;{\rm erg}$ (Fig.~\ref{fig:Ebind}), such that even a weak outgoing shock can quickly
lead to their ejection, potentially before large anisotropies can build up.
While these arguments hold best for the lowest mass NSs formed from our models, it is possible in principle that one can form even
rather massive NSs from ultra-stripped progenitors with low kicks, provided that the whole (most) of the ONeMg core collapses. However, these more
massive ONeMg cores are also expected to produce larger iron cores and may therefore lead to more 'normal' Fe~CCSNe with larger explosion
energies and larger kicks. We are unable to resolve this issue as this requires full core-collapse calculations. In addition, it will be left to
a future study to investigate how the final pre-SN cores in ultra-stripped progenitors differ from those of single stars of comparable
initial mass to confirm whether low-kick NSs preferentially occur in close binaries, as seems to be indicated by observations \citep{prps02,plp+04}.

While it is difficult to estimate kick values, these arguments taken together allow us to  
speculate that ultra-stripped SNe (EC~SNe and, at least, Fe~CCSNe with relatively small metal cores) generally lead to the formation of NSs with small kicks.
This would have important consequences for the formation of DNS systems and we discuss these implications in a future study. 

We recall that the metal core masses ($M_{\rm core,f}$) listed in Table~\ref{table:models} correspond to the mass inside the CO core boundary.  
Whereas this boundary is fairly stable towards the end of the nuclear evolution of our modelling, the boundary of the ONeMg core is quite
sensitive to the number of carbon burning shells which develop in the final phases of our calculations.
This explains the somewhat larger scattering in envelope boundary energies in the lower panel compared to the upper panel of Fig.~\ref{fig:Ebind}.

\subsection{Formation and properties of very wide-orbit DNSs}\label{subsec:wideDNS}
In the previous sections, we have discussed the evolution of tight orbit ($P_{\rm orb,i}\le 2\;{\rm days}$) helium star--NS systems, leading to the formation of 
fairly close orbit DNS systems via an ultra-stripped SN. In addition, we have investigated a number of wide-orbit helium star--NS systems 
with $5\le P_{\rm orb,i}\le 120 \;{\rm days}$, cf. Table~\ref{table:models}. 

A common feature of these wide-orbit systems is that by the time the helium star fills its Roche lobe (at an orbital period larger than
$P_{\rm orb,i}$ as a consequence of stellar wind mass loss), the helium star is quite evolved and is already undergoing carbon shell burning.
As a consequence of this, the remaining lifetime of the helium star is less than a few thousand years at the onset of RLO.
Therefore, by the time the helium star explodes in a SN (during RLO) it will still have a rather thick helium envelope of 
$M_{\rm He,f}^{\rm env}=0.2-0.8\;M_{\odot}$ (the wider the orbit, and the more massive the initial helium star, the larger 
envelope will remain at the onset of the core collapse). Such a relatively massive envelope ejected at the time of the SN will not only affect the
observational properties of the light curve. It also results in larger post-SN orbital periods and eccentricities of the DNS systems formed.
Typical values for the eccentricities in the case of symmetric SNe would be $0.2-0.5$. However, larger (and smaller) values are possible if kicks are added
(which indeed might be the case as a result of the larger ejecta mass of these stars, cf. discussion in Section~\ref{subsec:NSmass}).
We find that orbital periods of DNS systems with a recycled pulsar are expected out to about 200~days (again depending on the imparted kicks).

Given that the SN will terminate the recycling process of the accreting NS shortly after RLO is initiated, we expect the resulting wide-orbit DNS systems
to contain extremely mildly recycled pulsars, i.e. {\it marginally recycled pulsars}, with spin periods in excess of 100~ms and 
relatively large B-fields. These pulsars should be observed in the grey-zone in the $P\dot{P}$--diagram where it is difficult to 
distinguish such marginally recycled pulsars from the lower B-field population of isolated, old non-recycled pulsars. 
The recently discovered DNS system PSR~J1930$-$1852 \citep{srm+15} is an bona fide example of such a system.
It has a spin period of 185~ms, a spin period derivative of $\dot{P}=1.8\times 10^{-17}\;{\rm s}\,{\rm s}^{-1}$, 
an orbital period of 45~days and an eccentricity of 0.40.

\begin{figure}
\centering
\includegraphics[width=0.69\columnwidth,angle=-90]{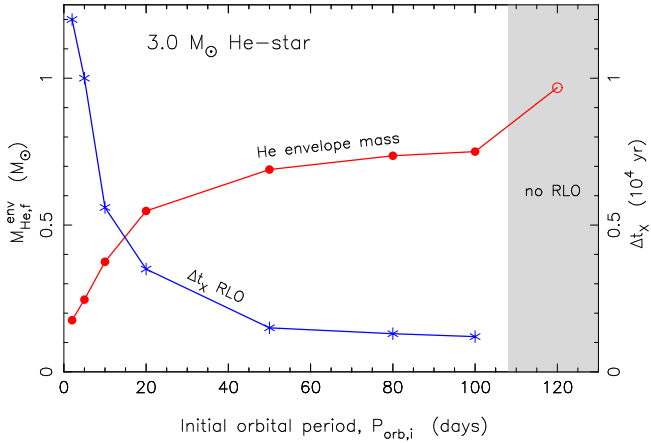}
\caption{
  Remaining helium envelope mass, $M_{\rm He,f}^{\rm env}$ (red curve) at the time of the SN as a function of 
  initial orbital period, $P_{\rm orb,i}$ of helium star--NS systems with $M_{\rm He,i}=3.0\;M_{\odot}$. Also plotted is the duration of the RLO
  prior to the explosion (blue curve), which determines the properties of the (marginally) recycled pulsar companion (see text). 
  In all cases shown, the metal core reaches a final mass of $\sim\!1.58\;M_{\odot}$ at the moment of the Fe~CCSN.
  Ultra-stripped SNe require $P_{\rm orb,i}\la 2.0\;{\rm days}$ for $M_{\rm He,i}=3.0\;M_{\odot}$.
  The grey-shaded region to the right indicates the location of systems which are too wide for post-HMXB/CE Case~BB RLO to occur. 
}
\label{fig:wideDNS}
\end{figure}
In Fig.~\ref{fig:wideDNS}, we have plotted the amount of helium left in the envelope at the time of the SN, as well as the
duration of the pulsar recycling phase via Case~BB (strictly speaking Case~BC) RLO, as a function of the initial orbital
period of the helium star--NS system. In this plot, calculations are shown for helium stars with $M_{\rm He,i}=3.0\;M_{\odot}$ -- see 
Table~\ref{table:models} for other systems with $P_{\rm orb,i}=20\;{\rm days}$.
It is clearly seen how the remaining helium envelope mass increases with $P_{\rm orb,i}$, and how the duration of the RLO ($\Delta t_x$) decreases
with $P_{\rm orb,i}$. Given that the growth of NS mass is expected to be limited by (possibly a few times) the Eddington-accretion limit
of $\dot{M}_{\rm Edd,NS}\sim\!4\times 10^{-8}\;M_{\odot}\,{\rm yr}^{-1}$ (for pure helium), we expect wide-orbit DNS systems to be (marginally)
recycled with $\Delta M_{\rm NS}\simeq 1-10\times 10^{-4}\;M_{\odot}$, typically resulting in
spin periods in excess of 100~ms [cf. eq.(14) in \citet{tlk12} for a rough estimate of the expected spin period after recycling].
 
The system with $P_{\rm orb,i}=120\;{\rm days}$ will not initiate RLO since the radius of the
helium star, reaching a maximum of $74\;R_{\odot}$, is unable to catch up with its Roche-lobe radius (which 
is increasing since the orbit is expanding due to stellar wind mass loss). Hence, its NS companion will avoid recycling. 

Finally, we note that in all cases, even for the widest systems which initiate RLO, the expanding helium envelope is stable
against convection (except in a very tiny layer near the surface). This means that: i) the mass-transfer rates
in these systems are not much different from those shown in Fig.~\ref{fig:Mdot_max} for helium star donors in closer orbits, 
and ii) the RLO is always dynamically stable such that these systems avoid evolving into yet another CE.

\subsection{Constraints on NS accretion prior to Case~BB RLO}\label{subsec:preNSacc}
Following the formation of the first born NS in a massive binary, there are four phases of evolution which could
potentially lead to accretion onto this NS: i) accretion from the stellar wind of the massive donor star prior to
the formation of a CE (i.e. standard HMXB accretion), ii) accretion during the CE-phase, iii) accretion from the stellar wind of
the exposed naked helium core (Wolf-Rayet star) of the former massive companion star, and iv) Case BB~RLO from the expanding helium star
as investigated in this paper.

Depending on the lifetime (mass) of the companion star, an upper limit to the amount of material accreted from a wind in the first phase 
is of the order $10^{-2}\;M_{\odot}$. 
The wind accretion from the naked helium star prior to Case~BB RLO (third phase) may reach a few $10^{-4}\;M_{\odot}$.
However, in both cases it is unclear to which extent the wind-accreting NS is able to efficiently spin up. This depends on, for example, whether or not an 
accretion disk is formed, and whether or not propeller effects are important \citep[e.g.][]{fkr02}. 

It was recently demonstrated by \citet{mr15a,mr15b} that a NS embedded in a CE only accretes a modest amount of material during its spiral-in
as a result of density gradients across the NS's accretion radius (which strongly limits accretion by imposing a net
angular momentum to the flow around the NS). They computed a firm upper limit of $0.1\;M_{\odot}$. 
Also in this phase, given the extreme conditions with a very high ram pressure and a squeezed magnetosphere, 
the efficiency of transporting angular momentum to the NS is likely to be very limited.

We therefore conclude that the properties of the first born NS in an observed DNS system might entirely be explained by the Case~BB RLO investigated here.
Further evidence for this comes from observations of the wide-orbit DNS systems: PSR~J1753$-$2240 \citep{kkl+09}, PSR~J1811$-$1736 \citep{lcm+00}
and PSR~J1930$-$1852 \citep{srm+15}, which have rather slow spin periods of 95, 144 and 185~ms, respectively,
in agreement with our calculated estimates from Case~BB RLO and indicating 
rather inefficient angular momentum transport during the other phases briefly discussed above.

\subsection{Merger time of DNS systems resulting from our models}\label{subsec:DNSmerger}
As previously mentioned, the final orbital period prior to an ultra-stripped SN, $P_{\rm orb,f}$, is typically between
1~hr and 2~days (cf.~Table~\ref{table:models}). In general, the dynamical effects of SNe may change the orbital period significantly \citep{hil83,bp95,tt98,wkk00}.
However, the core structure of a star depends strongly on whether it evolves in a close binary or in a wide binary/isolation \citep{plp+04}.
As suggested in Section~\ref{subsec:NSmass}, kicks imparted on NSs formed from ultra-stripped stars (i.e. during the second explosion forming DNS systems) 
are possibly small. If correct, this would imply that, on average, the post-SN orbital periods will not be much larger than the pre-SN orbital periods.

As a consequence of gravitational wave radiation \citep{pet64}, the merger timescale for such newly formed DNS systems with orbital periods between 1~hr and 2~days 
is between a few Myr and $\infty$; shortest for close binaries with high post-SN eccentricities (which may still be significant even from small kicks).
The maximum post-SN orbital period of a DNS system that will merge within a Hubble time (assuming NS masses of $1.4\;M_{\odot}$)  
is between 0.7 and 1.7~days for post-SN eccentricities of $0.1-0.7$.

\begin{figure}
\centering
\includegraphics[width=0.50\columnwidth,angle=-90]{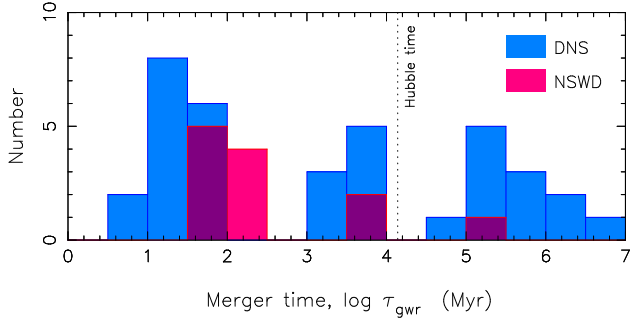}
\caption{
  Histogram of merger times caused by orbital decay via gravitational wave radiation of the final double degenerate
  systems resulting from the binaries modelled in this paper with $P_{\rm orb,i}=0.1$, 0.5 and 2.0~days (cf. Table~\ref{table:models}), 
  corresponding to the three peaks, respectively. The blue colour represents DNS systems; the red colour 
  represents NSWD systems (i.e. systems where the helium donor star left behind a massive WD orbiting the NS). See text for details.
}
\label{fig:merger}
\end{figure}
In Fig.~\ref{fig:merger}, we have plotted the merger times, $\tau _{\rm gwr}$, caused by orbital decay via gravitational wave radiation 
of the double degenerate systems resulting from our models presented in Table~\ref{table:models}.
In case the donor star terminated as a SN, creating a DNS system, we assumed, for simplicity, a symmetric explosion (i.e. without a kick).
The mass of the newborn NS was calculated using: $M_{\rm NS}=M_{\rm core,f}-M_{\rm NS}^{\rm bind}$, where the
released binding energy of the NS corresponds to a mass decrease of $M_{\rm NS}^{\rm bind}=0.084\;M_{\odot}\cdot (M_{\rm NS}/M_{\odot})^2$ \citep{ly89},
i.e. following the yellow line of NS masses in Fig.~\ref{fig:cross-section-SN}. 
Subsequently, we calculated the amount of mass ejected and the resulting post-SN orbital period and eccentricity \citep[e.g.][]{bv91}, before calculating
$\tau _{\rm gwr}$ via numerical integration \citep{pet64}. In case the donor star terminated as a WD, we assumed that $M_{\rm WD}=M_{\rm core,f}$ and
that the orbit remained circular with a resulting orbital period equal to $P_{\rm orb,f}$. (The slight widening of the orbit from losing the thin envelope
in a proto-WD wind is assumed to be compensated with orbital decay due to gravitational wave radiation during this epoch.)
The systems in the histogram are seen to cluster in three groups corresponding to $\tau_{\rm gwr}< 250\;{\rm Myr}$, $\tau_{\rm gwr}=1-10\;{\rm Gyr}$ and 
$\tau_{\rm gwr}> 85\;{\rm Gyr}$, reflecting our systems chosen initially with helium-star ZAMS orbital periods of $P_{\rm orb,i}\le 0.1$~days (left),
0.5~days (middle), and 2.0~days (right), cf. Table~\ref{table:models}. 
We emphasize that the distribution in Fig.~\ref{fig:merger} only represents examples within the expected range of
$\tau _{\rm gwr}$ and that the true distribution of $\tau _{\rm gwr}$ requires population synthesis modelling \citep[e.g.][]{vt03}. 
It is possible that some of the tight systems which end up in a CE may survive the spiral-in phase and produce very tight systems with even smaller values 
of $\tau _{\rm gwr}$ \citep{dp03}, see discussion in Section~\ref{subsec:comparison}.

In a future publication we analyse in more detail the observed distributions of orbital periods, spin periods, eccentricities and systemic velocities 
of DNS systems and confront them with simulations and expectations from ultra-stripped SNe.

\section{Discussions}\label{sec:discussions}

\subsection{Mapping initial parameter space to final outcome}
\begin{figure}
\centering
\includegraphics[width=0.72\columnwidth,angle=-90]{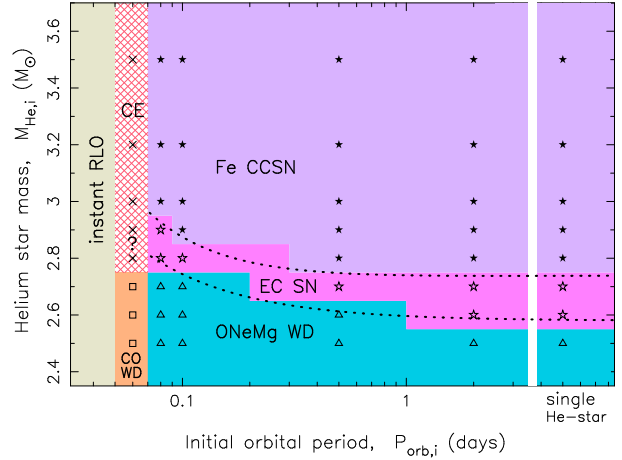}
\caption{
  Grid showing the outcome (the final fate of the helium star donors) of our modelling as a function of
  the initial values of orbital period and helium star mass -- see also Table~\ref{table:models}.
  The dotted lines (equations~\ref{eq:final_fate}--\ref{eq:ECSN}) separating the regions of different outcomes are useful for population synthesis. 
  Our modelled systems with $P_{\rm orb,i}$ of $5-120\;{\rm days}$ are not shown here, but fall in the region
  between $P_{\rm orb,i}=2.0\;{\rm days}$ and the single helium stars. 
}
\label{fig:outcome_grid}
\end{figure}

Fig.~\ref{fig:outcome_grid} shows the final fate of the $2.5-3.5\;M_{\odot}$ helium stars which we have modelled in binaries. 
Also shown to the right is the outcome of the same helium stars evolved in
isolation without a companion star.
As discussed in Section~\ref{subsec:FeCCSN}, the boundary between EC~SNe and Fe~CCSNe is perhaps somewhat less certain 
compared to the other boundaries shown.
Furthermore, along the sequence calculated for $P_{\rm orb,i}=0.06\;{\rm days}$ we could not determine (due to numerical instabilities)
the exact mass for which the mass transfer becomes dynamically unstable, leading to a CE evolution. Certainly, the helium stars
with $M_{\rm He,i}>3\;M_{\odot}$ end up in a CE for these short initial orbital periods. 
Systems with $P_{\rm orb,i}$ less than 0.06~days are excluded because the
helium ZAMS stars cannot fit inside such small Roche~lobes.
One can use this figure for determining the fate of helium stars in close binaries when performing population synthesis.

\subsubsection{Fitting formulae with applications to population synthesis}
Based on our results presented in Figs.~\ref{fig:Mcore_final}, \ref{fig:He-env}, \ref{fig:Mcore-Menv} and \ref{fig:outcome_grid}
we have fitted a number of relations which are, for example, useful for population synthesis modelling.

For the relation in Fig.~\ref{fig:Mcore_final} between final pre-SN metal core mass, $M_{\rm core,f}$ and initial values of helium star mass, $M_{\rm He,i}$ 
(prior to Case~BB RLO and wind mass loss), and orbital period, $P_{\rm orb,i}$ we find:
\begin{equation}\label{eq:MfMi}
  \displaystyle M_{\rm core,f}\simeq \left(\frac{1}{400\;P_{\rm orb,i}}+0.49 \right)\cdot M_{\rm He,i} - \left(\frac{0.016}{P_{\rm orb,i}}-0.106\right), 
\end{equation}
where $P_{\rm orb,i}$ is in units of days and masses are in $M_{\odot}$. For the investigated mass interval of $2.5 \le M_{\rm He,i}/M_{\odot} \le 3.5$ this fit is accurate to within about 1~per~cent (except for the very low-mass end 
where the discrepancy is 3~per~cent for the wider systems).

The remaining amount of helium in the pre-SN envelope, $M_{\rm He,f}^{\rm env}$ can be fitted as a function of final pre-SN metal core mass 
and initial orbital period as follows: 
\begin{equation}\label{eq:MHef}
 M_{\rm He,f}^{\rm env}\simeq\left\{ \begin{array}{ll}
   0.18\;P_{\rm orb,i}^{\,0.45}\cdot\big[\ln(M_{\rm core,f}^4)-1.05 \big],\\
   M_{\rm core,f}\cdot\big[\ln(P_{\rm orb,i}^{\,-0.2})+1\big]+\ln(P_{\rm orb,i}^{\,0.5})-1.5,\\
           \end{array}
         \right.  
\end{equation} 
where the upper expression is valid (generally within an error of $\pm 0.01\;M_{\odot}$) for $0.06<P_{\rm orb,i}\le 2.0\;{\rm days}$, and 
the lower expression is an approximation for $P_{\rm orb,i}> 2.0\;{\rm days}$ and single helium stars ($P_{\rm orb,i} \to \infty$).
By combining equations~(\ref{eq:MfMi}) and (\ref{eq:MHef}) one can calculate $M_{\rm He,f}^{\rm env}$ directly as a function of initial
orbital period and initial helium star mass.

The orbital evolution can easily be calculated analytically by separating the wind mass loss prior to Case~BB RLO
from the subsequent Roche-lobe mass transfer (during which the effect of wind mass loss is small compared to a mass-transfer rate of $\sim10^{-4}\;M_{\odot}\,{\rm yr}^{-1}$, cf. Section~\ref{sec:HeMdot}). 
For mass loss due to a direct fast stellar wind from the helium star \citep[i.e. Jeans's mode,][]{hua63,vdh94a}, the orbital period change is given by: 
\begin{equation}\label{eq:Jeans}
  \displaystyle \frac{P}{P_0}= \left( \frac{1+q_0}{1+q} \right) ^2,
\end{equation}
where $P_0$ and $P$ represent the initial helium ZAMS orbital period ($P_{\rm orb,i}$) and the orbital period at the onset of the RLO, respectively.
The mass ratio between the stellar components, $q\equiv M_{\rm He}/M_{\rm NS}$ at the same two epochs is denoted by $q_0$ and $q$.

By integrating the orbital angular momentum balance equation and applying the so-called isotropic re-emission model \citep{bv91}
one can find the change in orbital separation during non-conservative RLO. \citet{tau96} derived:
\begin{equation}\label{eq:tau96}
  \frac{a}{a_0}= \left( \frac{q_0(1-\beta)+1}{q(1-\beta)+1} \right) ^{\frac{3\beta-5}{1-\beta}}
                 \left( \frac{q_0+1}{q+1} \right) \left( \frac{q_0}{q} \right) ^2,
\end{equation}
where $a_0$ and $a$ refer to the orbital separation prior to RLO and after RLO, respectively, and where $q_0$ and $q$ now represent the
mass ratios at these two epochs. The parameter $\beta$ represents the (constant) fraction of transferred material which is
lost from the vicinity of the NS. Given that the mass-transfer rate of Case~BB RLO is much higher than the Eddington limit
for an accreting NS ($|\dot{M}_2| \gg \dot{M}_{\rm Edd,NS}$, and thus $\beta \to 1$) one can approximate the above equation and combine it with
Kepler's third law to yield \citep[see also][]{kskd01}:
\begin{equation}
\lim_{\beta \to 1} \left( \frac{P}{P_0} \right) = 
         \left( \frac{q_0+1}{q+1} \right) ^2 \left( \frac{q_0}{q} \right) ^3 \, e^{3(q-q_0)}. 
\end{equation}
As a numerical example, we can compare this expression with our BEC computation of Case~BB RLO for a $3.2\;M_{\odot}$ helium star
orbiting a $1.35\;M_{\odot}$ NS with an initial orbital period of 2.0~days, and for which stellar winds were not included.
As seen in Table~\ref{table:models}, the final mass of the donor star prior to the explosion is $1.98\;M_{\odot}$.
Hence, we get $q_0=2.37037$ and $q=1.46656$ for this system, and inserting $P_0=2.0\;{\rm days}$ in the above equation
we find that the final orbital period at the end of Case~BB RLO is $P=1.0476\;{\rm days}$ (exactly as calculated
with our binary stellar evolution code, cf. 1.05~days in Table~\ref{table:models}).
Of course, gravitational wave radiation included in the code also carries away orbital angular momentum. However, this effect is negligible during
the short lasting (typically a few $10^4$~yr) Case~BB RLO, cf. Fig.~\ref{fig:Mdot_3.0}.

For the final fate of a helium star in a close orbit with a NS (Fig.~\ref{fig:outcome_grid}, dotted lines) we find that the outcome is either:
\begin{equation}\label{eq:final_fate} 
  \begin{array}{lrl}
	{\rm Fe~CCSN}:	&	M_{\rm He,i}\!\!\!\! & \ge M_{\rm FeCCSN},\\ \nonumber
	{\rm EC~SN}:	&	M_{\rm ECSN}\!\!\!\! & \le M_{\rm He,i} < M_{\rm FeCCSN},\\ \nonumber
	{\rm ONeMg~WD}:	&       M_{\rm He,i}\!\!\!\! & < M_{\rm ECSN},
  \end{array}
\end{equation} 
where the threshold initial mass for producing an Fe~CCSN is:
\begin{equation}\label{eq:MFeCCSN}
  M_{\rm FeCCSN}=2.74\;P_{\rm orb,i}^{-0.00209/P_{\rm orb,i}},
\end{equation}
and the threshold initial mass for producing an EC~SN is:
\begin{equation}\label{eq:ECSN}
  M_{\rm ECSN}=\frac{1}{61\,P_{\rm orb,i}}+2.58,
\end{equation}
where $P_{\rm orb,i}$ is in days and the threshold masses are in $M_{\odot}$.
The above formulae are valid for $P_{\rm orb,i}\ga 0.07\;{\rm days}$ and $M_{\rm He,i}\ge 2.5\;M_{\odot}$.
For $P_{\rm orb,i}\simeq 0.06-0.07\;{\rm days}$, the outcome is a CO~WD if $M_{\rm He,i} \la 2.8\;M_{\odot}$,
and most likely a CE for more massive helium stars.
Furthermore, one should keep in mind that all calculations were performed for a $1.35\;M_{\odot}$ accreting NS.
Using other NS masses may slightly change the boundaries stated above.

\subsection{Comparison to previous work}\label{subsec:comparison}
Whereas ultra-stripped SNe as such have not been discussed much in the literature up to now (besides from Paper~I), 
we can compare our Case~BB RLO modelling to earlier work. 

The systematic investigations by \citet{dpsv02}, \citet{dp03} and 
\citet{ibk+03}, also considered Case~BB RLO from helium stars with masses of $1.5-6.7\;M_{\odot}$ in binaries with a NS. 
However, whereas these previous computations usually terminated their calculations after carbon burning,
we were often able to continue the binary stellar evolution until oxygen burning. 
Hence, we have been able to strip the envelopes of our donor stars further prior to the SN (i.e. ultra-stripped SNe), 
in some cases all the way down to $<0.01\;M_{\odot}$.

The main conclusion of \citet{dp03} is that helium stars with $M_{\rm He,i}=2.8-3.3\;M_{\odot}$, and those with $M_{\rm He,i}=3.3-3.8\;M_{\odot}$
in narrow systems of $P_{\rm orb,i}\la 0.25\;{\rm days}$, lead to unstable RLO and formation of a CE. They argued that the outcome 
of these systems is ejection of the helium envelope, and thus survival of such a (second) CE-phase, and eventually formation of 
tight orbit DNS systems with $P_{\rm orb}\sim 0.01\;{\rm days}$. For the more massive helium stars
with $M_{\rm He,i}>3.8\;M_{\odot}$, and those with $M_{\rm He,i}=3.3-3.8\;M_{\odot}$
in somewhat wider systems of $P_{\rm orb,i}\ga 0.25\;{\rm days}$, they found stable RLO leading to final DNS systems with
$P_{\rm orb}\simeq 0.1-1\;{\rm days}$.
The outcomes of our computations are quite different. From Fig.~\ref{fig:outcome_grid} and Table~\ref{table:models}
it is clearly seen that {\it all} our low-mass ($2.5-6.0\;M_{\odot}$) helium stars are dynamically stable
throughout the RLO. The only exceptions are for $M_{\rm He,i}\ge 3.0\,(2.8)\;M_{\odot}$ in very close initial
orbits of $P_{\rm orb,i}=0.06\;{\rm days}$ where mass transfer is initiated early on the main~sequence (Case~BA).
In some cases our models experience numerical instabilities near the end of the RLO, for example, triggered by a vigorous helium shell flash 
after ignition of oxygen (see Paper~I). However, as demonstrated here (cf. Section~\ref{subsec:CEtimescale} and the Appendix), 
even in case such events lead to a runaway mass transfer 
and the possibility of forming a (second) CE, the remaining lifetime of the donor stars until the SN explosion in these systems is much shorter than the
timescale of the spiral-in process in a dilute envelope with a mass of typically $<0.01\;M_{\odot}$.
In addition, the binding energy of such an envelope is very small, leading to rapid ejection without significant spiral-in.  
Therefore, we do not (directly) produce such extremely short $P_{\rm orb}$ DNS systems as predicted by \citet{dpsv02} and \citet{dp03}.
Of course, all our DNS systems formed with $P_{\rm orb}\la 0.5\;{\rm days}$ will also become extremely tight systems within a few Gyr due to
gravitational wave radiation (Fig.~\ref{fig:merger}), so this issue should not affect much the predictions for the detection rate of LIGO/VIRGO
which are based on population synthesis \citep[e.g.][]{bkb02,vt03}. However, this difference in characteristics for the
newly formed DNS systems is quite important for the numbers of observable DNS systems as radio pulsars with $P_{\rm orb}<1\;{\rm day}$.
 
The results of \citet{ibk+03} are somewhat closer to ours. A remaining key difference is that \citet{ibk+03} find that
their calculated mass-transfer rates for systems with initial orbital periods, $P_{\rm orb,i}<0.4\;{\rm days}$ 
sometime exceed a critical mass-transfer rate, $\dot{M}_{\rm crit}$ 
related to the location of the so-called trapping radius \citep[e.g.][]{beg79,che93,mr15a}. 
In systems with very large super-Eddington mass-transfer rates, matter presumably piles up around the NS and forms a growing,
bloated cloud engulfing a large fraction of the accretion disk. A system will only avoid a CE if it manages to evaporate the bulk of
the transferred matter via the liberated accretion energy at a distance from the NS larger than the trapping radius. Otherwise, 
the incoming material has too much negative binding energy to be ejected. At the same time,
this trapping radius must be located inside the Roche~lobe of the NS in order to avoid a CE \citep{kb99}.
The exact location of the trapping radius, and thus the value of $\dot{M}_{\rm crit}$, is difficult to calculate because it also
depends on the cooling processes of the infalling gas \citep{ny95,bb99}. 
Nevertheless, in all of our models presented in Table~\ref{table:models} with $P_{\rm orb,i}>0.06\;{\rm days}$
we find that $|\dot{M}_{\rm He}^{\rm max}|\le \dot{M}_{\rm crit}$,
where $\dot{M}_{\rm crit}$ is calculated from equation~15 in \citet{ibk+03}, and varies from about $3.7\times 10^{-3}\;M_{\odot}\,{\rm yr}^{-1}$
(for $P_{\rm orb,i}=2.0\;{\rm days}$) to $4.3\times 10^{-4}\;M_{\odot}\,{\rm yr}^{-1}$ (for $P_{\rm orb,i}=0.08\;{\rm days}$). 
Only for $P_{\rm orb,i}=0.06\;{\rm days}$ we find a couple of systems where $|\dot{M}_{\rm He}^{\rm max}|> \dot{M}_{\rm crit}$.
However, these systems are anyway found to result in a runaway mass transfer and thus formation of a CE.
Therefore, our calculated mass-transfer rates from the helium stars donors must be slightly smaller than those calculated by
\citet{ibk+03}. The reason for this, besides from using different stellar evolution codes, is perhaps that our mass-transfer rates 
are calculated using the prescription by \citet{rit88} whereas \citet{ibk+03} adopted the prescription by \citet{te88}.
To summarize, we caution that all numerical calculations of Case~BB RLO could potentially be affected by uncertain accretion processes at 
high mass-transfer rates exceeding the Eddington limit by $\sim\!4$~orders of magnitude.

\citet{dpsv02} evolved a number of helium star donors to become CO~WDs or ONeMg~WDs. For Case~BA RLO
they found that all helium stars with $M_{\rm He,i}=1.5-2.9\;M_{\odot}$ form CO~WDs. This is in good agreement
with our calculations. We find that helium stars with a mass up to (at least, possibly slightly higher)
$2.7\;M_{\odot}$ will produce CO~WDs with a mass of $0.78-0.87\;M_{\odot}$.
For Case~BB RLO they find that $2.1-2.5\;M_{\odot}$ helium stars produce ONeMg~WDs. This is also in fine accordance
with our calculations where we find that ONeMg~WDs (of masses $1.21-1.35\;M_{\odot}$) are produced from
helium stars with an initial mass of up to $2.6-2.7\;M_{\odot}$ for Case~BB RLO (cf. Fig.~\ref{fig:outcome_grid}).

To determine the outcome of our models at the borderline between EC~SNe and Fe~CCSNe we can compare our computations 
to the work of \citet{uyt12} and \citet{jhn+13}. This is necessary at this stage since we could not calculate through
silicon burning until the onset of the core collapse. 
For example, we notice that the computed structure of our $2.9\;M_{\odot}$ helium star with $P_{\rm orb,i}=0.1\;{\rm days}$
(discussed in detail in Paper~I) resembles very well the core of $9.5-11\;M_{\odot}$ single stars computed by these authors and
which were found to produce Fe~CC SNe.

\subsection{Varying the mixing length and the metallicity}\label{subsec:mixing_length}
In our default models, we assumed a mixing-length parameter of $\alpha=l/H_{\rm p}=2.0$ and therefore we
tested a number of models using $\alpha=1.5$ \citep{lan91}.
For close-orbit systems ($P_{\rm orb,i}=0.1\;{\rm days}$), we did not find any differences, whereas
for very wide systems ($P_{\rm orb,i}=20\;{\rm days}$) the stripping of the helium envelope was more efficient 
when applying $\alpha=1.5$, resulting in reduced envelope masses of about 25~per~cent. 
However, in these cases the final metal core masses were only slightly smaller by $< 2\;{\rm per~cent}$, and thus the value of
the mixing-length parameter is not very crucial for the results presented in this study.

We computed models with a helium star metallicity of $Z=0.02$. Given the possibility for long delay times between formation and
merger events, it is possible that a fair fraction of the merging DNS systems that LIGO/VIRGO will detect were formed in an 
evolutionary phase of the local Universe where the metallicity was lower. A decrease in metallicity results in smaller (and bluer)
stars with a weaker stellar wind mass loss. Hence, compared to more metal-rich environments, the outcome is smaller stars in systems which 
widen less due to stellar winds. These two competing effects can, in general, be mimicked for Case~BB RLO by a more metal-rich star in a 
different initial orbit.

\subsection{The ZAMS progenitor masses of our naked helium stars}
The ZAMS mass interval of the original hydrogen-rich progenitor stars, of the naked helium stars studied here, is  
difficult to determine precisely 
since it depends on the initial ZAMS binary separation and the modelling of convective core-overshooting, rotation, chemical mixing, metallicity, as well
as the physics of the ejection process of the CE.
From a density profile analysis of the stellar models evolved by \citet{bdc+11}, we find that a $10\;M_{\odot}$ ZAMS star ($Z=0.02$, non-rotating) produces a
helium core of $2.4-3.2\;M_{\odot}$ at the tip of the red giant branch, for a convective core-overshooting parameter in the interval $\delta_{\rm OV}=0.1-0.335$.
Similarly, a $20\;M_{\odot}$ ZAMS star produces a helium core in the interval $7.7-8.9\;M_{\odot}$. 
These helium cores will be exposed via the CE and spiral-in phase as soon as the progenitor star fills its Roche lobe in the HMXB system.
The thickness of the remaining outer hydrogen-rich layer at this point is somewhat uncertain.
The location of the hydrogen shell source in these progenitor stars is initially located at smaller mass coordinates and thus the final (pure) helium star
masses are slightly smaller. Furthermore, these (now) naked helium cores, or Wolf-Rayet stars, will subsequently reduce their masses further 
by wind mass loss \citep{wl99}. 

The initial masses of our binary stars which produce EC~SNe are most likely found in the interval 
$8-11\;M_{\odot}$ \citep[][and references therein]{plp+04,uyt12,jhn+13,jhn14,tyu13,dgs+15}. 
It has previously been demonstrated that the initial mass range for producing EC~SNe is smaller in single stars \citep{phlh08}
compared to binary stars in non-degenerate systems \citep{plp+04}. 

\subsection{Stripping by a black~hole or a WD companion star}\label{subsec:WDBH-stripping}
Replacing the accreting NS with an accreting black~hole (BH) should also lead to ultra-stripped SNe in close binaries.
As a consequence of their larger masses, we expect accreting BHs would give rise to slightly smaller mass-transfer rates (more stable Case~BB RLO) and also  
lead to wider pre-SN orbits. However, given the initial mass function, we expect less ultra-stripped SNe to occur in binaries with BHs. This is also
reflected in the hitherto missing detection of a radio pulsar--BH binary (the descendent system of such a helium~star--BH binary).

Whether or not a massive WD is able to strip a helium star efficiently prior to a SN is more questionable.
We know of two WD--NS systems, PSR~B2303+46 \citep[$P_{\rm orb}=12\;{\rm days}$,][]{std85,kk99} and
PSR~J1141$-$6545 \citep[$P_{\rm orb}=0.20\;{\rm days}$,][]{klm+00,abw+11} where the observed radio pulsar was created
{\it after} the WD \citep{ts00}. Hence, it is quite likely that these systems evolved through a post-CE Case~BB RLO phase  
where a massive WD was indeed accreting from an evolved helium star prior to its explosion. It is also clear in
these two cases that the systems remained dynamically stable and that the WDs were able to survive the presumably highly super-Eddington mass-transfer rate 
($\sim\!10^{-4}\;M_{\odot}\,{\rm yr}^{-1}$, Section~\ref{sec:HeMdot}) from their helium star companion without merging.
How the excess material was ejected from the system in these cases is not well understood. However, it is possible that
significant stripping of the helium star prior to the SN was at work too in both of these systems.

\subsection{Effects of a dilute envelope on the pre-SN evolution}\label{subsec:CEtimescale}
As also discussed in detail by \citet{dp03} and \citet{ibk+03}, runaway mass transfer in helium star--NS binaries may lead to the onset of a (second) CE evolution. 
The final outcome is determined by the competition between the timescale for spiral-in of the NS and the remaining lifetime 
of the pre-SN donor star before gravitational collapse.
In general, the duration of the CE and spiral-in phase is found to be $<10^3\;{\rm yr}$ \citep[e.g.][]{pod01},
which is long compared to the estimated final lifetime of our models (in some cases just a few days). 
In the Appendix we estimate the timescale of NS spiral-in caused by a helium shell flash following ignition of oxygen burning.
We find that the spiral-in timescale is often a factor of (at least) 100 longer than the remaining lifetime
of the evolved helium star until the SN explosion. Hence, it is safe to ignore the expanding envelope at the late
stages of evolution when determining the final outcome of the models.

\section{Summary}\label{sec:summary}
We have performed a systematic study of the evolution of close binaries containing a helium star and a NS.
These systems are the descendants of HMXBs which undergo CE and spiral-in evolution and expose 
the core of an OB-star as a naked helium star. The subsequent evolution of these helium stars causes
an additional mass-transfer towards the NS, often during helium shell burning (i.e. so-called Case~BB RLO) when 
they expand to become giant stars again.
In particular, we have investigated the expected nature of the resultant SN explosion (EC~SN vs Fe~CCSN) as well as the amount of stripping 
of the helium star donors prior to the explosion in order to predict the observational properties of these events as SN~Ic or SN~Ib. 

We studied helium stars with masses between $2.5-10.0\;M_{\odot}$ with an emphasis on the lighter ones with
masses of $2.5-3.5\;M_{\odot}$ which expand more during their evolution.
In all cases we assumed an accreting NS with an initial mass of $1.35\;M_{\odot}$ and varied the initial
orbital period between $0.06-120~\;{\rm days}$, leading to either Case~BA, Case~BB or Case~BC RLO. 
For comparison, we also evolved a sequence of isolated helium stars.
The final fates of our helium stars span from CO~WDs and ONeMg~WDs to NSs produced via EC~SNe and Fe~CCSNe.
In the latter case, we were able to compute the evolution all the way to the onset of silicon burning,
a few days prior to the gravitational collapse.
The results of our study are presented in a table (Table~\ref{table:models}) with several parameters and a chart (Fig.~\ref{fig:outcome_grid}) 
with the mapping of initial parameter space to final outcome. In addition, fitting formulae are derived with easy
applications to, for example, future population synthesis studies. 

We define ultra-stripped SNe as exploding stars which contain envelope masses $\la 0.2\;M_{\odot}$ (Fig~\ref{fig:Mcore-Menv}) and having a compact star companion. 
The compact nature of their companions allows for extreme stripping in a tight binaries which is not possible otherwise. 
We find several cases of Case~BB RLO with $P_{\rm orb,i}<2\;{\rm days}$ leading to ultra-stripped SNe, i.e. exploding stars which are almost pure naked metal cores
with only a tiny amount of helium envelope left, in some cases only $\sim\!0.005\;M_{\odot}$ for EC~SNe and $\sim\!0.02\;M_{\odot}$ for Fe~CCSNe. 
In general, we find that the progenitors of EC~SNe are stripped slightly further than those of Fe~CCSNe, meaning that
our EC~SNe are expected to be mainly Type~Ic SNe, whereas the Fe~CCSNe from these systems can be either of Type~Ic or Type~Ib.
The amount of helium remaining in the envelope is found to be correlated with the initial orbital period (Figs.~\ref{fig:He-env} and \ref{fig:wideDNS}), and thus also
the final pre-SN orbital period, of the binary system.
In addition, there is a correlation of remaining total envelope mass with the final core mass of the exploding star (Fig.~\ref{fig:Mcore-Menv}).  

From our modelling we expect ultra-stripped SNe to produce NSs (the second formed NS in DNS systems) within a relatively wide range of masses 
(possibly even in the entire range of $1.10-1.80\;M_{\odot}$, Fig.~\ref{fig:cross-section-SN}).
Besides forming low-mass NSs of $1.10-1.40\;M_{\odot}$ from almost naked metal cores barely above the Chandrasekhar mass, we find cases 
where we expect ultra-stripped SNe to produce more massive NSs in tight orbits, originating from metal cores of evolved helium stars with initial masses of $4-6\;M_{\odot}$. 

On the basis of our modelling, we conclude that EC~SNe only occur for a narrow interval of initial helium star masses of $M_{\rm He,i}=2.60-2.95\;M_{\odot}$ 
depending on $P_{\rm orb,i}$ (Fig.~\ref{fig:Mcore_final}).
The general outcome of our investigated binaries, however, is an Fe~CCSN above this mass interval 
and an ONeMg~WD or a CO~WD for smaller masses. 
We discuss one peculiar model where a $2.60\;M_{\odot}$ helium star donor completely shuts off core nuclear burning as a consequence of extreme mass transfer 
(Figs.~\ref{fig:kippenhahn_COWD}--\ref{fig:Mdot_COWD}) and we followed its subsequent evolution as a CO~WD to the 
UCXB-phase where it filled its Roche~lobe with an orbital period of about 30~seconds.

The maximum mass-transfer rate during stable RLO from all the binary helium stars is close to $\sim\!10^{-4}\;M_{\odot}\,{\rm yr}^{-1}$
within a factor of a few (cf. Figs.~\ref{fig:Mdot_max}--\ref{fig:Mdot_3.0}). 
The duration of the mass-transfer phase is anti-correlated with the initial $P_{\rm orb}$, which results in the prediction of 
a correlation between $P_{\rm orb}$ and $P_{\rm spin}$ for the mildly recycled pulsars in DNS systems.
The reason is that in closer orbits the duration of the mass-transfer phase is longer (Figs.~\ref{fig:Mdot_3.0} and \ref{fig:wideDNS}), which results in more mass 
being accreted by the NS, leading to more efficient recycling and thus a smaller value of $P_{\rm spin}$ (and assuming this relation to 
survive the dynamical effects of the SN explosion).

The main channel for producing DNS systems which will merge within a Hubble time is via ultra-stripped SNe.
It will therefore be useful to compare upcoming DNS merger rates from LIGO/VIRGO with future observational limits on the detection rate of ultra-stripped SNe.
The result of this comparison may help to answer whether, and to which extent, ultra-stripped SNe also occur in binaries with a WD or a black hole companion star.
However, this step requires a clear method to distinguish the light curves of ultra-stripped SNe from other fast decaying and relatively weak
light curves, e.g. resulting from partial explosions \citep['.Ia' SNe,][]{bswn07} or the accretion-induced collapse (AIC) of a WD \citep{dbo+06}
-- see also Section~\ref{subsec:spectra} for discussions and references to spectroscopic signatures of diverse SNe.  

Ultra-stripped SNe have small amounts of ejecta mass, often with a low binding energy (Fig.~\ref{fig:Ebind}). We therefore speculate that ultra-stripped SNe 
(both EC~SNe and, at least, Fe~CCSNe with relatively small iron cores) lead to fast explosions and often produce small kicks. 

We also performed a brief investigation of helium~star--NS binaries with $P_{\rm orb,i}>10\;{\rm days}$ (Fig.~\ref{fig:wideDNS}) 
leading to the formation of wide-orbit DNS systems after the second SN explosion. 
These DNS systems often have $P_{\rm orb}>20\;{\rm days}$ and are characterized by inefficient recycling,
often leading to spin periods of the marginally recycled pulsar in excess of 100~ms. Furthermore, their eccentricities are expected to be large
(0.2--0.5, even for symmetric SNe with no kicks) because such exploding stars eject fairly thick envelopes and their pre-SN orbits are relatively wide,  
and thus weakly bound. 
In a future publication (Tauris et al., in~prep.) we analyse in more detail the observed distributions of orbital periods, spin periods, eccentricities 
and systemic velocities of DNS systems and confront them with simulations and expectations from ultra-stripped SNe. 

In a few cases, envelope expansion triggered by a helium shell flash following ignition of oxygen burning 
leads to a CE-like situation with the NS embedded in the outer helium envelope of the donor star.
However, we demonstrate that such events will not lead to a runaway mass-transfer (and a dynamical instability)
which would cause the NS to merge with the core of its companion star. The reason is that the gas in the
expanded envelope is so dilute that the timescale of spiral-in (cf. Appendix)   
often exceeds the remaining lifetime of the pre-SN star by a factor of (at least) 100.
In addition, the binding energy of this expanded envelope is much smaller than the release of orbital energy
from a slight spiral-in in these tight orbits.
Thus we conclude that it is safe to ignore the expanding envelope at the very late
stages of evolution when determining the final outcome of such models.

\section*{Acknowledgements}
We thank the anonymous referee for many valuable suggestions, and
also Paulo Freire for discussions on DNS systems. 
TMT acknowledges the receipt of DFG Grant: TA 964/1-1.

\bibliographystyle{mn2e}
\bibliography{tauris_refs}

\begin{thebibliography}{}

\bibitem[\protect\citeauthoryear{{Aasi}, {Abadie}, {Abbott}, {Abbott},
  {Abbott}, {Abernathy}, {Accadia}, {Acernese} \& et al.}{{Aasi}
  et~al.}{2013}]{aaa+13}
{Aasi} J.,  {Abadie} J.,  {Abbott} B.~P.,  {Abbott} R.,  {Abbott} T.~D.,
  {Abernathy} M.,  {Accadia} T.,  {Acernese} F.,    et al. 2013, LIGO-P1200087,
  VIR-0288A-12, ArXiv astro-ph:1304.0670

\bibitem[\protect\citeauthoryear{{Abramowicz}, {Calvani} \&
  {Nobili}}{{Abramowicz} et~al.}{1980}]{acn80}
{Abramowicz} M.~A.,  {Calvani} M.,    {Nobili} L.,  1980, \apj, 242, 772

\bibitem[\protect\citeauthoryear{{Antoniadis}, {Bassa}, {Wex}, {Kramer} \&
  {Napiwotzki}}{{Antoniadis} et~al.}{2011}]{abw+11}
{Antoniadis} J.,  {Bassa} C.~G.,  {Wex} N.,  {Kramer} M.,    {Napiwotzki} R.,
  2011, \mnras, 412, 580

\bibitem[\protect\citeauthoryear{{Antoniadis}, {Freire}, {Wex}, {Tauris},
  {Lynch}, {van Kerkwijk}, {Kramer} \& {et al.}}{{Antoniadis}
  et~al.}{2013}]{afw+13}
{Antoniadis} J.,  {Freire} P.~C.~C.,  {Wex} N.,  {Tauris} T.~M.,  {Lynch}
  R.~S.,  {van Kerkwijk} M.~H.,  {Kramer} M.,    {et al.} 2013, Science, 340,
  448

\bibitem[\protect\citeauthoryear{{Arnett}}{{Arnett}}{1979}]{arn79}
{Arnett} W.~D.,  1979, \apjl, 230, L37

\bibitem[\protect\citeauthoryear{{Arnett}}{{Arnett}}{1982}]{arn82}
{Arnett} W.~D.,  1982, \apj, 253, 785

\bibitem[\protect\citeauthoryear{{Basko} \& {Sunyaev}}{{Basko} \&
  {Sunyaev}}{1976}]{bs76}
{Basko} M.~M.,  {Sunyaev} R.~A.,  1976, \mnras, 175, 395

\bibitem[\protect\citeauthoryear{{Begelman}}{{Begelman}}{1979}]{beg79}
{Begelman} M.~C.,  1979, \mnras, 187, 237

\bibitem[\protect\citeauthoryear{{Belczynski}, {Kalogera} \&
  {Bulik}}{{Belczynski} et~al.}{2002}]{bkb02}
{Belczynski} K.,  {Kalogera} V.,    {Bulik} T.,  2002, \apj, 572, 407

\bibitem[\protect\citeauthoryear{{Benvenuto}, {De Vito} \&
  {Horvath}}{{Benvenuto} et~al.}{2015}]{bdh15}
{Benvenuto} O.~G.,  {De Vito} M.~A.,    {Horvath} J.~E.,  2015, \apj, 798, 44

\bibitem[\protect\citeauthoryear{Bhattacharya \& {van den Heuvel}}{Bhattacharya
  \& {van den Heuvel}}{1991}]{bv91}
Bhattacharya D.,  {van den Heuvel} E. P.~J.,  1991, Physics Reports, 203, 1

\bibitem[\protect\citeauthoryear{{Bildsten}, {Shen}, {Weinberg} \&
  {Nelemans}}{{Bildsten} et~al.}{2007}]{bswn07}
{Bildsten} L.,  {Shen} K.~J.,  {Weinberg} N.~N.,    {Nelemans} G.,  2007,
  \apjl, 662, L95

\bibitem[\protect\citeauthoryear{{Blandford} \& {Begelman}}{{Blandford} \&
  {Begelman}}{1999}]{bb99}
{Blandford} R.~D.,  {Begelman} M.~C.,  1999, \mnras, 303, L1

\bibitem[\protect\citeauthoryear{{Blondin} \& {Mezzacappa}}{{Blondin} \&
  {Mezzacappa}}{2006}]{bm06}
{Blondin} J.~M.,  {Mezzacappa} A.,  2006, \apj, 642, 401

\bibitem[\protect\citeauthoryear{{Bondi} \& {Hoyle}}{{Bondi} \&
  {Hoyle}}{1944}]{bh44}
{Bondi} H.,  {Hoyle} F.,  1944, \mnras, 104, 273

\bibitem[\protect\citeauthoryear{{Brandt} \& {Podsiadlowski}}{{Brandt} \&
  {Podsiadlowski}}{1995}]{bp95}
{Brandt} N.,  {Podsiadlowski} P.,  1995, \mnras, 274, 461

\bibitem[\protect\citeauthoryear{{Brott}, {de Mink}, {Cantiello}, {Langer}, {de
  Koter}, {Evans}, {Hunter}, {Trundle} \& {Vink}}{{Brott}
  et~al.}{2011}]{bdc+11}
{Brott} I.,  {de Mink} S.~E.,  {Cantiello} M.,  {Langer} N.,  {de Koter} A.,
  {Evans} C.~J.,  {Hunter} I.,  {Trundle} C.,    {Vink} J.~S.,  2011, \aap,
  530, A115

\bibitem[\protect\citeauthoryear{{Brown}, {Heger}, {Langer}, {Lee}, {Wellstein}
  \& {Bethe}}{{Brown} et~al.}{2001}]{bhl+01}
{Brown} G.~E.,  {Heger} A.,  {Langer} N.,  {Lee} C.,  {Wellstein} S.,
  {Bethe} H.~A.,  2001, New Astronomy, 6, 457

\bibitem[\protect\citeauthoryear{{Bruenn}, {Lentz}, {Hix}, {Mezzacappa},
  {Harris}, {Bronson Messer}, {Endeve}, {Blondin}, {Chertkow}, {Lingerfelt},
  {Marronetti} \& {Yakunin}}{{Bruenn} et~al.}{2014}]{blh+15}
{Bruenn} S.~W.,  {Lentz} E.~J.,  {Hix} W.~R.,  {Mezzacappa} A.,  {Harris}
  J.~A.,  {Bronson Messer} O.~E.,  {Endeve} E.,  {Blondin} J.~M.,  {Chertkow}
  M.~A.,  {Lingerfelt} E.~J.,  {Marronetti} P.,    {Yakunin} K.~N.,  2014,
  ArXiv astro-ph:1409.5779

\bibitem[\protect\citeauthoryear{{Burgay}, {D'Amico}, {Possenti}, {Manchester},
  {Lyne}, {Joshi}, {McLaughlin}, {Kramer}, {Sarkissian}, {Camilo}, {Kalogera},
  {Kim} \& {Lorimer}}{{Burgay} et~al.}{2003}]{bdp+03}
{Burgay} M.,  {D'Amico} N.,  {Possenti} A.,  {Manchester} R.~N.,  {Lyne} A.~G.,
   {Joshi} B.~C.,  {McLaughlin} M.~A.,  {Kramer} M.,  {Sarkissian} J.~M.,
  {Camilo} F.,  {Kalogera} V.,  {Kim} C.,    {Lorimer} D.~R.,  2003, \nat, 426,
  531

\bibitem[\protect\citeauthoryear{{Canuto}}{{Canuto}}{1970}]{can70}
{Canuto} V.,  1970, \apj, 159, 641

\bibitem[\protect\citeauthoryear{{Chaurasia} \& {Bailes}}{{Chaurasia} \&
  {Bailes}}{2005}]{cb05}
{Chaurasia} H.~K.,  {Bailes} M.,  2005, \apj, 632, 1054

\bibitem[\protect\citeauthoryear{Chevalier}{Chevalier}{1993}]{che93}
Chevalier R.~A.,  1993, \apj, 411, L33

\bibitem[\protect\citeauthoryear{{Chini}, {Hoffmeister}, {Nasseri}, {Stahl} \&
  {Zinnecker}}{{Chini} et~al.}{2012}]{chn+12}
{Chini} R.,  {Hoffmeister} V.~H.,  {Nasseri} A.,  {Stahl} O.,    {Zinnecker}
  H.,  2012, \mnras, 424, 1925

\bibitem[\protect\citeauthoryear{{Coleiro} \& {Chaty}}{{Coleiro} \&
  {Chaty}}{2013}]{cc13}
{Coleiro} A.,  {Chaty} S.,  2013, \apj, 764, 185

\bibitem[\protect\citeauthoryear{{Crowther}}{{Crowther}}{2013}]{cro13}
{Crowther} P.~A.,  2013, \mnras, 428, 1927

\bibitem[\protect\citeauthoryear{{De Greve} \& {De Loore}}{{De Greve} \& {De
  Loore}}{1977}]{dd77}
{De Greve} J.-P.,  {De Loore} C.,  1977, \apss, 50, 75

\bibitem[\protect\citeauthoryear{{de Mink}}{{de Mink}}{2010}]{deM10}
{de Mink} S.~E.,  2010, PhD thesis, Utrecht University

\bibitem[\protect\citeauthoryear{{Delgado} \& {Thomas}}{{Delgado} \&
  {Thomas}}{1981}]{dt81}
{Delgado} A.~J.,  {Thomas} H.-C.,  1981, \aap, 96, 142

\bibitem[\protect\citeauthoryear{{Dermer} \& {Atoyan}}{{Dermer} \&
  {Atoyan}}{2006}]{da06}
{Dermer} C.~D.,  {Atoyan} A.,  2006, \apjl, 643, L13

\bibitem[\protect\citeauthoryear{{Dessart}, {Burrows}, {Ott}, {Livne}, {Yoon}
  \& {Langer}}{{Dessart} et~al.}{2006}]{dbo+06}
{Dessart} L.,  {Burrows} A.,  {Ott} C.~D.,  {Livne} E.,  {Yoon} S.-C.,
  {Langer} N.,  2006, \apj, 644, 1063

\bibitem[\protect\citeauthoryear{{Dessart} \& {Hillier}}{{Dessart} \&
  {Hillier}}{2015}]{dh15}
{Dessart} L.,  {Hillier} D.~J.,  2015, \mnras, 447, 1370

\bibitem[\protect\citeauthoryear{{Dessart}, {Hillier}, {Livne}, {Yoon},
  {Woosley}, {Waldman} \& {Langer}}{{Dessart} et~al.}{2011}]{dhl+11}
{Dessart} L.,  {Hillier} D.~J.,  {Livne} E.,  {Yoon} S.-C.,  {Woosley} S.,
  {Waldman} R.,    {Langer} N.,  2011, \mnras, 414, 2985

\bibitem[\protect\citeauthoryear{{Detmers}, {Langer}, {Podsiadlowski} \&
  {Izzard}}{{Detmers} et~al.}{2008}]{dlpi08}
{Detmers} R.~G.,  {Langer} N.,  {Podsiadlowski} P.,    {Izzard} R.~G.,  2008,
  \aap, 484, 831

\bibitem[\protect\citeauthoryear{{Dewi}, {Podsiadlowski} \& {Pols}}{{Dewi}
  et~al.}{2005}]{dpp05}
{Dewi} J.~D.~M.,  {Podsiadlowski} P.,    {Pols} O.~R.,  2005, \mnras, 363, L71

\bibitem[\protect\citeauthoryear{{Dewi} \& {Pols}}{{Dewi} \&
  {Pols}}{2003}]{dp03}
{Dewi} J.~D.~M.,  {Pols} O.~R.,  2003, \mnras, 344, 629

\bibitem[\protect\citeauthoryear{{Dewi}, {Pols}, {Savonije} \& {van den
  Heuvel}}{{Dewi} et~al.}{2002}]{dpsv02}
{Dewi} J.~D.~M.,  {Pols} O.~R.,  {Savonije} G.~J.,    {van den Heuvel}
  E.~P.~J.,  2002, \mnras, 331, 1027

\bibitem[\protect\citeauthoryear{{Dewi} \& {Tauris}}{{Dewi} \&
  {Tauris}}{2000}]{dt00}
{Dewi} J. D.~M.,  {Tauris} T.~M.,  2000, \aap, 360, 1043

\bibitem[\protect\citeauthoryear{{Dewi} \& {Tauris}}{{Dewi} \&
  {Tauris}}{2001}]{dt01}
{Dewi} J.~D.~M.,  {Tauris} T.~M.,  2001, in {P.~Podsiadlowski, S.~Rappaport,
  A.~R.~King, F.~D'Antona, \& L.~Burderi } ed., Evolution of Binary and
  Multiple Star Systems Vol.~229 of Astronomical Society of the Pacific
  Conference Series, {On the {$\lambda$}-Parameter of the Common Envelope
  Evolution}.
p.~255

\bibitem[\protect\citeauthoryear{{Doherty}, {Gil-Pons}, {Siess}, {Lattanzio} \&
  {Lau}}{{Doherty} et~al.}{2015}]{dgs+15}
{Doherty} C.~L.,  {Gil-Pons} P.,  {Siess} L.,  {Lattanzio} J.~C.,    {Lau}
  H.~H.~B.,  2015, \mnras, 446, 2599

\bibitem[\protect\citeauthoryear{{Drout}, {Chornock}, {Soderberg} \& {et
  al.}}{{Drout} et~al.}{2014}]{dcs+14}
{Drout} M.~R.,  {Chornock} R.,  {Soderberg} A.~M.,    {et al.} 2014, \apj, 794,
  23

\bibitem[\protect\citeauthoryear{{Drout}, {Soderberg}, {Mazzali} \& {et
  al.}}{{Drout} et~al.}{2013}]{dsm+13}
{Drout} M.~R.,  {Soderberg} A.~M.,  {Mazzali} P.~A.,    {et al.} 2013, \apj,
  774, 58

\bibitem[\protect\citeauthoryear{Eggleton}{Eggleton}{1983}]{egg83}
Eggleton P.~P.,  1983, \apj, 268, 368

\bibitem[\protect\citeauthoryear{{Eldridge}, {Izzard} \& {Tout}}{{Eldridge}
  et~al.}{2008}]{eit08}
{Eldridge} J.~J.,  {Izzard} R.~G.,    {Tout} C.~A.,  2008, \mnras, 384, 1109

\bibitem[\protect\citeauthoryear{{Foglizzo}, {Galletti}, {Scheck} \&
  {Janka}}{{Foglizzo} et~al.}{2007}]{fgsj07}
{Foglizzo} T.,  {Galletti} P.,  {Scheck} L.,    {Janka} H.-T.,  2007, \apj,
  654, 1006

\bibitem[\protect\citeauthoryear{{Foglizzo}, {Kazeroni}, {Guilet}, {Masset},
  {Gonz{\'a}lez}, {Krueger}, {Novak}, {Oertel}, {Margueron}, {Faure}, {Martin},
  {Blottiau}, {Peres} \& {Durand}}{{Foglizzo} et~al.}{2015}]{fkg+15}
{Foglizzo} T.,  {Kazeroni} R.,  {Guilet} J.,  {Masset} F.,  {Gonz{\'a}lez} M.,
  {Krueger} B.~K.,  {Novak} J.,  {Oertel} M.,  {Margueron} J.,  {Faure} J.,
  {Martin} N.,  {Blottiau} P.,  {Peres} B.,    {Durand} G.,  2015, ArXiv
  astro-ph:1501.01334

\bibitem[\protect\citeauthoryear{{Frank}, {King} \& {Raine}}{{Frank}
  et~al.}{2002}]{fkr02}
{Frank} J.,  {King} A.,    {Raine} D.~J.,  2002, {Accretion Power in
  Astrophysics: Third Edition}.
Cambridge University Press

\bibitem[\protect\citeauthoryear{{Fryer}, {Belczynski}, {Wiktorowicz},
  {Dominik}, {Kalogera} \& {Holz}}{{Fryer} et~al.}{2012}]{fbw+12}
{Fryer} C.~L.,  {Belczynski} K.,  {Wiktorowicz} G.,  {Dominik} M.,  {Kalogera}
  V.,    {Holz} D.~E.,  2012, \apj, 749, 91

\bibitem[\protect\citeauthoryear{{Habets}}{{Habets}}{1986}]{hab86a}
{Habets} G.~M.~H.~J.,  1986, \aap, 165, 95

\bibitem[\protect\citeauthoryear{{Hachinger}, {Mazzali}, {Taubenberger},
  {Hillebrandt}, {Nomoto} \& {Sauer}}{{Hachinger} et~al.}{2012}]{hmt+12}
{Hachinger} S.,  {Mazzali} P.~A.,  {Taubenberger} S.,  {Hillebrandt} W.,
  {Nomoto} K.,    {Sauer} D.~N.,  2012, \mnras, 422, 70

\bibitem[\protect\citeauthoryear{{Hamann}, {Koesterke} \&
  {Wessolowski}}{{Hamann} et~al.}{1995}]{hkw95}
{Hamann} W.-R.,  {Koesterke} L.,    {Wessolowski} U.,  1995, \aap, 299, 151

\bibitem[\protect\citeauthoryear{{Hamann}, {Schoenberner} \& {Heber}}{{Hamann}
  et~al.}{1982}]{hsh82}
{Hamann} W.-R.,  {Schoenberner} D.,    {Heber} U.,  1982, \aap, 116, 273

\bibitem[\protect\citeauthoryear{{Heger}, {Langer} \& {Woosley}}{{Heger}
  et~al.}{2000}]{hlw00}
{Heger} A.,  {Langer} N.,    {Woosley} S.~E.,  2000, \apj, 528, 368

\bibitem[\protect\citeauthoryear{{Hills}}{{Hills}}{1983}]{hil83}
{Hills} J.~G.,  1983, \apj, 267, 322

\bibitem[\protect\citeauthoryear{{Hobbs}, {Lorimer}, {Lyne} \&
  {Kramer}}{{Hobbs} et~al.}{2005}]{hllk05}
{Hobbs} G.,  {Lorimer} D.~R.,  {Lyne} A.~G.,    {Kramer} M.,  2005, \mnras,
  360, 974

\bibitem[\protect\citeauthoryear{{Huang}}{{Huang}}{1963}]{hua63}
{Huang} S.-S.,  1963, \apj, 138, 471

\bibitem[\protect\citeauthoryear{{Hubbard} \& {Lampe}}{{Hubbard} \&
  {Lampe}}{1969}]{hl69}
{Hubbard} W.~B.,  {Lampe} M.,  1969, \apjs, 18, 297

\bibitem[\protect\citeauthoryear{{Iglesias} \& {Rogers}}{{Iglesias} \&
  {Rogers}}{1996}]{ir96}
{Iglesias} C.~A.,  {Rogers} F.~J.,  1996, \apj, 464, 943

\bibitem[\protect\citeauthoryear{{Istrate}, {Tauris} \& {Langer}}{{Istrate}
  et~al.}{2014}]{itl14}
{Istrate} A.~G.,  {Tauris} T.~M.,    {Langer} N.,  2014, \aap, 571, A45

\bibitem[\protect\citeauthoryear{{Ivanova}, {Belczynski}, {Kalogera}, {Rasio}
  \& {Taam}}{{Ivanova} et~al.}{2003}]{ibk+03}
{Ivanova} N.,  {Belczynski} K.,  {Kalogera} V.,  {Rasio} F.~A.,    {Taam}
  R.~E.,  2003, \apj, 592, 475

\bibitem[\protect\citeauthoryear{{Ivanova}, {Justham}, {Chen}, {De Marco},
  {Fryer}, {Gaburov}, {Ge}, {Glebbeek}, {Han}, {Li}, {Lu}, {Marsh},
  {Podsiadlowski}, {Potter}, {Soker}, {Taam}, {Tauris}, {van den Heuvel} \&
  {Webbink}}{{Ivanova} et~al.}{2013}]{ijc+13}
{Ivanova} N.,  {Justham} S.,  {Chen} X.,  {De Marco} O.,  {Fryer} C.~L.,
  {Gaburov} E.,  {Ge} H.,  {Glebbeek} E.,  {Han} Z.,  {Li} X.-D.,  {Lu} G.,
  {Marsh} T.,  {Podsiadlowski} P.,  {Potter} A.,  {Soker} N.,  {Taam} R.,
  {Tauris} T.~M.,  {van den Heuvel} E.~P.~J.,    {Webbink} R.~F.,  2013, \aapr,
  21, 59

\bibitem[\protect\citeauthoryear{{Janka}}{{Janka}}{2012}]{jan12}
{Janka} H.-T.,  2012, Annual Review of Nuclear and Particle Science, 62, 407

\bibitem[\protect\citeauthoryear{{Janka}}{{Janka}}{2013}]{jan13}
{Janka} H.-T.,  2013, \mnras, 434, 1355

\bibitem[\protect\citeauthoryear{{Jones}, {Hirschi} \& {Nomoto}}{{Jones}
  et~al.}{2014}]{jhn14}
{Jones} S.,  {Hirschi} R.,    {Nomoto} K.,  2014, \apj, 797, 83

\bibitem[\protect\citeauthoryear{{Jones}, {Hirschi}, {Nomoto}, {Fischer},
  {Timmes}, {Herwig}, {Paxton}, {Toki}, {Suzuki}, {Mart{\'{\i}}nez-Pinedo},
  {Lam} \& {Bertolli}}{{Jones} et~al.}{2013}]{jhn+13}
{Jones} S.,  {Hirschi} R.,  {Nomoto} K.,  {Fischer} T.,  {Timmes} F.~X.,
  {Herwig} F.,  {Paxton} B.,  {Toki} H.,  {Suzuki} T.,
  {Mart{\'{\i}}nez-Pinedo} G.,  {Lam} Y.~H.,    {Bertolli} M.~G.,  2013, \apj,
  772, 150

\bibitem[\protect\citeauthoryear{{Kasliwal}, {Kulkarni}, {Gal-Yam} \& {et
  al.}}{{Kasliwal} et~al.}{2010}]{kkg+10}
{Kasliwal} M.~M.,  {Kulkarni} S.~R.,  {Gal-Yam} A.,    {et al.} 2010, \apjl,
  723, L98

\bibitem[\protect\citeauthoryear{{Kasliwal} \& {Nissanke}}{{Kasliwal} \&
  {Nissanke}}{2014}]{kn14}
{Kasliwal} M.~M.,  {Nissanke} S.,  2014, \apjl, 789, L5

\bibitem[\protect\citeauthoryear{{Kaspi}, {Lyne}, {Manchester}, {Crawford},
  {Camilo}, {Bell}, {D'Amico}, {Stairs}, {McKay}, {Morris} \&
  {Possenti}}{{Kaspi} et~al.}{2000}]{klm+00}
{Kaspi} V.~M.,  {Lyne} A.~G.,  {Manchester} R.~N.,  {Crawford} F.,  {Camilo}
  F.,  {Bell} J.~F.,  {D'Amico} N.,  {Stairs} I.~H.,  {McKay} N.~P.~F.,
  {Morris} D.~J.,    {Possenti} A.,  2000, \apj, 543, 321

\bibitem[\protect\citeauthoryear{{Keith}, {Kramer}, {Lyne}, {Eatough},
  {Stairs}, {Possenti}, {Camilo} \& {Manchester}}{{Keith}
  et~al.}{2009}]{kkl+09}
{Keith} M.~J.,  {Kramer} M.,  {Lyne} A.~G.,  {Eatough} R.~P.,  {Stairs} I.~H.,
  {Possenti} A.,  {Camilo} F.,    {Manchester} R.~N.,  2009, \mnras, 393, 623

\bibitem[\protect\citeauthoryear{{King} \& {Begelman}}{{King} \&
  {Begelman}}{1999}]{kb99}
{King} A.~R.,  {Begelman} M.~C.,  1999, \apjl, 519, L169

\bibitem[\protect\citeauthoryear{{King}, {Schenker}, {Kolb} \& {Davies}}{{King}
  et~al.}{2001}]{kskd01}
{King} A.~R.,  {Schenker} K.,  {Kolb} U.,    {Davies} M.~B.,  2001, \mnras,
  321, 327

\bibitem[\protect\citeauthoryear{{Kippenhahn} \& {Weigert}}{{Kippenhahn} \&
  {Weigert}}{1990}]{kw90}
{Kippenhahn} R.,  {Weigert} A.,  1990, {Stellar Structure and Evolution}.
Springer, Berlin

\bibitem[\protect\citeauthoryear{{Kitaura}, {Janka} \& {Hillebrandt}}{{Kitaura}
  et~al.}{2006}]{kjh06}
{Kitaura} F.~S.,  {Janka} H.-T.,    {Hillebrandt} W.,  2006, \aap, 450, 345

\bibitem[\protect\citeauthoryear{{Kleiser} \& {Kasen}}{{Kleiser} \&
  {Kasen}}{2014}]{kk14}
{Kleiser} I.~K.~W.,  {Kasen} D.,  2014, \mnras, 438, 318

\bibitem[\protect\citeauthoryear{{Kozyreva}, {Yoon} \& {Langer}}{{Kozyreva}
  et~al.}{2014}]{kyl14}
{Kozyreva} A.,  {Yoon} S.-C.,    {Langer} N.,  2014, \aap, 566, A146

\bibitem[\protect\citeauthoryear{{Lamb}, {Pethick} \& {Pines}}{{Lamb}
  et~al.}{1973}]{lpp73}
{Lamb} F.~K.,  {Pethick} C.~J.,    {Pines} D.,  1973, \apj, 184, 271

\bibitem[\protect\citeauthoryear{{Langer}}{{Langer}}{1991}]{lan91}
{Langer} N.,  1991, \aap, 252, 669

\bibitem[\protect\citeauthoryear{{Langer}}{{Langer}}{1998}]{lan98}
{Langer} N.,  1998, \aap, 329, 551

\bibitem[\protect\citeauthoryear{{Langer}}{{Langer}}{2012}]{lan12}
{Langer} N.,  2012, \araa, 50, 107

\bibitem[\protect\citeauthoryear{{Langer}, {Fricke} \& {Sugimoto}}{{Langer}
  et~al.}{1983}]{lfs83}
{Langer} N.,  {Fricke} K.~J.,    {Sugimoto} D.,  1983, \aap, 126, 207

\bibitem[\protect\citeauthoryear{{Lattimer} \& {Yahil}}{{Lattimer} \&
  {Yahil}}{1989}]{ly89}
{Lattimer} J.~M.,  {Yahil} A.,  1989, \apj, 340, 426

\bibitem[\protect\citeauthoryear{{Lazarus}, {Tauris}, {Knispel}, {Freire},
  {Deneva}, {Kaspi}, {Allen}, {Bogdanov}, {Chatterjee}, {Stairs} \&
  {Zhu}}{{Lazarus} et~al.}{2014}]{ltk+14}
{Lazarus} P.,  {Tauris} T.~M.,  {Knispel} B.,  {Freire} P.~C.~C.,  {Deneva}
  J.~S.,  {Kaspi} V.~M.,  {Allen} B.,  {Bogdanov} S.,  {Chatterjee} S.,
  {Stairs} I.~H.,    {Zhu} W.~W.,  2014, \mnras, 437, 1485

\bibitem[\protect\citeauthoryear{{Livio} \& {Soker}}{{Livio} \&
  {Soker}}{1988}]{ls88}
{Livio} M.,  {Soker} N.,  1988, \apj, 329, 764

\bibitem[\protect\citeauthoryear{{Lyne}, {Camilo}, {Manchester}, {Bell},
  {Kaspi}, {D'Amico}, {McKay}, {Crawford}, {Morris}, {Sheppard} \&
  {Stairs}}{{Lyne} et~al.}{2000}]{lcm+00}
{Lyne} A.~G.,  {Camilo} F.,  {Manchester} R.~N.,  {Bell} J.~F.,  {Kaspi} V.~M.,
   {D'Amico} N.,  {McKay} N.~P.~F.,  {Crawford} F.,  {Morris} D.~J.,
  {Sheppard} D.~C.,    {Stairs} I.~H.,  2000, \mnras, 312, 698

\bibitem[\protect\citeauthoryear{Lyne \& Lorimer}{Lyne \& Lorimer}{1994}]{ll94}
Lyne A.~G.,  Lorimer D.~R.,  1994, \nat, 369, 127

\bibitem[\protect\citeauthoryear{{MacLeod} \& {Ramirez-Ruiz}}{{MacLeod} \&
  {Ramirez-Ruiz}}{2015a}]{mr15a}
{MacLeod} M.,  {Ramirez-Ruiz} E.,  2015a, \apj, 803, 41

\bibitem[\protect\citeauthoryear{{MacLeod} \& {Ramirez-Ruiz}}{{MacLeod} \&
  {Ramirez-Ruiz}}{2015b}]{mr15b}
{MacLeod} M.,  {Ramirez-Ruiz} E.,  2015b, \apjl, 798, L19

\bibitem[\protect\citeauthoryear{{Marek} \& {Janka}}{{Marek} \&
  {Janka}}{2009}]{mj09}
{Marek} A.,  {Janka} H.-T.,  2009, \apj, 694, 664

\bibitem[\protect\citeauthoryear{{Martinez}, {}, {} \& {et al.}}{{Martinez}
  et~al.}{2015}]{mar15}
{Martinez} J.~G.,  {} {}   {et al.} .,  2015, ApJL, submitted

\bibitem[\protect\citeauthoryear{{Metzger}}{{Metzger}}{2012}]{met12}
{Metzger} B.~D.,  2012, \mnras, 419, 827

\bibitem[\protect\citeauthoryear{{Narayan} \& {Yi}}{{Narayan} \&
  {Yi}}{1995}]{ny95}
{Narayan} R.,  {Yi} I.,  1995, \apj, 444, 231

\bibitem[\protect\citeauthoryear{{Nomoto}}{{Nomoto}}{1987}]{nom87}
{Nomoto} K.,  1987, \apj, 322, 206

\bibitem[\protect\citeauthoryear{{Nomoto}, {Yamaoka}, {Pols}, {van den Heuvel},
  {Iwamoto}, {Kumagai} \& {Shigeyama}}{{Nomoto} et~al.}{1994}]{nyp+94}
{Nomoto} K.,  {Yamaoka} H.,  {Pols} O.~R.,  {van den Heuvel} E.~P.~J.,
  {Iwamoto} K.,  {Kumagai} S.,    {Shigeyama} T.,  1994, \nat, 371, 227

\bibitem[\protect\citeauthoryear{{Nugis} \& {Lamers}}{{Nugis} \&
  {Lamers}}{2000}]{nl00}
{Nugis} T.,  {Lamers} H.~J.~G.~L.~M.,  2000, \aap, 360, 227

\bibitem[\protect\citeauthoryear{{{\"O}zel}, {Psaltis}, {Narayan} \& {Santos
  Villarreal}}{{{\"O}zel} et~al.}{2012}]{opns12}
{{\"O}zel} F.,  {Psaltis} D.,  {Narayan} R.,    {Santos Villarreal} A.,  2012,
  \apj, 757, 55

\bibitem[\protect\citeauthoryear{{Paczynski}}{{Paczynski}}{1992}]{pac92}
{Paczynski} B.,  1992, \actaa, 42, 145

\bibitem[\protect\citeauthoryear{{Paschalidis}, {Liu}, {Etienne} \&
  {Shapiro}}{{Paschalidis} et~al.}{2011}]{ples11}
{Paschalidis} V.,  {Liu} Y.~T.,  {Etienne} Z.,    {Shapiro} S.~L.,  2011, \prd,
  84, 104032

\bibitem[\protect\citeauthoryear{{Perets}, {Gal-Yam}, {Mazzali} \& {et
  al.}}{{Perets} et~al.}{2010}]{pgm+10}
{Perets} H.~B.,  {Gal-Yam} A.,  {Mazzali} P.~A.,    {et al.} 2010, \nat, 465,
  322

\bibitem[\protect\citeauthoryear{{Peters}}{{Peters}}{1964}]{pet64}
{Peters} P.~C.,  1964, Physical Review, 136, 1224

\bibitem[\protect\citeauthoryear{{Pfahl}, {Rappaport}, {Podsiadlowski} \&
  {Spruit}}{{Pfahl} et~al.}{2002}]{prps02}
{Pfahl} E.,  {Rappaport} S.,  {Podsiadlowski} P.,    {Spruit} H.,  2002, \apj,
  574, 364

\bibitem[\protect\citeauthoryear{{Piran} \& {Shaviv}}{{Piran} \&
  {Shaviv}}{2004}]{ps04}
{Piran} T.,  {Shaviv} N.~J.,  2004, ArXiv astro-ph:0401553

\bibitem[\protect\citeauthoryear{{Piran} \& {Shaviv}}{{Piran} \&
  {Shaviv}}{2005}]{ps05}
{Piran} T.,  {Shaviv} N.~J.,  2005, Physical Review Letters, 94, 051102

\bibitem[\protect\citeauthoryear{{Podsiadlowski}}{{Podsiadlowski}}{2001}]{pod01}
{Podsiadlowski} P.,  2001, in {P.~Podsiadlowski, S.~Rappaport, A.~R.~King,
  F.~D'Antona, \& L.~Burderi } ed., Evolution of Binary and Multiple Star
  Systems Vol.~229 of Astronomical Society of the Pacific Conference Series,
  {Common-Envelope Evolution and Stellar Mergers}.
pp 239--+

\bibitem[\protect\citeauthoryear{{Podsiadlowski}, {Dewi}, {Lesaffre}, {Miller},
  {Newton} \& {Stone}}{{Podsiadlowski} et~al.}{2005}]{pdl+05}
{Podsiadlowski} P.,  {Dewi} J.~D.~M.,  {Lesaffre} P.,  {Miller} J.~C.,
  {Newton} W.~G.,    {Stone} J.~R.,  2005, \mnras, 361, 1243

\bibitem[\protect\citeauthoryear{{Podsiadlowski}, {Joss} \&
  {Hsu}}{{Podsiadlowski} et~al.}{1992}]{pjh92}
{Podsiadlowski} P.,  {Joss} P.~C.,    {Hsu} J.~J.~L.,  1992, \apj, 391, 246

\bibitem[\protect\citeauthoryear{{Podsiadlowski}, {Langer}, {Poelarends},
  {Rappaport}, {Heger} \& {Pfahl}}{{Podsiadlowski} et~al.}{2004}]{plp+04}
{Podsiadlowski} P.,  {Langer} N.,  {Poelarends} A.~J.~T.,  {Rappaport} S.,
  {Heger} A.,    {Pfahl} E.,  2004, \apj, 612, 1044

\bibitem[\protect\citeauthoryear{{Poelarends}, {Herwig}, {Langer} \&
  {Heger}}{{Poelarends} et~al.}{2008}]{phlh08}
{Poelarends} A.~J.~T.,  {Herwig} F.,  {Langer} N.,    {Heger} A.,  2008, \apj,
  675, 614

\bibitem[\protect\citeauthoryear{{Pringle} \& {Rees}}{{Pringle} \&
  {Rees}}{1972}]{pr72}
{Pringle} J.~E.,  {Rees} M.~J.,  1972, \aap, 21, 1

\bibitem[\protect\citeauthoryear{{Ritter}}{{Ritter}}{1988}]{rit88}
{Ritter} H.,  1988, \aap, 202, 93

\bibitem[\protect\citeauthoryear{{Sana}, {de Mink}, {de Koter}, {Langer},
  {Evans}, {Gieles}, {Gosset}, {Izzard}, {Le Bouquin} \& {Schneider}}{{Sana}
  et~al.}{2012}]{sdd+12}
{Sana} H.,  {de Mink} S.~E.,  {de Koter} A.,  {Langer} N.,  {Evans} C.~J.,
  {Gieles} M.,  {Gosset} E.,  {Izzard} R.~G.,  {Le Bouquin} J.-B.,
  {Schneider} F.~R.~N.,  2012, Science, 337, 444

\bibitem[\protect\citeauthoryear{{Sauer}, {Mazzali}, {Deng}, {Valenti},
  {Nomoto} \& {Filippenko}}{{Sauer} et~al.}{2006}]{smd+06}
{Sauer} D.~N.,  {Mazzali} P.~A.,  {Deng} J.,  {Valenti} S.,  {Nomoto} K.,
  {Filippenko} A.~V.,  2006, \mnras, 369, 1939

\bibitem[\protect\citeauthoryear{{Savonije} \& {Takens}}{{Savonije} \&
  {Takens}}{1976}]{st76}
{Savonije} G.~J.,  {Takens} R.~J.,  1976, \aap, 47, 231

\bibitem[\protect\citeauthoryear{{Scheck}, {Kifonidis}, {Janka} \&
  {M{\"u}ller}}{{Scheck} et~al.}{2006}]{skjm06}
{Scheck} L.,  {Kifonidis} K.,  {Janka} H.-T.,    {M{\"u}ller} E.,  2006, \aap,
  457, 963

\bibitem[\protect\citeauthoryear{{Shima}, {Matsuda}, {Takeda} \&
  {Sawada}}{{Shima} et~al.}{1985}]{smts85}
{Shima} E.,  {Matsuda} T.,  {Takeda} H.,    {Sawada} K.,  1985, \mnras, 217,
  367

\bibitem[\protect\citeauthoryear{{Stokes}, {Taylor} \& {Dewey}}{{Stokes}
  et~al.}{1985}]{std85}
{Stokes} G.~H.,  {Taylor} J.~H.,    {Dewey} R.~J.,  1985, \apjl, 294, L21

\bibitem[\protect\citeauthoryear{{Swiggum}, {Rosen}, {McLaughlin} \&
  {et~al.}}{{Swiggum} et~al.}{2015}]{srm+15}
{Swiggum} J.,  {Rosen} R.,  {McLaughlin} M.,    {et~al.} 2015, ArXiv e-prints,
  astro-ph:1503.06276

\bibitem[\protect\citeauthoryear{{Takahashi}, {Yoshida} \& {Umeda}}{{Takahashi}
  et~al.}{2013}]{tyu13}
{Takahashi} K.,  {Yoshida} T.,    {Umeda} H.,  2013, \apj, 771, 28

\bibitem[\protect\citeauthoryear{{Tauris}}{{Tauris}}{1996}]{tau96}
{Tauris} T.~M.,  1996, \aap, 315, 453

\bibitem[\protect\citeauthoryear{{Tauris}, {Langer} \& {Kramer}}{{Tauris}
  et~al.}{2011}]{tlk11}
{Tauris} T.~M.,  {Langer} N.,    {Kramer} M.,  2011, \mnras, 416, 2130

\bibitem[\protect\citeauthoryear{{Tauris}, {Langer} \& {Kramer}}{{Tauris}
  et~al.}{2012}]{tlk12}
{Tauris} T.~M.,  {Langer} N.,    {Kramer} M.,  2012, \mnras, 425, 1601

\bibitem[\protect\citeauthoryear{{Tauris}, {Langer}, {Moriya}, {Podsiadlowski},
  {Yoon} \& {Blinnikov}}{{Tauris} et~al.}{2013}]{tlm+13}
{Tauris} T.~M.,  {Langer} N.,  {Moriya} T.~J.,  {Podsiadlowski} P.,  {Yoon}
  S.-C.,    {Blinnikov} S.~I.,  2013, \apjl, 778, L23

\bibitem[\protect\citeauthoryear{{Tauris} \& {Sennels}}{{Tauris} \&
  {Sennels}}{2000}]{ts00}
{Tauris} T.~M.,  {Sennels} T.,  2000, \aap, 355, 236

\bibitem[\protect\citeauthoryear{{Tauris} \& {Takens}}{{Tauris} \&
  {Takens}}{1998}]{tt98}
{Tauris} T.~M.,  {Takens} R.~J.,  1998, \aap, 330, 1047

\bibitem[\protect\citeauthoryear{{Tauris} \& {van den Heuvel}}{{Tauris} \& {van
  den Heuvel}}{2006}]{tv06}
{Tauris} T.~M.,  {van den Heuvel} E.~P.~J.,  2006, {Formation and evolution of
  compact stellar X-ray sources}.
Cambridge University Press, pp 623--665

\bibitem[\protect\citeauthoryear{{Timmes}, {Woosley} \& {Taam}}{{Timmes}
  et~al.}{1994}]{twt94}
{Timmes} F.~X.,  {Woosley} S.~E.,    {Taam} R.~E.,  1994, \apj, 420, 348

\bibitem[\protect\citeauthoryear{{Timmes}, {Woosley} \& {Weaver}}{{Timmes}
  et~al.}{1996}]{tww96}
{Timmes} F.~X.,  {Woosley} S.~E.,    {Weaver} T.~A.,  1996, \apj, 457, 834

\bibitem[\protect\citeauthoryear{{Tout} \& {Eggleton}}{{Tout} \&
  {Eggleton}}{1988}]{te88}
{Tout} C.~A.,  {Eggleton} P.~P.,  1988, \apj, 334, 357

\bibitem[\protect\citeauthoryear{{Umeda}, {Yoshida} \& {Takahashi}}{{Umeda}
  et~al.}{2012}]{uyt12}
{Umeda} H.,  {Yoshida} T.,    {Takahashi} K.,  2012, Prog. Theor. Exp. Phys.,
  DOI: 10.1093/ptep/pts017

\bibitem[\protect\citeauthoryear{{van den Heuvel}}{{van den
  Heuvel}}{1994}]{vdh94a}
{van den Heuvel} E.~P.~J.,  1994, in {Shore} S.~N.,  {Livio} M.,  {van den
  Heuvel} E.~P.~J.,  {Nussbaumer} H.,   {Orr} A.,  eds, Saas-Fee Advanced
  Course 22: Interacting Binaries {Interacting binaries: topics in close binary
  evolution.}.
pp 263--474

\bibitem[\protect\citeauthoryear{{van Haaften}, {Nelemans}, {Voss}, {Wood} \&
  {Kuijpers}}{{van Haaften} et~al.}{2012}]{vnv+12}
{van Haaften} L.~M.,  {Nelemans} G.,  {Voss} R.,  {Wood} M.~A.,    {Kuijpers}
  J.,  2012, \aap, 537, A104

\bibitem[\protect\citeauthoryear{{van Kerkwijk}, {Charles}, {Geballe}, {King},
  {Miley}, {Molnar}, {van den Heuvel}, {van der Klis} \& {van Paradijs}}{{van
  Kerkwijk} et~al.}{1992}]{vcg+92}
{van Kerkwijk} M.~H.,  {Charles} P.~A.,  {Geballe} T.~R.,  {King} D.~L.,
  {Miley} G.~K.,  {Molnar} L.~A.,  {van den Heuvel} E.~P.~J.,  {van der Klis}
  M.,    {van Paradijs} J.,  1992, \nat, 355, 703

\bibitem[\protect\citeauthoryear{{van Kerkwijk} \& {Kulkarni}}{{van Kerkwijk}
  \& {Kulkarni}}{1999}]{kk99}
{van Kerkwijk} M.~H.,  {Kulkarni} S.~R.,  1999, \apjl, 516, L25

\bibitem[\protect\citeauthoryear{{Voss} \& {Tauris}}{{Voss} \&
  {Tauris}}{2003}]{vt03}
{Voss} R.,  {Tauris} T.~M.,  2003, \mnras, 342, 1169

\bibitem[\protect\citeauthoryear{{Wellstein} \& {Langer}}{{Wellstein} \&
  {Langer}}{1999}]{wl99}
{Wellstein} S.,  {Langer} N.,  1999, \aap, 350, 148

\bibitem[\protect\citeauthoryear{{Wellstein}, {Langer} \& {Braun}}{{Wellstein}
  et~al.}{2001}]{wlb01}
{Wellstein} S.,  {Langer} N.,    {Braun} H.,  2001, \aap, 369, 939

\bibitem[\protect\citeauthoryear{{Wex}, {Kalogera} \& {Kramer}}{{Wex}
  et~al.}{2000}]{wkk00}
{Wex} N.,  {Kalogera} V.,    {Kramer} M.,  2000, \apj, 528, 401

\bibitem[\protect\citeauthoryear{{Winget} \& {Kepler}}{{Winget} \&
  {Kepler}}{2008}]{wk08}
{Winget} D.~E.,  {Kepler} S.~O.,  2008, \araa, 46, 157

\bibitem[\protect\citeauthoryear{{Wongwathanarat}, {Janka} \&
  {M{\"u}ller}}{{Wongwathanarat} et~al.}{2013}]{wjm13}
{Wongwathanarat} A.,  {Janka} H.-T.,    {M{\"u}ller} E.,  2013, \aap, 552, A126

\bibitem[\protect\citeauthoryear{{Woosley}, {Heger} \& {Weaver}}{{Woosley}
  et~al.}{2002}]{whw02}
{Woosley} S.~E.,  {Heger} A.,    {Weaver} T.~A.,  2002, Reviews of Modern
  Physics, 74, 1015

\bibitem[\protect\citeauthoryear{{Woosley} \& {Weaver}}{{Woosley} \&
  {Weaver}}{1995}]{ww95}
{Woosley} S.~E.,  {Weaver} T.~A.,  1995, \apjs, 101, 181

\bibitem[\protect\citeauthoryear{{Yoon}, {Woosley} \& {Langer}}{{Yoon}
  et~al.}{2010}]{ywl10}
{Yoon} S.,  {Woosley} S.~E.,    {Langer} N.,  2010, \apj, 725, 940

\bibitem[\protect\citeauthoryear{{Zdziarski}, {Miko{\l}ajewska} \&
  {Belczy{\'n}ski}}{{Zdziarski} et~al.}{2013}]{zmb13}
{Zdziarski} A.~A.,  {Miko{\l}ajewska} J.,    {Belczy{\'n}ski} K.,  2013,
  \mnras, 429, L104

\end{thebibliography}

\newpage
\appendix
\section{Timescale of neutron star spiral-in caused by a helium flash}\label{appendix}
For the purpose of determining the fate of binaries in which the donor star experiences a vigorous helium shell flash
after (or close to) oxygen burning has been ignited, we briefly outline an analytical investigation to demonstrate
that the timescale of spiral-in is much longer than remaining lifetime of the pre-SN star.

As an example, we shall analyze the situation for the system with initial parameters: $M_{\rm He,i}=2.9\;M_{\odot}$,
$P_{\rm orb,i}=0.1\;{\rm days}$ and $M_{\rm NS}=1.35\;M_{\odot}$. This systems was studied in detail in Paper~I \citep{tlm+13}.
The vigorous shell flash gave rise to numerical problems for our stellar code and we alleviated this problem by
evolving the donor star as an isolated star during the last few years of evolution\footnote{Note, there is a typo 
in the second paragraph of Section~2 in Paper~I. The off-centre oxygen burning was (re)ignited at $t=1.854\,353\;{\rm Myr}$ 
(not $t=1.854\,553\;{\rm Myr}$), i.e. only some 16~yr in evolutionary time after we took the star out of the
binary and evolved it as an isolated star to investigate its expansion due to the flash. When $t$ had increased by an additional few years (3--4~yr) 
the computations finally broke down at $t=1.854\,356\;{\rm Myr}$.}.
Using this method, we could follow the expansion of the donor star during the flash. 
The donor star expanded from $R=0.40$ to $2.41\;R_{\odot}$, which would engulf the NS inside the dilute helium-rich envelope
of the donor. As we briefly argued in Paper~I, this star would undergo an Fe~CCSN within only 10~yr and thus we could neglect 
the spiral-in caused by the frictional drag acting on the NS prior to the explosion. Here we elaborate a bit more on this argument and follow the
method suggested by \citet{dp03} to evaluate the timescale of the spiral-in of the NS.

The loss rate of orbital energy, $|\dot{E}_{\rm orb}|$ as a result of the drag force, $F_{\rm drag}$ acting on the NS 
inside the dilute envelope of the evolved helium star is roughly given by \citep{bh44,smts85}:
\begin{equation}\label{eq:Eorb_dot1}
  |\dot{E}_{\rm orb}| = F_{\rm drag}\,v = A\,P_{\rm drag}\,v = \xi(\mu)\,\pi R_{\rm acc}^2\,\rho\,v^3 ,
\end{equation}
where the NS accretion radius is:
\begin{equation}
  R_{\rm acc}=\frac{2GM_{\rm NS}}{v^2+v_s^2}.
\end{equation}
Here $G$ is the gravitational constant, $v$ is the relative velocity between the NS and the CE, $v_s=\sqrt{\gamma \,P/\rho}$
is the speed of sound, $\gamma$ being the adiabatic index (5/3 for a monatomic gas), $P$ and $\rho$ are the local pressure and mass density
provided by our stellar models, cf. Fig.~A1. 
$\xi (\mu)$ is a scaling factor determining the dissipation rate of $E_{\rm orb}$ due to the drag. 
Here we chose $\xi (\mu)=2.5$ for the supersonic case (in our models $v$ is larger than $v_s$ by a factor of $3-10$).   
Similarly, the rate at which the orbital energy, $E_{\rm orb}= -GM_{\rm He}M_{\rm NS}/(2a)$ is dissipated is given by:
\begin{equation}\label{eq:Eorb_dot2}
  |\dot{E}_{\rm orb}| = - \frac{G\,M_{\rm He}\,M_{\rm NS}}{2\,a^2}\;\dot{a},  
\end{equation}
where $a$ is the orbital separation between the NS and the centre of the evolved helium star. 
Assuming Keplerian motion for the inspiralling NS (and neglecting for a moment corotation of the CE):  
$v(a)=\sqrt{G\,(M_{\rm He}+M_{\rm NS})/a}$, and this
enables us to equate the expressions in equations~(\ref{eq:Eorb_dot1}) and (\ref{eq:Eorb_dot2}). 

Integration then yields the timescale of the spiral-in, for example, from the current location of the NS, $a_0$ until it reaches the metal core boundary
of the evolved helium star at radius, $R_{\rm core}$:
\begin{equation}
  t_{\rm decay} = \int_{a_0}^{R_{\rm core}} -\frac{G\,M_{\rm He}\,M_{\rm NS}}
                  {2\,a^2\;\xi(\mu)\,\pi\,R_{\rm acc}(a)^2\;\rho(a)\,v(a)^3}\;da .
\end{equation}
In Fig.~A2, we have plotted $t_{\rm decay}$ for the system discussed in Paper~I. The blue line corresponds to $R_{\rm core}=0$ 
(i.e. the spiral-in timescale for complete merger to
the centre of the star). The red line (similar to the blue line for the majority of the computation) 
shows the timescale for spiral-in until reaching the boundary of the metal core, $R_{\rm core}=0.024\;R_{\odot}$.
The orange line simply shows a local characteristic orbital decay timescale defined as: 
$\tau \equiv E_{\rm orb}/\dot{E}_{\rm orb}$.
One can see that for the present location of the NS, $\tau \approx 3500\;{\rm yr}$ ($3.8\times 10^{48}\;{\rm erg}/3.2\times 10^{37}\;{\rm erg\,s^{-1}}$).
However, $\tau$ is underestimating the rate of spiral-in, because the density increases inward, and therefore
$\tau$ is an overestimate of the more realistic orbital decay time. This is seen in comparison with the blue line which yields 
$t_{\rm decay}=600\;{\rm yr}$ for the location of the NS in our final model ($a_{\rm f}=1.02\;R_{\odot}$ and $P_{\rm orb,f}=0.070\;{\rm days}$). 

During the helium shell flash it is only the outermost part of the star which expands. 
In the specific model discussed here it is the outer $2.58\times 10^{-3}\;M_{\odot}$ of material. This outer envelope is expanded by a factor~10 in radius during the flash 
(reaching $2.41\;R_{\odot}$) and explains the relatively weak drag on the NS due to the dilute gas. 

To conclude, it is safe to assume that $t_{\rm decay} \gg t_{\rm collapse}$, where $t_{\rm collapse}$ is the
remaining lifetime of the evolved helium star until the SN explosion. Therefore, our calculated models will lead to
ultra-stripped stars exploding without the NS companion being part of their core region\footnote{In case the NS would have merged with the core
of the pre-SN star it would presumably undergo hypercritical accretion, leading to formation of a black hole and possibly
a $\gamma$-burst like event \citep{da06}.}. This conclusion is further strengthened if one includes  
the deposition of orbital angular momentum into the envelope which would cause it to corotate and thus
reduce the drag force on the NS \citep{ls88}, leading to larger values of $t_{\rm decay}$. 
In addition, one must bear in mind that in the example demonstrated here, we considered a very conservative case with $t_{\rm collapse}\simeq 10\;{\rm yr}$
(which is still small compared to $t_{\rm decay}=600\;{\rm yr}$).
However, in other models we even have $t_{\rm collapse}\sim$ a~few~days, cf. Section~\ref{subsec:FeCCSN}.

Even in case the NS would experience a slight drag from spiral-in, the released orbital binding energy in this tight system is more than sufficient
to effectively remove the loosely bound envelope ($\Delta E_{\rm orb} \gg E_{\rm bind}$, by more than two orders of magnitude).  

\begin{figure}
\centering
\includegraphics[width=0.72\columnwidth,angle=-90]{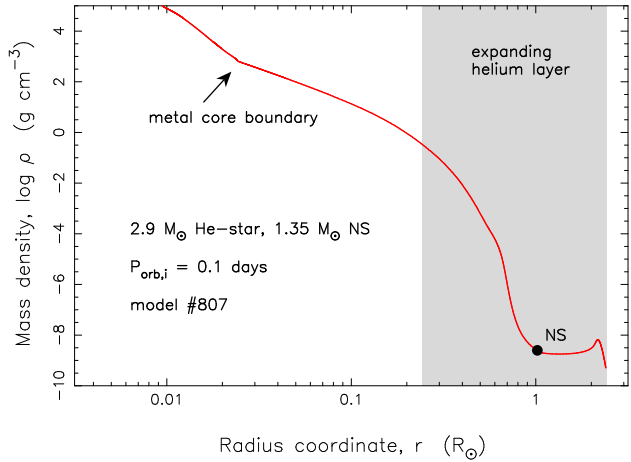}\label{fig:appendix_rho}
\caption{The mass-density profile of the evolved $2.9\;M_{\odot}$ helium star ($P_{\rm orb,i}=0.1\;{\rm days}$ and $M_{\rm NS}=1.35\;M_{\odot}$) 
investigated in \citet{tlm+13}, see also Table~\ref{table:models} for
further information on this system. The total mass of the expanding outer envelope is $2.58\times 10^{-3}\;M_{\odot}$ and its average
mass density is $2.6\times 10^{-4}\;{\rm g}\,{\rm cm}^{-3}$. The location of the NS is shown.}
\end{figure}
\begin{figure}
\centering
\includegraphics[width=0.72\columnwidth,angle=-90]{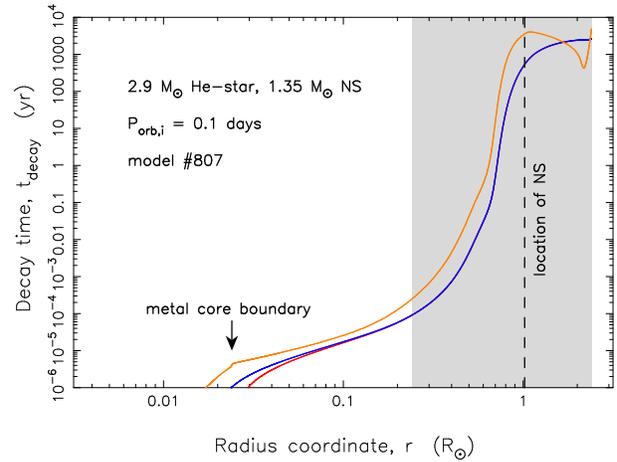}\label{fig:appendix_t_decay}
\caption{The orbital decay time, $t_{\rm decay}$ as a function radius coordinate. Blue curve: $t_{\rm decay}$ is calculated assuming that 
the NS spirals in all the way to the centre ($r=0$) of the evolved helium star. Red curve: NS in-spiral until the
boundary of the metal core ($r=0.024\;R_{\odot}$). Orange curve: the local characteristic orbital decay timescale, $\tau \equiv E_{\rm orb}/\dot{E}_{\rm orb}$. See text.}
\end{figure}

\end{document}